\documentclass[11pt]{article}
\usepackage{xspace,amsmath,amssymb}
\usepackage{color}
\usepackage{mathrsfs}
\usepackage[colorlinks=true,linkcolor=blue,filecolor=blue,citecolor=blue,urlcolor=blue]{hyperref}
\newcommand{\namedref}[2]{\hyperref[#2]{#1~\ref*{#2}}}
\renewcommand{\eqref}[1]{\hyperref[#1]{~(\ref*{#1})}}
\usepackage{chicagor}

\newtheorem{theorem}{Theorem}[section]
\newtheorem{lemma}[theorem]{Lemma}

\newcommand{\commentout}[1]{}
\newcommand{\IN}{\mbox{$I\!\!N$}}
\newcommand{\IR}{\mbox{$I\!\!R$}}

\newcounter{deff}
\newcommand{\authnote}[2]{{ \textbf{ [#1's Note:} {\em #2} \textbf{]} }}

\newcommand{\Rnote}[1]{{\authnote{Rafael}{#1}}}
\newcommand{\Jnote}[1]{{\authnote{Joe}{#1}}}

\newcommand{\bit}{\ensuremath{\{0,1\}}}

\newcommand{\from}{\stackrel{{\scriptscriptstyle R}}{\leftarrow}}

\newenvironment{sketch}{\noindent\emph{Proof Sketch:}}{\qed}
\newenvironment{proof}{\noindent\emph{Proof:}}{$\quad \Box$}

\newcommand{\hide}[1]{}
\newcommand{\para}[1]{\bigskip \noindent {\em #1} }
\def\qed{\vrule height7pt width4pt depth1pt\par}

\newenvironment{gproof}{\noindent{\bf Proof Sketch:~~}}{\qed}
\newcommand{\BPF}{\begin{gproof}} \newcommand {\EPF}{\end{gproof}}
\newenvironment{fproof}{\noindent{\bf Proof:~~}}{\qed}
\newcommand{\BPRF}{\begin{fproof}} \newcommand {\EPRF}{\end{fproof}}
\newenvironment{hproof}{\noindent{\bf Proof~}}{\qed}
\newcommand{\BPR}{\begin{hproof}} \newcommand {\EPR}{\end{hproof}}

\newcommand{\BI}{\begin{itemize}}
\newcommand{\EI}{\end{itemize}}
\newcommand{\BE}{\begin{enumerate}}
\newcommand{\EE}{\end{enumerate}}

\newtheorem{thm}{Theorem}[section]      %
\newcommand{\BT}{\begin{thm}}   \newcommand{\ET}{\end{thm}}
\newtheorem{dfn}[thm]{Definition}      %
\newcommand{\BD}{\begin{dfn}}   \newcommand{\ED}{\end{dfn}}
\newtheorem{corr}[theorem]{Corollary}      %
\newcommand{\BCR}{\begin{corr}} \newcommand{\ECR}{\end{corr}}
\newtheorem{constr}[thm]{Construction}
\newcommand{\BCT}{\begin{constr}} \newcommand{\ECT}{\end{constr}}
\newtheorem{prop}[thm]{Proposition}
\newcommand{\BP}{\begin{prop}}   \newcommand{\EP}{\end{prop}}
\newtheorem{lemm}[thm]{Lemma}   %
\newcommand{\BL}{\begin{lemm}}   \newcommand{\EL}{\end{lemm}}
\newtheorem{clm}[thm]{Claim}            %
\newcommand{\BCM}{\begin{clm}}   \newcommand{\ECM}{\end{clm}}
\newtheorem{sclm}[thm]{Sub-Claim}            %
\newcommand{\BSCM}{\begin{sclm}}   \newcommand{\ESCM}{\end{sclm}}
\newtheorem{assumption}[thm]{Assumption}            %
\newcommand{\BA}{\begin{assumption}}   \newcommand{\EA}{\end{assumption}}
\newtheorem{remark}[thm]{Remark}            %
\newcommand{\BR}{\begin{remark}}   \newcommand{\ER}{\end{ER}}
\newtheorem{example}[thm]{Example}

\newcommand{\timec}{\textsc{steps}}
\newcommand{\size}{\textsc{size}}

\newcommand{\complex}{\canon}
\newcommand{\complexity}{\complex}

\newcommand{\real}{\textsc{real}}
\newcommand{\ideal}{\textsc{ideal}}

\newcommand{\M}{{\cal M}}
\newcommand{\Mcanon}{\Lambda}

\newcommand{\G}{{\cal G}}
\newcommand{\T}{{T}}

\newcommand{\C}{{\sf comm}}
\newcommand{\worstcase}{{\sf worstcase}}

\newcommand{\F}{{\cal F}}
\newcommand{\Z}{{\cal Z}}
\newcommand{\Play}{{\cal P}}
\newcommand{\Exp}{{\mathbf{E}}}

\newcommand{\bitset}{\{0,1\}}

\newcommand{\N}{\mbox{N}}

\DeclareRobustCommand*{\slashfracstyle}[1]{%
  {\ensuremath{\mbox{\fontsize\sf@size\z@\selectfont #1}}}}

\DeclareRobustCommand*{\slashfrac}[2]{\leavevmode
  \raise.5ex\hbox{\scriptsize #1}\kern-.13em/%
  \kern-.15em\lower.25ex\hbox{\scriptsize #2}}

\def\*{\Z^*}

\def\B{\{0,1\}}
\def\B*{\B^*}

\def\union{\cup}
\def\cross{\times}

\newcommand{\view}{{\sf view}}
\newcommand{\iview}{\mathit{view}}
\newcommand{\iprecise}{\mathit{precise}}

\newcommand{\canon}{\cC}

\newcommand{\pnatural}{homogeneous}
\newcommand{\cnatural}{output-invariant}

\newcommand{\gnatural}{natural}

\def\cC{\mathscr{C}}

\newcommand{\barw}{h}

\newcommand{\fullv}[1]{#1}
\newcommand{\shortv}[1]{\commentout{#1}}
\newcommand{\nshortv}[1]{\commentout{#1}}
\newcommand{\nfullv}[1]{#1}
\newcommand{\ashortv}[1]{\commentout{#1}}
\newcommand{\afullv}[1]{#1}

\def\beginsmall#1{\begin{#1}}
\def\endsmall#1{\end{#1}}

\newenvironment{RETHM}[2]{\trivlist \item[\hskip 10pt\hskip\labelsep{\bf
#1\hskip 5pt\relax\ref{#2}.}]\it}{\endtrivlist}
\newcommand{\rethm}[1]{\begin{RETHM}{Theorem}{#1}}
\newcommand{\repro}[1]{\begin{RETHM}{Proposition}{#1}}
\newcommand{\relem}[1]{\begin{RETHM}{Lemma}{#1}}
\newcommand{\recor}[1]{\begin{RETHM}{Corollary}{#1}}
\newcommand{\erethm}{\end{RETHM}}
\newcommand{\erepro}{\end{RETHM}}
\newcommand{\erelem}{\end{RETHM}}
\newcommand{\erecor}{\end{RETHM}}

\shortv{
\newcommand{\citeyear}{\cite}}
\begin{document}
\date{}
\title{Game Theory with Costly Computation}
\author{
Joseph Halpern\\
Cornell University\\
halpern@cs.cornell.edu 
\and Rafael Pass\\
Cornell University\\
rafael@cs.cornell.edu}
\date{First Draft: April 2007\\This Draft: August 2008}
\maketitle

\begin{abstract}
We develop a general game-theoretic framework for reasoning about
strategic agents performing possibly costly computation. 
In this framework, many traditional game-theoretic results (such as the
existence of a Nash equilibrium) no longer hold.  Nevertheless, we can
use the framework to
provide psychologically appealing explanations to observed behavior in
well-studied games (such as finitely repeated prisoner's dilemma and
rock-paper-scissors). 
Furthermore, we provide natural conditions on games sufficient to guarantee that 
equilibria exist.  
As an application of this framework, we
consider a notion of game-theoretic implementation of 
mediators in computational games. 
\iffalse
%
%
%
%
an entirely 
%
%
game-theoretic definition of protocol security
that takes both computation and incentives into account. 
%
\fi
We 
show
that a special case of 
this notion
is equivalent to a variant of 
the traditional cryptographic definition of protocol security;
this result 
shows that, when taking computation into account,
the two approaches used for dealing with ``deviating'' players
in two different communities---\emph{Nash equilibrium} in game theory,
and \emph{zero-knowledge ``simulation''} in cryptography---are intimately
connected.
\end{abstract}
\thispagestyle{empty}
\newpage
\pagenumbering{arabic}

\tableofcontents
\newpage

\section{Introduction}
\subsection{Game Theory with Costly Computation}
Consider the following game. You are given a random odd $n$-bit number $x$ and you are supposed
to decide whether $x$ is prime or composite. If you guess correctly you receive \$2, if you guess incorrectly you instead have to pay a penalty of $\$1000$. Additionally you have the choice of ``playing safe'' by giving up, in which case you receive $\$1$.
In traditional
game theory, computation is considered ``costless''; in other words,
players are allowed to perform an unbounded amount of computation without it affecting 
their utility. Traditional game theory suggests that you should compute
whether $x$ is prime or composite and output the correct answer; this
is the only Nash equilibrium of the one-person game, no matter what $n$
(the size of the prime) is.
Although for small $n$ this
seems reasonable, when $n$ grows larger most people would probably
decide to ``play safe''---as eventually the cost of computing the answer
(e.g., by buying powerful enough computers) outweighs the possible gain of
$\$1$. 

The importance of considering such computational issues in game theory
has been recognized since at least the  work of Simon \citeyear{Simon55}.
There have been a number of attempts to capture various aspects of
computation. Two major lines of research can be identified.
The first line, initiated by Neyman \citeyear{Ney85}, tries to model
the fact that players can do only bounded computation, typically by
modeling players as finite automata.  
(See \cite{PY94} and the references
therein for more recent work on modeling players as finite automata;
a more recent approach, first considered by Urbano and Villa
\citeyear{UV04}
and formalized by Dodis, Halevi and Rabin \citeyear{DHR00}, instead models
players as polynomially bounded Turing machine.) 
The second line, initiated
by Rubinstein \citeyear{Rub85}, tries to capture the fact that doing
costly computation affects an agent's utility. 
Rubinstein assumed that players choose a finite automaton to play the 
game rather than choosing a strategy directly; a player's utility
depends both  
on the move made by the automaton and the complexity of the automaton
(identified with the number of states of the automaton).
Intuitively, automata that use more states are seen as representing more
complicated procedures.  
(See \cite{Kalai90} for an overview of the work in
this area in the 1980s, and \cite{BKK07} for more recent work.)

%
%
%
\iffalse
Our focus is on the second approach.  
%
%
%
%
%
%
We provide
a general game-theoretic framework for reasoning
about agents performing costly computation.   All the earlier 
frameworks
work in the second line of 
research 
can be seen as a special case of our
framework.  
\fi
%
Our focus is on providing a general game-theoretic framework for reasoning
about agents performing costly computation.   
As in Rubinstein's work, we view players as choosing a machine, but for
us the machine is a Turing machine, rather than a finite automaton.  We
associate a complexity, not just with a machine, but with the machine
and its input.
The complexity 
could represent the running time of or space used by the machine on that
input.  The complexity can also be used to capture the complexity of the
machine itself (e.g., the number of states, as in Rubinstein's case) or to
model the cost of searching for a new strategy to replace one that the
player already has.  For example, if a mechanism designer recommends
that player $i$ use a particular strategy (machine) $M$, then there is a
cost for searching for a better strategy; switching to another strategy
may also entail a psychological cost.  By allowing the complexity to
depend on the machine \emph{and} the input, we can deal with the fact 
that machines run much longer on  some inputs than on others. 
A player's utility depends both
on the actions chosen by all the players' machines and the complexity of
these machines.  Note that, in general, unlike earlier papers, player
$i$'s utility may depend not just on the complexity of $i$'s machine,
but also on the complexity of the machines of other players.  For
example, it may be important to player 1 to compute an answer to a
problem before player 2.
In this setting, we can define Nash equilibrium in the obvious way. 
However, as we show by a simple example (a rock-paper-scissors game 
where randomization is costly), a Nash
equilibrium 
may not always exist.  
Other standard results in the game theory, such as the \emph{revelation
principle} (which, roughly speaking, says that there is always an
equilibrium where players truthfully report their types, i.e., their
private information \cite{Myerson79,F86}) also do not hold. 
We view this as a feature.  We believe that taking computation
into account should force us to rethink a number of basic notions.  
To this end,
we introduce refinements of Nash equilibrium that take
into account the
computational aspects of games.
We also show that the non-existence of Nash equilibrium is not such a
significant problem.  A Nash equilibrium does exist for many
computational games of interest, and can help
explain experimentally-observed phenomena in games such as
repeated prisoner's dilemma in a psychologically appealing way.
Moreover, there are natural conditions (such as the assumption that
randomizing is free) that guarantee the existence of Nash equilibrium in
computational games.
\subsection{Implementing Mediators in Computational Games}
It is often simpler to design, and analyze, mechanisms when assuming
that players have access to a trusted mediator through which they can
communicate. 
Equally often, however, such a trusted mediator is hard to find.
A central question in both cryptography and game theory is investigating
under what circumstances mediators can be replaced---or
\emph{implemented}---by 
simple ``unmediated'' communication between the players.  
There are some significant differences between the  approaches used by
the two communities to formalize this question.

The cryptographic notion of a \emph{secure computation} \cite{GMW} 
considers two types of players: \emph{honest} players and
\emph{malicious} players.  
Roughly speaking, a protocol $\Pi$ is said to securely implement the
mediator $\F$ if (1) the malicious players cannot influence the output
of the communication phase any more than they could have by
communicating directly with the mediator; this is called
\emph{correctness}, and (2) the malicious players cannot ``learn'' more
than what can be efficiently computed from only the output of
mediator; this is called \emph{privacy}. These properties are
formalized through the \emph{zero-knowledge simulation paradigm}
\cite{GMR}: roughly, we require that any ``harm'' done by an adversary
in the protocol execution could be simulated by a
polynomially-bounded Turing machine, called the \emph{simulator}, that
communicates only with the mediator. Three levels of security are
usually considered: \emph{perfect}, \emph{statistical}, and
\emph{computational}.  Roughly
speaking, perfect security guarantees that correctness and
privacy hold with probability 1; statistical security allows for a
``negligible'' error probability; and computational security 
considers only adversaries that can be implemented by
polynomially-bounded Turing machines. 

The traditional game-theoretic notion of implementation (see
\cite{F86,F90}) does not explicitly consider properties such as privacy and
correctness, but instead requires that the implementation 
preserve a given Nash equilibrium of the mediated game. Roughly
speaking, a protocol 
$\Pi$ 
game-theoretically 
implements a mediator $\F$ if, given \emph{any} game $G$ for which 
it is an equilibrium for the players to provide their types to $\F$
and output what $\F$ recommends,
running $\Pi$ on the same inputs is also an equilibrium. In other words,
whenever a set of parties have incentives to truthfully provide their
types to the mediator, they also have incentives to honestly run the
protocol $\Pi$ using the same inputs.%
\footnote{While the definitions of implementation in the game-theory
literature (e.g., \cite{F86,F90}) do not stress the uniformity of the
implementation---that is, the fact that it works for all games---the
implementations provided are in fact uniform in this sense.}

Roughly speaking, the key differences between the notions are that the
game-theoretic notion does not consider computational issues and the
cryptographic notion does not consider incentives.  The game-theoretic
notion thus talks about preserving Nash equilibria (which cannot be done
in the cryptographic notion, since there are no incentives), while the
cryptographic notion talks about security against malicious adversaries.  

Although the cryptographic notion does not consider incentives, it is
nonetheless stronger than the game-theoretic notion.  More precisely, 
it is easily seen that perfectly-secure implementation implies 
the game-theoretic notion of implementation; that is, 
all perfectly-secure implementations are also game-theoretic
implementations.  
\commentout{
\footnote{Specifically, the cryptographic notion of perfectly-secure
implementation requires that for every player $i$ and every strategy
$\sigma_i$, there exists a ``simulator'' strategy $\sigma'_i$, such that
the outputs of all players in the following two experiments are
identically distributed: The first experiment considers a scenario where
all players except $i$ honestly execute the protocol using their correct
input; player $i$ instead runs $\sigma_i$. In the second experiment, the
players have access to a trusted party and all players except $i$ submit
their input to the trusted party and output whatever they hear back;
player $i$ instead runs $\sigma'_i$. In game-theoretic terms, this means
that every ``deviating'' strategy $\sigma_i$ in the communication game
can be mapped into a deviating strategy $\sigma'_i$ in the mediated game
with the same utility. Since no deviations in the mediated
with the same output distribution for each type (and, hence, the same
utility, since 
the utility depends only on the type and the output distribution). Since
no deviations in the mediated 
game can give higher utility than the Nash equilibrium strategy,
following the protocol in the communication game must also be a Nash
equilibrium.} 
}%
A corresponding implication holds for statistically- and
computationally-secure implementations if we consider appropriate
variants of 
game-theoretic implementation that require only that running $\Pi$ is
an $\epsilon$-Nash equilibrium, resp., a ``computational''
$\epsilon$-Nash equilibrium, where players are
restricted to using polynomially-bounded Turing machines%
; see \cite{DHR00,DR07,LMPS04}.%
\footnote{\cite{DHR00,LMPS04} consider only implementations 
of correlated equilibrium, but the same proof extends to arbitrary
mediators as well.} 
%
\iffalse
A special case of the latter result is stated informally 
by Dodis and Rabin \citeyear{DR07} (they state a ``meta-theorem'' saying
that a computational-secure correlated equilibrium can be implemented;
this result is proved in \cite{DHR00}).  A result similar in spirit is
proved by Lepinski et al.~\citeyear{LMPS04}.  For completeness, we state
%
%
%
and prove our version of the results here (see Theorems~\ref{??}; STILL
TO COME).
\fi
%
%
%
%
%
%
%
%
%
%
%
%
%
%
%
%
%
%
%

%
The converse implication does not hold.  
Since the traditional game-theoretic notion of implementation does 
not consider computational cost, it cannot take into account
issues like computational efficiency, or the computational advantages
possibly gained by using $\Pi$, issues which are critical in the
cryptographic notion.  
Our notion of computational games lets us take these issues into account.
We define a notion of implementation, called \emph{universal
implementation}, that extends the traditional game-theoretic notion of
implementation by also considering games where computation is costly. In
essence, a protocol $\Pi$ \emph{universally implements} $\F$ if, given
any game $G$ (including games where computation is costly) for which  
it is an equilibrium for the players to provide their inputs to $\F$
and output what $\F$ recommends,
running $\Pi$ on the same inputs is also an equilibrium.

Note that, like the traditional game-theoretic notion of
implementation, our notion of universal implementation requires only
that a 
Nash equilibrium in the mediated game be preserved when moving to the
unmediated game.  
Thus, our definition does not capture  many of the stronger
desiderata (such as preserving other types of
equilibria besides Nash equilibria, and ensuring that the communication game does not introduce new equilibria not present in the mediated game) considered in the recent notion of
\emph{perfect implementation} of Izmalkov, Lepinski and Micali
\citeyear{ILM08}; see 
Section~\ref{sec:security} for further discussion of this issue. 
\iffalse
lacks two important features of cryptographic definitions: 1) implementations are not immune against coalitions---i.e., a colluding set of parties might have incentives to deviate, 2) a game theoretic implementation does not necessarily provide privacy: the protocol $\pi$ might very well make it possible for a player to learn the type of one of the other players, although the player could not have \emph{efficiently} obtained this information using only the mediator.
\fi

%
%
%
Also note that although our notion of universal 
implementation does not explicitly
consider the privacy of players' inputs, it nevertheless captures privacy 
requirements.  For suppose
that, when using $\Pi$, some information about $i$'s input is revealed
to $j$.  We consider a game $G$ where a player $j$ gains some significant
utility by having this information.   In this game, $i$ will not want to
use $\Pi$.  However, universal implementation requires
that, even with the utilities in $G$, $i$ should want to use $\Pi$ if
$i$ is willing to use the mediator $\F$.  
(This argument depends on the fact that we consider games where
computation is costly; 
the fact that $j$ gains information about $i$'s input may mean that $j$
can do some computation faster with this information than without it.)
As a consequence, our definition gives 
a relatively simple (and strong) way of formalizing the security of
protocols, relying only on basic notions from game theory.
Our main result shows that, under some minimal assumptions on the
utility of 
players,
our notion of universal implementation  
is equivalent to a 
variant of
the cryptographic notion of 
\emph{precise secure computation}, 
recently introduced by Micali and Pass \citeyear{MP06}.
Roughly speaking, 
the notion of 
precise secure
computation requires that any harm done by an adversary in a protocol
execution could have been done also by a simulator, using the same
complexity distribution as the adversary. In contrast, the traditional
definition of secure computation requires only that the simulator's
complexity preserves the \emph{worst-case} complexity of the
adversary.
By imposing more restrictions on the utilities of players, we can also obtain
a game-theoretic characterization of the traditional (i.e., ``non-precise'')
notion of secure computation. 

This result 
shows that the two approaches used for dealing with ``deviating'' players
in two different communities---\emph{Nash equilibrium} in game theory,
and \emph{zero-knowledge ``simulation''} in cryptography---are intimately
connected; indeed, they are essentially equivalent in the context of
implementing mediators.
It follows immediately from our result that known
protocols (\cite{MP06,MP07,gmw87,bgw,ILM08,UC}) can be used to obtain
universal implementations. 
Moreover, 
lower bounds for the traditional
notion of 
secure computation immediately yield lower bounds for universal
implementation.

At first sight, our equivalence result might seem like a negative
result: it demonstrates that considering only rational players 
(as opposed to arbitrary malicious players, as in cryptographic
notions of implementation) 
does not
facilitate protocol design. 
We emphasize, however, that for the
equivalence to hold, we must consider implementations
universal with respect to essentially \emph{all} computational games.  
In many settings it might be reasonable to
consider implementations universal with respect to only certain
subclasses of games; in such scenarios, universal implementations may be
significantly simpler or more efficient, and may also circumvent
traditional lower bounds. 
For instance, in some settings it might be reasonable to assume that
players strictly prefer to compute less, that players do not want to 
be caught ``cheating'', or that players might not be concerned about the
privacy of part of their inputs; these different assumptions can be
captured by specifying appropriate subclasses of games
(see Section \ref{sec:utilityassumptions} for more details).
We believe that this extra generality is an important advantage of a
fully game-theoretic definition, which does not rely on the traditional
cryptographic simulation paradigm. 
\nfullv{
\subsection{Outline}
The rest of this paper is organized as follows.  In
Section~\ref{sec:framework} we describe our framework.  In
Section~\ref{sec:security}, 
we apply the framework to 
obtain a computational notion of implementations of mediators,
and in Section~\ref{sec:mainthe} we demonstrate how this notion is
related to the cryptographic notion of secure computation.
We conclude in Section~\ref{sec:conclusion} with
potential new directions of research made possible by our framework.
}%

\section{A Computational Game-Theoretic Framework}\label{sec:framework}
\subsection{Bayesian Games}\label{sec:Bayesian games}
We model costly computation using \emph{Bayesian machine games}.  To
explain our approach, we first
review the standard notion of a \emph{Bayesian game}.
A Bayesian game is a
game of incomplete information, where each player makes a single move.
The ``incomplete information'' is captured by assuming that nature makes
an initial move, and chooses for each player $i$ a \emph{type}
in some set $T_i$.  Player $i$'s type can be viewed as describing $i$'s
private information.  For ease of exposition, we assume in this paper
that the set $N$ of players is always $[m] = \{1,\ldots, m\}$, for some $m$.
If $N = [m]$, the set $T =
T_1 \times \ldots \times T_m$ is the \emph{type space}.  
As is standard, 
we assume that
there is a commonly-known 
probability distribution $\Pr$ on the type space $T$.  Each player $i$ 
must choose an action from a space $A_i$ of actions.  Let $A = A_1
\times \ldots \times A_n$
be the set of action profiles.  A Bayesian game is
characterized by the tuple $([m],T,A,\Pr, \vec{u})$, where $[m]$ is the
set of players, $T$ is the type space, $A$ is the set of joint actions,
and $\vec{u}$ is the utility function, where $u_i(\vec{t},\vec{a})$ is
player $i$'s utility (or payoff) if the type profile is $\vec{t}$ and
action profile $\vec{a}$ is played.  

In general, a player's choice of action will depend on his type.
A \emph{strategy} for player $i$ is a function from $T_i$ to
$\Delta(A_i)$ (where, as usual, we denote by $\Delta(X)$ the set of
distributions on the set $X$).  If $\sigma$ is a strategy for player $i$, $t \in T_i$ and $a \in A_i$, then  $\sigma(t)(a)$ denotes the
probability of action $a$ according to the distribution on acts
induced by $\sigma(t)$.  Given a joint strategy
$\vec{\sigma}$, we can take $u_i^{\vec{\sigma}}$ to be the random
variable on the type space $T$ defined by taking
$u_i^{\vec{\sigma}}(\vec{t}) = \sum_{\vec{a} \in A}
(\sigma_1(t_1)(a_1) \times \ldots \times \sigma_m(t_m)(a_m))u_i(\vec{t},
\vec{a})$. Player $i$'s expected utility if $\vec{\sigma}$ is
played, denoted $U_i(\vec{\sigma})$, 
is then just $\Exp_{\Pr}[u_i^{\vec{\sigma}}] 
= \sum_{\vec{t} \in T} \Pr(\vec{t})u_i^{\vec{\sigma}}(\vec{t})$.

%
\iffalse
%
%
%
We assume familiarity with \emph{Bayesian games}, that is, games of
incomplete information where each player makes a single move;
``incomplete information'' means that each of the player first receives
a \emph{type} (chosen according to some distribution) as private
%
%
information. See Appendix \ref{sec:Bayesian games} a definition. 
%
\fi

%
%
\nfullv{\subsection{Bayesian Machine Games}}
\nshortv{\paragraph{Bayesian machine games}}
In a Bayesian game, it is implicitly assumed that computing a
strategy---that is, computing what move to make given a type---is free.  
We want to take the cost of computation into account here.  
\fullv{To this end,
we consider what we call \emph{Bayesian machine games}, where we
replace strategies by \emph{machines}.}
\shortv{In a Bayesian machine game, we replace strategies by machines.}
For definiteness, we take the machines to be Turing machines, although 
the exact choice of computing formalism is not
significant for our purposes.  
\iffalse
What does matter is that we can
talk about \emph{complexity} of a machine.
, where complexity for instance
can be thought of as 
the number of steps taken by the machine and the size of 
(the representation of) the
machine, where the size can be thought of as the number of states
in the Turing machine or finite automaton, or the length of the program,
if we think of a machine as a program in a programming language.
\fi
%
\fullv{
Given a type, a strategy in a Bayesian game returns a distribution over
actions.  Similarly, given as input a type, the machine returns a
distribution over actions.  As is standard, we model the distribution by
assuming that the machine actually gets as input not only the type, but
a random string of 0s and 1s (which can be thought of as the sequence of
heads and tails), and then (deterministically) outputs an action.
}%
Just as we talk about the 
expected
utility of a strategy profile in a Bayesian
game, we can talk about the expected utility of a machine profile in a
Bayesian machine game.  However, we can no longer compute the
expected utility by just taking the expectation over the action profiles
that result from playing the game.  
A player's utility depends not only on the type profile
and action profile played by the machine, but also on the
``complexity'' of the machine 
given an input.
The complexity of a machine can represent, for example, the 
running time or space usage of the machine on that input, the size
of the
program description, or  some combination of these factors.  For
simplicity, we describe the complexity by a single number, although,
since a number of factors may be relevant, it may be more appropriate to
represent it by a tuple of numbers in some cases.  
\fullv{
(We can, of course,
always encode the tuple as a single number, 
but in that case, ``higher''
complexity is not necessarily worse.)  
}
Note that when determining player $i$'s
utility, we consider the complexity of 
all machines in the profile, not just that of $i$'s machine.  
For example, $i$ might be happy as long as his
machine takes fewer steps than $j$'s.  
We assume that nature has a type in $\{0,1\}^*$.  While there is no
need to include a type for nature in standard Bayesian games (we can
effectively incorporate nature's type into the type of the players), once
we take computation into account, we obtain a more expressive class of
games by allowing nature to have a type.
(since the complexity of computing the utility may depend on nature's type).
We assume that machines take as input strings of 0s and 1s and 
output strings of 0s and 1s.  
Thus, we assume that both types and actions can be represented as
elements of $\{0,1\}^*$.
We allow machines to randomize, so given a type as input, we actually
get a distribution over strings.  
To capture this, 
we assume that the input to a machine is not only a type, but 
also a string chosen with uniform probability from 
$\{0,1\}^\infty$ (which we can view as the outcome of an infinite
sequence of coin tosses).   
Implicit in the representation above is the assumption that machines 
terminate with probability 1, so the output is a finite
string.\footnote{For ease of presentation, our notation ignores the
possibility that a machine does not terminate. Technically, we
also need to assign a utility to inputs where this happens, 
but since it happens with probability 0, as long as this 
utility is finite, these outcomes can be ignored in our calculations.} 
We define a \emph{view} to be a pair $(t,r)$ 
of two \emph{finite} bitstrings; we think of $t$ as that part of the
type that is read, and $r$ is the string of random bits used. 
(Our definition is slightly different from the
traditional way of defining a view, in that we include only the parts of
the type and the random sequence \emph{actually} read. 
If computation is not taken into account, there is no loss in generality
in including the full type and the full random sequence, and this is
what has traditionally been done in the literature.
However, when computation is costly, this might no longer be
the case.) 
We denote by $t;r$ a string in
$\{0,1\}^*; \{0,1\}^*$ representing the view.  Note that here and
elsewhere, we use ``;'' as a special symbol that acts as a separator
between strings in $\bit^*$. 
If $v = (t;r)$, we take $M(v)$ to be the output of $M$ given input type
$t$ and random string $r\cdot 0^\infty$.

We use a \emph{complexity function} $\complex: {\bf M}\times \bitset^*
; \bitset^*
\rightarrow \IN$, where ${\bf M}$ denotes the set of Turing Machines
that terminate with probability 1, to 
describe the complexity of a machine given a view. 
Given our intuition that the only machines that can have a complexity of
0 are those that ``do nothing'', we require 
that, for all complexity functions $\complex$, $\complex(M,v) = 0$ 
for some view $v$ iff $M = \bot$ iff $\complex(M,v) = 0$ for all views $v$, 
where $\bot$ is a canonical represent of the TM that does
nothing: it does not read its input, has no state changes, and writes
nothing.  
If $t \in \{0,1\}^*$ and $r \in \{0,1\}^\infty$, we identify 
$\complex(M, t;r)$ with $\complex(M,t;r')$, where $r'$ is the finite
prefix of $r$ actually used by $M$ when running on input $t$ with random
string $r$.
\fullv{
For now, we assume that machines run in isolation, so the output and
complexity of a machine does not depend on the machine profile.  We
remove this restriction in the next section.
}

\BD [Bayesian machine game]
\label{bmg.def}
A \emph{Bayesian machine game $G$} is described by a tuple 
$([m], \M,T, \Pr,$ $\complexity_1, \ldots, \complexity_m,
u_1, \ldots, u_m)$, where 
\beginsmall{itemize}
\item $[m] = \{1, \ldots, m\}$ is the set of players;
\item $\M$ is the set of possible
machines;
\item $\T \subseteq (\{0,1\}^{*})^{m+1}$ is the set of type 
profiles, where the $(m+1)$st element in the profile corresponds to
nature's type;%
\footnote{We may want to restrict the type space to be a finite subset
of $\{0,1\}^*$ (or to restrict $\Pr$ to having finite support, as is
often done in the game-theory literature, so
that the type space is effectively finite) or to restrict the output
space to being finite, although these assumptions are not 
essential for any of our results.}
\item $\Pr$ is a distribution on $\T$;
\item $\complexity_i$ is a complexity function;
\item 
$u_i: \T \times (\{0,1\}^*)^m \times \IN^m
\rightarrow \IR$ is player $i$'s utility function.   
Intuitively, $u_i(\vec{t},\vec{a},\vec{c})$ is the
utility of player $i$ if $\vec{t}$ is the type profile, $\vec{a}$ is the
action profile (where we identify $i$'s action with $M_i$'s output), and
$\vec{c}$ is the profile of machine complexities.  
\endsmall{itemize}
\ED

We can now define the expected utility of a machine profile.
Given a Bayesian machine game $G = ([m],
\M,\Pr, T, \vec{\complexity}, \vec{u}), 
\vec{t} \in \T$,
and $\vec{M} \in \M^m$, define the random variable $u^{G,\vec{M}}_i$ on 
$\T \times (\{0,1\}^{\infty})^m$ (i.e., the space of type
profiles and 
sequences of random strings) by taking
$$
u_i^{G,\vec{M}} (\vec{t}, \vec{r}) = u_i(\vec{t},
M_1(t_1;r_1), \ldots, M_m(t_m;r_m), \complexity_1(M_1, t_1; r_1),
\ldots, \complexity_m(M_m, t_m)).  
$$
Note that there are two sources of uncertainty in computing the expected
utility: the type $t$ and realization of the random coin tosses of the
players, which is an element of $(\bit^\infty)^k$.  
Let $\Pr_U^k$ denote the uniform distribution on $(\{0,1\}^\infty)^k$.
Given an arbitrary distribution $\Pr_X$ on a space $X$, we
write $\Pr_X^{+k}$ to denote the distribution $\Pr_X \times \Pr_U^k$ 
on $X \times (\{0,1\}^\infty)^k$.  If $k$ is clear from context or not
relevant, we often omit it, writing $\Pr_U$ and $\Pr_X^+$.
Thus, given the probability $\Pr$ on $T$, the expected utility 
of player $i$ in game $G$ if $\vec{M}$ is used is the expectation of the
random variable  $u^{G,\vec{M}}_i$
with respect to the distribution $\Pr^+$ (technically, $\Pr^{+m}$):
$$U^G_i(\vec{M}) = \mathbf{E}_{\Pr^+ }[u_i^{G,\vec{M}}].$$ 
Note that this notion of utility allows an agent to prefer a machine
that runs faster to one that runs slower, even if they give the same
output, or to prefer a machine that has faster running time to one that
gives a better output.
Because we allow the utility to depend on the whole profile of
complexities, we can capture a situation where $i$ can be ``happy'' as
long as his machine runs faster than $j$'s machine.  Of course, an
important special case is where $i$'s utility depends only on his own
complexity.  All of our technical results continue to hold if we make
this restriction.

\subsection{Nash Equilibrium in Machine Games}
Given the definition of utility above, we can now define
($\epsilon$-) Nash equilibrium in the standard way.

\BD [Nash equilibrium in machine games]
Given a Bayesian machine game $G$, a machine profile $\vec{M}$, 
and $\epsilon \ge 0$, 
$M_i$ is an $\epsilon$-best response to $\vec{M}_{-i}$ 
if, for every $M_i' \in \M$, 
$$U_i^G[(M_i, \vec{M}_{-i}] \geq U_i^G[(M_i', \vec{M}_{-i}] -
\epsilon.$$
(As usual, $\vec{M}_{-i}$ denotes the tuple consisting
of all machines in $\vec{M}$ other than $M_i$.)
$\vec{M}$ is an \emph{$\epsilon$-Nash equilibrium} of $G$ if, 
for all players $i$, $M_i$ is an $\epsilon$-best response to
$\vec{M}_{-i}$. A \emph{Nash equilibrium} is a 0-Nash equilibrium.
\ED

There is an important conceptual point that must be stressed with regard
to this definition.  Because we are implicitly assuming, as is standard
in the game theory literature, that the game is common knowledge, we are
assume that the agents understand the costs associated with each Turing
machine.  That is, they do not have to do any ``exploration'' to compute
the costs.  In addition, we do not charge the players for
computing which machine is the best one to use in a given setting; we
assume that this too is known.  This model is appropriate in settings
where the players have enough experience to understand the behavior of
all the machines on all relevant inputs, either through experimentation
or theoretical analysis.   We can easily extend the model to incorporate
uncertainty, by allowing the complexity function to depend on the state
(type) of nature as well as the machine and the input; see
Example~\ref{primality.ex} for further discussion of this point.

\fullv{
One immediate advantage of taking computation into account is that we
can formalize the intuition that $\epsilon$-Nash equilibria are
reasonable, because players will not bother changing strategies for a gain
of $\epsilon$.  Intuitively, the complexity function can ``charge''
$\epsilon$ for switching strategies.  Specifically,
an $\epsilon$-Nash
equilibrium $\vec{M}$ can be converted to a Nash equilibrium by 
modifying player $i$'s complexity function to incorporate the overhead
of switching from 
$M_i$ to some 
other strategy, and having player $i$'s utility function decrease by
$\epsilon' > \epsilon$ if the switching cost is nonzero; we omit the
formal details here.  Thus, the framework lets us incorporate explicitly
the reasons that players might be willing to play an $\epsilon$-Nash
equilibrium.
}%
Although the notion of Nash equilibrium in Bayesian machine games is
defined in the same way as Nash equilibrium in standard Bayesian games,
the introduction of complexity leads to some significant differences in
their properties. 
We highlight 
a few 
of them here.  
First, note that our definition of a Nash equilibrium considers 
\emph{behavioral strategies}, which in particular might be
randomized. 
It is somewhat more common in the game-theory literature to consider
\emph{mixed strategies}, which are probability
distributions over deterministic strategies.   As long as agents have
perfect recall, 
mixed strategies and behavioral strategies are essentially equivalent
\cite{Kuhn53}. 
However, in our setting, since we might want to charge  
for the randomness used in a computation, such an equivalence 
does not necessarily hold.

Mixed strategies are needed to show that Nash equilibrium always exists
in standard Bayesian games.
As the following example shows,
since we can charge for randomization,
a Nash equilibrium may not exist in 
a Bayesian machine game,
even if we restrict our attention to games where the type space and the
output space
are finite.
\begin{example}
\label{roshambo}
 {\rm
Consider the 2-player Bayesian game of roshambo (rock-paper-scissors).
Here the type space has size 1 (the players have no private information).  
We model playing rock, paper, and scissors as playing 0, 1, and 2,
respectively.  The payoff to player 1 of the outcome $(i, j)$ is 1 if $i
= j \oplus 1$ (where $\oplus$ denotes addition mod 3), $-1$ if $j = i
\oplus 1$, and 0 if $i = j$.  Player 2's playoffs are the negative of those
of player 1; the game is a zero-sum game.  As is well known, the unique
Nash equilibrium of this game has the players randomizing uniformly
between 0, 1, and 2.  }

{\rm Now consider a machine game version of roshambo.  Suppose that we 
take the complexity of a deterministic strategy to be 1, and the
complexity of 
strategy that uses randomization to be 2, and take player $i$'s
utility to be 
his payoff in the underlying Bayesian game minus the complexity
of his 
strategy. 
Intuitively, programs involving randomization are more
complicated than those that do not randomize.  With this utility
function, it is easy to see that there is no Nash equilibrium.  For
suppose that $(M_1, M_2)$ is an equilibrium.  
If $M_1$ is uses randomization, 
then 1 can do better by playing the deterministic strategy $j\oplus 1$,
where $j$ is the action that gets the highest probability according to
$M_2$ (or is the deterministic choice of player 2 if $M_2$ does not use
randomization).  Similarly, $M_2$ cannot use randomization.  But it is well
known (and easy to check) that there is no equilibrium for roshambo with 
deterministic strategies. 
(Of course, there is nothing special about the 
costs of 1 and 2 for deterministic vs.~randomized strategies.  This
argument works as long as all three deterministic strategies have the
same cost, and it is less than that of a randomized strategy.)

Now consider the variant where we do not charge for the
first, say, 10,000 steps of computation, and after that 
there is a positive cost of computation.
It is not hard to show that, in finite
time, using coin tossing with equal likelihood of heads and tails, we
cannot exactly compute a uniform distribution over the
three choices, although we can approximate it 
closely.\footnote{Consider a probabilistic Turing machine $M$ with
running time bounded by $T$ that outputs $0$ (resp $1,2$) with
probability $1/3$. Since $M$'s running time 
is bounded by $T$, $M$ can use at most $T$ of its random bits; there thus exists some natural number $p_0$ such that $M$ outputs $0$ for $p_0$ out of the $2^T$ possible random strings it receives as input. But, since $2^T$ is not divisible by 3, this is a contradiction.}
(For example, if we toss a coin twice, playing rock if we get heads
twice, paper with heads and tails, scissors with tails and heads, and
try again with two tails, we do get a uniform distribution, except for
the small probability of nontermination.)
From this
observation it easily follows, as above, that there is no Nash
equilibrium in this game either.
As a corollary, it follows that that there are computational
games without a Nash
equilibrium where all constant-time strategies are taken to be free.
It is well known that people have difficulty simulating randomization;
we can think of the cost for
randomizing as capturing this difficulty.  Interestingly, there are
roshambo tournaments (indeed, even a Rock Paper Scissors World
Championship), and books written on roshambo strategies 
\cite{Walker04}.
Championship players are clearly
not randomizing uniformly (they could not hope to get a higher payoff
than an opponent by randomizing).  Our framework provides a
psychologically plausible account of this lack of randomization.
The key point here is that,
in standard Bayesian games, to guarantee equilibrium requires using
randomization.  Here, we allow randomization, but we charge for it.
This charge may prevent there from being an equilibrium.
\qed 
}
\end{example}

Example~\ref{roshambo} shows that sometimes there is no Nash
equilibrium.  It is also trivial to show that given any standard
Bayesian game $G$ (without computational costs) and a \emph{computable}
strategy profile  $\vec{\sigma}$ in $G$ (where $\vec{\sigma}$ is
computable if, for each player $i$ and
type $t$ of player $i$, there exists a Turing machine $M$ 
that outputs $a$ with probability $\sigma_i(t)(a)$),
 we can choose computational costs and modify the utility
function in $G$ in such a way as to make $\vec{\sigma}$ an equilibrium
of the modified game:  
we simply make the cost to
player $i$ of implementing a strategy other than $\sigma_i$ sufficiently
high.

One might be tempted to conclude from these examples that Bayesian
machine games are uninteresting.  They are not useful prescriptively,
since they do not always provide a prescription as to the right thing to
do.  Nor are they useful descriptively, since we can always ``explain''
the use of a particular strategy in our framework as the result of a
high cost of switching to another strategy.  With regard to the first
point, 
as shown by our definition of security, our framework can provide useful
prescriptive 
insights by making minimal assumptions regarding the form of the 
complexity
function.  
Moreover, although there may not always be a Nash equilibrium, 
it is easy to see that 
there is always an $\epsilon$-Nash equilibrium for some $\epsilon$;
this $\epsilon$-Nash can give useful
guidance into how the game should be played.  For example, in the second
variant of 
the roshambo example above, we can get an $\epsilon$-equilibrium
for a small $\epsilon$ by attempting to simulate the uniform
distribution by tossing the coin 10,000 times, and then just playing
rock if 10,000 tails are tossed.  Whether it is worth continuing the
simulation after 10,000 tails depends on the cost of the additional
computation.
Finally, as we show below, there are natural classes of games where a
Nash equilibrium is guaranteed to exist (see Section~\ref{sec:suff}).
With regard to the second point, we
would argue that, in fact, it is the case that people continue to play
certain strategies that they know are not optimal because of the
overhead of switching to a different strategy; that is, our model
captures a real-world phenomenon.  
A third property of (standard) Bayesian games that does not hold for
Bayesian machine games is the following.
Given a Nash equilibrium $\vec{\sigma}$ in a Bayesian game,
not only is $\sigma_i$ 
a best response to $\vec{\sigma}_{-i}$, 
but it continues to be a best response conditional on $i$'s type.
That is, if $\Pr$ is the probability distribution
on types, and $\Pr(t_i) > 0$,
then $U_i(\sigma_i,\vec{\sigma}_{-i} \mid t_i) \ge U_i(\sigma'_i,
\vec{\sigma}_{-i} \mid t_i)$ for all strategies $\sigma'_i$ for player $i$, where 
$U_i(\sigma_i,\vec{\sigma}_{-i} \mid t_i)$ is the expected utility of 
$\vec{\sigma}$ conditional on player $i$ having type $i$.  Intuitively,
if player $i$ could do better than playing $\sigma_i$ if his type were
$t_i$ by playing $\sigma'_i$, then $\sigma_i$ would not be a best
response to $\vec{\sigma}_{-i}$; 
we could just modify $\sigma_i$ to
agree with $\sigma_i'$ when $i$'s type is $t_i$ to get a strategy that
does better against $\vec{\sigma}_{-i}$ than $\sigma_i$.
This is no longer true with Bayesian machine games, as the following
simple example shows.

\begin{example}
\label{primality.ex}
 {\rm Suppose that the probability on the type space assigns
uniform 
probability to all $2^{100}$ odd numbers between $2^{100}$ and $2^{101}$
(represented as bit strings of length 100).
Suppose that a player $i$ wants to compute if its input (i.e.,
its type) is prime.  Specifically, $i$ gets a utility of 2  minus the
costs of its 
running time (explained below) if $t$ is prime and it outputs 1 or if $t$
is composite and it outputs 0; on the other hand, if it outputs either 0
or 1 and gets the wrong answer, then it gets a utility of $-1000$.
But $i$ also has the option of ``playing safe'';
if $i$ outputs 2, then $i$ gets a utility of 1 no matter what the input
is.  The 
running-time cost is taken to be 0 if $i$'s machine takes less than 2 units
of time and otherwise is $-2$.  We assume that outputting a constant
function takes 1 unit of time.  Note that although testing for primality
is in polynomial time, it will take more than 2 units of time on all 
inputs that have positive probability.  Since $i$'s utility is
independent of what other players do, $i$'s best response is to always
output 2.  However, if $t$ is actually a prime, $i$'s best response
conditional on $t$ is to output 1; similarly, if $t$ is not a prime,
$i$'s best response conditional on $t$ is to output 0.  The key point is
that the
machine that outputs $i$'s best response conditional on a type does not
do any computation; it just outputs the appropriate value.}

{\rm Note that here we are strongly using the assumption that $i$
understands the utility of outputting 0 or 1 conditional on type $t$.  
This amounts to saying that if he is playing the game conditional
on $t$, then he has enough experience with the game to know whether $t$
is prime.  If we wanted to capture a more general setting where the player
did not understand the game, even after learning $t$, then we could do
this by considering two types (states) of nature, $s_0$ and $s_1$, where,
intuitively, $t$ is composite if the state (nature's type) is $s_0$ and
prime if it is $s_1$.  Of course, 
$t$ is either prime or it is not.  We can avoid this problem by simply
having the utility of outputting 1 in $s_0$ or 0 in $s_1$ being $-2$
(because, intuitively, in state $s_0$, $t$ is composite and in $s_1$ it
is prime) and the utility of outputting 0 in $s_0$ or 1 in $s_1$ being
2.  The relative probability of $s_0$ and $s_1$ would reflect the
player $i$'s prior probability that $t$ is prime.}

{\rm In this case, there was no uncertainty about the complexity; there
was simply uncertainty about whether the type satisfied a certain
property.  As we have seen, we can already model the latter type of
uncertainty in our framework.  To model uncertainty about complexity, we
simply allow the complexity function to depend on nature's type, as well
as the machine and the input.  We leave the straightforward details to
the reader.}
\qed 
\end{example}

A common criticism (see e.g., \cite{Aumann85}) of Nash equilibria 
that use randomized strategies is that such equilibria cannot be
\emph{strict} (i.e., it cannot be the case that each player's
equilibrium strategy gives a strictly better payoff than any other
strategy, given the other players' strategies).
This follows since any pure strategy in the support of the randomized
strategy must give the same payoff as the randomized strategy. As the
example below shows, this 
is no longer the case when considering games with computational
costs. 

\begin{example}
{\rm Consider the same game as in Example~\ref{primality.ex}, except that
all machines with running time less than or equal to $T$ have a
cost of 0, and machines that take time  greater than $T$ have a cost of
$-2$. It might very well be the case that, for some 
values of $T$, there might be a probabilistic primality testing
algorithms that runs in time $T$ and determines with high probability 
determine whether a given input $x$ is prime or composite, whereas all
deterministic algorithms take too much time. 
(Indeed, although deterministic polynomial-time algorithms for primality
testing are known \cite{AKS02}, in practice, randomized algorithms are
used because they run significantly faster.)}
\end{example}

\subsection{Sufficient Conditions for the Existence of Nash
Equilibrium 
}\label{sec:suff}
Example~\ref{roshambo} shows, Nash equilibrium does not always exist in
machine games.  The complexity function in this example 
charged for randomization.  Our goal in this section is to show that
this is essentially the reason that Nash equilibrium did not exist; if
randomization were free (as it is, for example, in the model of
\cite{BKK07}), then Nash equilibrium would always exist.  

This result turns out to be surprisingly subtle.  To prove it, we first
consider machine games where \emph{all} computation is free,
that is, the utility of a player depends only on the type and action
profiles (and not the complexity profiles). 
Formally, 
a machine game $G = ([m],\M,\Pr, T, \vec{\complex}, \vec{u})$ is 
\emph{computationally cheap}
if $\vec{u}$ depends only on the type and action profiles, i.e., if
there exists $\vec{u'}$ such that $\vec{u}(\vec{t},\vec{a},\vec{c}) =
\vec{u}'(\vec{t},\vec{a})$ for all $\vec{t},\vec{a},\vec{c}$.  

We would like to show that every computationally cheap Bayesian
machine game has a Nash equilibrium.
But this is too much to hope for.  The first problem is that the game
may have infinitely many possible actions, and may not be compact in any
reasonable topology.  This problem is easily solved; we will simply
require that the type space and the set of possible actions be finite.
Given a \emph{bounding function} $B: \N \rightarrow \N$
be a function,  a \emph{$B$-bounded Turing machine} $M$ is one that
terminates with probability 1 on each input and, 
for each $x\in\bitset^n$, the output of $M(x)$ has length 
at most $B(n)$.  If we restrict our attention to games with a finite type space where
only $B$-bounded machines can be used for some bounding function $B$,
then we are guaranteed to have only finitely many types and actions.

With this restriction, since we do not charge for computation in a
computationally cheap game, it may seem that this result should follow
trivially from the fact that every finite game has a Nash equilibrium.  
But this is false.  The problem is that the game itself might involve
non-computable features, so we cannot hope that that a Turing machine
will be able to play a Nash equilibrium, even if it exists.

Recall that a real number $r$ is \emph{computable}
\cite{Turing37} if there exists a Turing machine that on input $n$
outputs a number $r'$ such that $|r-r'| < 2^{-n}$. 
A game $G = ([m],\M,\Pr, T, \vec{\complexity}, \vec{u})$ is
\emph{computable} if (1) for every $\vec{t} \in T$, $\Pr[\vec{t}]$ is
computable, and (2) for every $\vec{t},\vec{a},\vec{c}$,
$u(\vec{t},\vec{a},\vec{c})$ is computable. 
As we now show, every computationally cheap \emph{computable} Bayesian
machine game has a Nash equilibrium.  Even this result is not
immediate.  Although the game itself is computable, a priori, there may
not be a computable Nash equilibrium.  Moreover, even if there is, 
a Turing machine may not be able to simulate it.  Our proof deals with
both of these problems.

To deal with the first problem, we follow lines similar to those of
Lipton and Markakis \citeyear{LM04}, who used the Tarski-Seidenberg 
\emph{transfer principle} \cite{Tar1} to prove the existence
of \emph{algebraic} Nash equilibria in finite normal norm games with integer valued utilities. 
We briefly review the relevant details here.

\BD
An ordered field $R$ is a \emph{real closed field} if every positive
element $x$ is a square 
(i.e., there exists a $y \in R$ such that $y^2 = x$), 
and every univariate polynomial of odd degree
with coefficients in $R$ has a root in $R$ 
\ED

Of course, the real numbers are a real closed field.  It is not hard to
check that the computable numbers are a real closed field as well.

\begin{theorem} [Tarski-Seidenberg \cite{Tar1}]
Let $R$ and $R'$ be real closed fields such that $R \subseteq R'$, and let
$\bar{P}$ be a finite set of (multivariate) polynomial inequalities with
coefficients in $R$. Then $\bar{P}$ has a solution in $R$ if and only if
it has a solution in $R'$. 
\end{theorem}

With this background, we can state and prove the theorem.

\BT
\label{nashexist.the}
If $T$ is a finite type space, $B$ is a bounding function, $\M$ is a
set of $B$-bounded machines, then  
$G=([m], \M,\Pr,\vec{\complex},\vec{u})$ is a computable, 
computationally cheap Bayesian machine game, 
then there exists a
Nash equilibrium in $G$. 
\ET

\begin{proof}
Note that since in $G$, (1) the type set is finite, (2) the machine set
contains only machines with bounded output length, and thus the action
set $A$ is finite, and (3) computation is free, 
there exists a finite (standard) Bayesian game $G'$ with the same type space,
action space, and utility functions as $G$.  
Thus, $G'$ has a Nash equilibrium.

Although $G'$ has a Nash equilibrium, some equilibria of $G'$ might 
not be implementable
by a randomized Turing machine; indeed, Nash \citeyear{Nash51} showed
that even if all utilities are rational, there exist normal-form games
where all Nash equilibria involve 
mixtures over actions with irrational probabilities. 
To deal with this problem we use the transfer principle.

Let $R'$ be the real numbers and $R$ be the
computable numbers.  Clearly $R \subset R'$.
We use the approach of Lipton and Markakis to show that a Nash
equilibrium in $G'$ must be the
solution to a set of polynomial inequalities with
coefficients in $R$ (i.e., with computable coefficients). 
Then, by combining the Tarski-Seidenberg transfer principle with the
fact that $G'$ has a Nash equilibrium,
it follows that there is a computable Nash equilibrium.

The polynomial equations characterizing the Nash equilibria of $G'$ are
easy to characterize.
By definition, $\vec{\sigma}$ is a Nash equilibrium of $G'$ if and only if
(1) for each player $i$, each type $t_i \in T_i$, and  $a_i \in A_i$,
$\sigma(t_i,a_i) \geq 
0$, (2) for each player $i$ and 
$t_i$, $\sum_{a_i \in A_i} \sigma(t_i,a_i) = 1$, and (3) for each player
$i$ $t_i \in T$, and action $a'_i \in A$, 
$$\sum_{\vec{t}_{-i} \in T_{-i}} \sum_{\vec{a} \in A} \Pr (\vec{t})
u'_i(\vec{t},\vec{a}) \prod_{j\in [m]} \sigma_j(t_j,a_j) \geq  
\sum_{\vec{t}_{-i} \in T_{-i}} \sum_{\vec{a}_{-i} \in A_{-i}} \Pr (\vec{t})
u'_i(\vec{t},(a'_i, \vec{a}_{-i})) \prod_{j\in [m]\backslash i}
\sigma_j(t_j,a_j). 
$$
Here we are using the fact that a Nash equilibrium must continue to be a
Nash equilibrium conditional on each type.

Let $P$ be the set of polynomial equations that result by replacing 
$\sigma_j(t_j,a_j)$ by the variable  $x_{j,t_j,a_j}$.  
Since both the type set and action set is finite, and since both the
type distribution and utilities are computable, this is a finite set of
polynomial inequalities with computable coefficients,
whose solutions are the Nash equilibria of $G'$.
It now follows from the transfer theorem that
$G'$ has a Nash equilibrium where all the probabilities
$\sigma_i(t_i,a_i)$ are computable. 

It remains only to show that this equilibrium can be implemented by a
randomized Turing machine. We show that, for each player $i$, and each
type $t_i$,  
there exists a randomized machine that samples according to the
distribution $\sigma_i(t_i)$; since the type set is finite, this
implies that there exists a machine that implements the strategy
$\sigma_i$. 

Let $a_1, \ldots, a_N$ denote the actions for player $i$, and let
$0=s_0\leq s_1 \leq \ldots \leq s_N=1$ be a sequence of numbers such
that $\sigma_i(t_i,a_j) = s_j - s_{j-1}$. Note that since $\sigma$ is
computable, each number $s_j$ is computable too.  
That means that there exists a machine that, on input $n$, computes an
approximation $\tilde{s}_{n_j}$ to $s_j$, such that $\tilde{s}_{n_j} -
2_{-n} \leq s_j \leq 
\tilde{s}_{n_j} + 2_{-n}$. 
Consider now the machine $M_i^{t_i}$ that proceeds 
as follows.  
The machine constructs a binary decimal $.r_1r_2r_3\ldots$ 
bit by bit.  After the $n$th step of the construction, the machine
checks if the decimal constructed thus far ($.r_1\ldots r_n$) is
guaranteed to be a unique interval $(s_k,s_{k+1}]$.  (Since $s_0, \ldots, s_N$
are computable, it can do this by approximating each one to within
$2_{-n}$.)  With probability 1, after a finite number of steps, the
decimal expansion will be known to lie in a unique interval
$(s_k,s_{k+1}]$.  When this happens, action $a_k$ is performed.
\end{proof}

\commentout{
\paragraph{Games with cheap pre-processing}
We next consider games where computation is costly but pre-processing is
free---namely, the players are allowed to perform any arbitrary
computation before looking at their type; thereafter computation becomes
costly. 
Cheap pre-processing can be viewed as way to model the ``populational"
aspect of a strategy: a player's ``cultural'' heritage is cheap, but the
actual computation it has to perform costly. 

Let $\M'$ be a finite set of machines with finite running-time.
In a \emph{game with cheap pre-processing}, we consider a set of
machines $M$ that terminate with probability 1 and that, on input the 
random tape $r$ and  
type $t$, proceed in two stages: 
in the first stage $M(r,t)$ performs some
computation---without reading the type $t$---and outputs $M' = M(r_1)$
where $r_1$ is a prefix of $r$ and  
$M'\in \M'$; in the second stage $M'$
is executed on input the 
the remaining part $r_2$ of the random tape $r$ and the
type $t$. The complexity of such a machine $M$
in a view $(r_1,r_2);t$ 
is only a function of $M'=M(r_1)$, $t$ and the random tape $r_2$ used by
$M'$. 
}
We now want to prove that a Nash equilibrium is guaranteed to exist
provided that randomization is free.  Thus, we assume that we start with
a 
finite
set $\M_0$ of \emph{deterministic} Turing machines and a finite set
$T$ of types (and continue to assume that all the machines in $\M_0$
terminate).  $\M$ is the
\emph{computable convex closure of $\M_0$} if $\M$ consists of machines
$M$ that, on input $(t,r)$, first perform some computation that depends
only on the random string $r$ and not on the type $t$, 
that with probability 1,
after some finite
time and after reading a finite prefix $r_1$ of $r$, choose a machine $M' \in
\M_0$, then run $M'$ on input $(t,r_2)$, where $r_2$ is the remaining
part of the random tape $r$. 
Intuitively, $M$ is randomizing
over the machines in $\M_0$.  
It
is easy to see that there must be some $B$ such that all the
machines in $\M$ are $B$-bounded.
\emph{Randomization is free} in a machine game $G = ([m],\M,\Pr, T,
\vec{\complexity}, \vec{u})$ where $\M$ is the computable convex closure
of $\M_0$ if $\complexity_i(M,t,r)$ is $\complexity_i(M',t,r_2)$ (using
the notation from above).  

\BT
\label{nashpreproc.the}
If $\M$ is the computable convex closure of some finite set $\M_0$ of
deterministic Turing machines, $T$ is a finite type space, and 
$G=([m], \M,\T, Pr,\vec{\complex},\vec{u})$ is a game where
randomization is free,
then there exists a Nash
equilibrium in $G$. 
\ET
\begin{sketch}
First consider the normal-form game where the agents choose a machine in
$\M_0$, and the payoff of player $i$ if $\vec{M}$ is chosen is the
expected payoff in $G$ (where the expectation is taken with respect to
the probability $\Pr$ on $T$).  
By Nash's theorem, it follows that
there exists a mixed strategy Nash equilibrium in this game. Using the
same argument as in the proof of Theorem \ref{nashexist.the}, it follows
that there exists a machine in $\M$ that samples according to
the mixed distribution over machines, as long as the game is computable
(i.e., the type distribution and utilities are computable)
and the type and action spaces finite. 
(The fact that the action space is finite again follows from the fact that type
space is finite and that there exist some $B$ such that all machine in $\M$ are $B$-bounded.)
The desired result follows.
\end{sketch}

We remark that if we take Aumann's \citeyear{Aumann87} view of a
mixed-strategy equilibrium as representing an equilibrium in players'
beliefs---that is, each player is actually 
using a deterministic strategy in $\M_0$,  and the probability that
player $i$ plays a strategy $M' \in \M_0$ in equilibrium represents all
the other players' beliefs about the probability that $M'$ will be
played---then we can justify randomization being free, since players are
not actually randomizing.
The fact that the randomization is computable here amounts to the
assumption that players' beliefs are computable (a requirement advocated by
Megiddo \citeyear{Meg2}) 
and that the population players are chosen from can be sampled by a
Turing machine. 
More generally, there may be settings where
randomization devices are essentially freely available (although, even
then, it may not be so easy to create an arbitrary computable
distribution).  

Theorem~\ref{nashpreproc.the} shows that if randomization is free, a
Nash equilibrium in machine games is guaranteed to exist.  We can
generalize this argument to show that, to guarantee the existence of 
$\epsilon$-Nash equilibrium (for arbitrarily small $\epsilon$) it is
enough to assume that 
``polynomial-time'' randomization is free.  
Lipton, Markakis and Mehta \citeyear{LMM03} show
that 
every finite game with action space $A$ has an $\epsilon$-Nash equilibrium
with support on only $poly(\log |A| + 1/\epsilon)$ actions; furthermore
the probability of each action is a rational number of length $poly(\log
|A| + 1/\epsilon)$. In our setting, it follows that there exists an
$\epsilon$-Nash equilibrium where the randomization can be computed by a
Turing machine with size and running-time bounded by
size $O(\log |\M'| + 1/\epsilon)$. 
We omit the details here.

\nfullv{\subsection{Computationally Robust Nash Equilibrium}}
\nshortv{\paragraph{Computationally Robust Nash Equilibrium}}
Computers get faster, cheaper, and more powerful every year.  Since
utility in a 
Bayesian machine game takes computational complexity into account, this
suggests that an agent's utility function will change when he 
replaces one computer by a newer computer.  We are thus interested in
\emph{robust} equilibria, intuitively, ones that continue to be
equilibria 
(or, more precisely, $\epsilon$-equilibria for some appropriate $\epsilon$)
even if agents' utilities change as a result of upgrading
computers.  

\BD [Computationally robust Nash equilibrium]
Let $p: \IN \rightarrow \IN$.
The complexity function $\complex'$ is \emph{at most a $p$-speedup} of
the complexity function $\complex$ if, for  
all machines $M$ and views $v$, 
$$ \complex'(M,v) \le \complex(M,v) \le  p(\complex'(M,v)).$$
Game $G'=([m'],\M',\Pr',\vec{\complexity'}, \vec{u'})$ is \emph{at most a $p$-speedup} of game $G=([m],\M,\Pr,\vec{\complexity}, \vec{u})$ if
$m'=m$, $\Pr=\Pr'$ and $\vec{u}=\vec{u'}$ (i.e., $G'$ and $G'$ differ only in their complexity and machine profiles),
and $\complex'_i$ is at most a $p$-speedup of $\complex_i$, for 
$i = 1, \ldots, m$. 
$\vec{M}$ is a \emph{$p$-robust 
$\epsilon$-equilibrium} for $G$ 
if, for every game $G'$ that is at most a $p$-speedup of $G$,
$\vec{M}$ is an $\epsilon$-Nash equilibrium of $G'$.
\ED
We also say that $M_i$ is a \emph{$p$-robust $\epsilon$-best response}
to $\vec{M}_{-i}$ in $G$, if for every game $\tilde{G}$ that is at most
a $p$-speedup of $G$, $M_i$ is  
an $\epsilon$-best response to $\vec{M}_{-i}$. Note that $\vec{M}$ is a
$p$-robust $\epsilon$-equilibrium iff, 
for $i = 1, \ldots, m$, $M_i$ is a $p$-robust $\epsilon$-best response
to $\vec{M}_{-i}$. 
\fullv{
Intuitively, if we think of complexity as denoting running time and
$\complex$ describes the running time of machines (i.e., programs) on an
older computer, then $\complex'$ describes the running time of machines
on an upgraded computer. 
For instance, if the upgraded computer runs at most twice as fast as the
older one (but never slower), then $\complex'$ is $\bar{2}$ speedup of
$\complex$, where $\bar{k}$ denotes the constant function $k$. 
Clearly, if $\vec{M}$ is a Nash equilibrium of $G$, then it is a 
$\bar{1}$-robust equilibrium.
  We can think of
$p$-robust equilibrium as a refinement of Nash equilibrium for machine games,
just like \emph{sequential equilibrium} \cite{KW82} or \emph{perfect
equilibrium} \cite{Selten75}; it provides a principled way of ignoring
``bad'' Nash equilibria.
}

Note that in games where computation is free, every Nash equilibrium is
also computationally robust.  

\nfullv{\subsection{Coalition Machine Games}}
\nshortv{\paragraph{Coalition Machine Games}}
We strengthen the notion of Nash equilibrium to 
allow for deviating coalitions.
Towards this goal, we consider a
generalization of Bayesian machine games called \emph{coalition machine
games}, where,
in the spirit of \emph{coalitional games} \cite{vNM},
 each \emph{subset} of players has a complexity function and
utility function associated with it. 
In analogy with the traditional
notion of Nash equilibrium, which considers only ``single-player''
deviations, we consider only ``single-coalition'' deviations. 
 
More precisely, given a subset $Z$ of $[m]$, we let $-Z$ denote the set
$[m]/Z$. We say that a machine $M'_Z$ \emph{controls} the players in $Z$
if $M'_Z$ 
controls the input and output tapes of the players in set $Z$
(and thus can coordinate their outputs).
In addition, the adversary that controls $Z$ has its own input and
output tape. 
A \emph{coalition machine game $G$} is described by a tuple 
$([m], \M,\Pr,$ $\vec{\complexity},\vec{u})$, where 
$\vec{\complex}$ and $\vec{u}$ are sequences of 
complexity functions $\complex_Z$ and utility functions $u_Z$, respectively,
one for each 
subset $Z$ of $[m]$; $m,\M,\Pr, \complex_Z$ are defined as in Definition
\ref{bmg.def}. In contrast, the utility function 
$u_Z$
for the set $Z$ 
is a function $T \times (\{0,1\}^*)^{m} \times (\IN \times \IN^{m-|Z|+1}) 
\rightarrow \IR$,
where $u_Z(\vec{t},\vec{a},(c_Z, \vec{c}_{-Z}))$ is the
utility of the coalition $Z$ if $\vec{t}$ is the (length $m+1$) type
profile, $\vec{a}$ is the 
(length $m$) action profile (where we identify $i$'s action as player
$i$ output), $c_Z$ is the complexity of the coalition $Z$, and
$\vec{c}_{-Z}$ is the (length $m-|Z|$) profile of machine complexities
for the players in $-Z$.   
The complexity $c_Z$ is a measure of the complexity according to whoever
controls coalition $Z$ of running the coalition.  
Note that even if the coalition is controlled by a machine $M'_{Z}$
that lets 
each of the players in $Z$ perform independent computations, the complexity
of $M'_{Z}$ is not necessarily some function of the
complexities $c_i$ of the players $i 
\in Z$ (such as the sum or the max). 
Moreover, while cooperative game theory tends to focus on
\emph{superadditive} utility functions, where the utility of a coalition
is at least the sum of the utilities of any partition of the
coalition  into sub-coalitions or individual players, we make no such
restrictions; indeed when taking complexity into account, it might very
well be the case that larger coalitions are more expensive than smaller
ones. 
Also note that,
in our calculations, we assume that, other than the coalition $Z$,
all the other players play individually (so that we use $c_i$ for $i\notin Z$);
there is at most one coalition in the picture.
Having defined $u_Z$, we can define the expected utility of the group
$Z$ in the obvious way.  
The \emph{benign machine for coalition $Z$}, denoted $M^b_Z$, is
the one where that gives each player $i \in Z$ its 
true input, and each player $i\in Z$ outputs the output of $M_i$;
$M^b_Z$ write nothing on its output tape.
Essentially, the benign machine does exactly what all the players in
the coalition would have done anyway.
We now extend the notion of Nash equilibrium to deal
with coalitions; it requires that in an equilibrium $\vec{M}$, no
coalition does (much) better than it would using the benign machine,
according to the utility function for that coalition.
\BD [Nash equilibrium in coalition machine games]
Given an $m$-player coalition machine game $G$, a machine profile $\vec{M}$, 
a subset $Z$ of $[m]$
and $\epsilon \ge 0$, 
$M^b_Z$ is an $\epsilon$-best response to $\vec{M}_{-Z}$ 
if, for every coalition machine $M_Z' \in \M$, 
$$U_Z^G[(M^b_Z, \vec{M}_{-Z})] \geq U_Z^G[(M_Z', \vec{M}_{-Z})] -
\epsilon.$$
Given a set $\Z$ of subsets of $[m]$,
$\vec{M}$ is a \emph{$\Z$-safe $\epsilon$-Nash equilibrium} 
for $G$ if, 
for all $Z \in \Z$.
$M^b_Z$ is an $\epsilon$-best response to
$\vec{M}_{-Z}$. 
\ED

Our notion of coalition games is quite general.
In particular, if we disregard the costs of computation, it allows us to
capture some standard notions of 
coalition resistance in 
the literature, by choosing $u_Z$ appropriately.  For example, 
Aumann's \citeyear{Aumann59} notion of \emph{strong equilibrium} requires
that, for all coalitions, it is not the case that there is a deviation that makes
everyone in the coalition strictly better off.  To capture this, fix a profile
$\vec{M}$, and define $u^{\vec{M}}_Z(M'_Z,\vec{M}'_{-Z}) =
\min_{i \in Z}u_i(M'_Z,\vec{M}'_{-Z}) - u_i(\vec{M})$.%
\footnote{Note that if we do not disregard the cost of computation, it
is not clear how to define 
the individual complexity of a player that is controlled by $M'_{\Z}$.}
We can capture the notion of \emph{$k$-resilient equilibrium}
\cite{ADGH06,ADH07}, where the only deviations allowed are by coalitions
of size at most $k$, by restricting $\Z$ to consist of sets of cardinality at
most $k$ (so a 1-resilient equilibrium is just a Nash equilibrium).  
\fullv{
Abraham et al.~\citeyear{ADGH06,ADH07} also consider a notion of \emph{strong}
$k$-resilient equilibrium, where there is no deviation by the coalition
that makes even one coalition member strictly better off.  We can
capture this by replacing the min in the definition of $u^{\vec{M}}_Z$
by max.
}
\nfullv{\subsection{Machine Games with Mediators}\label{sec:mediator}}
\nshortv{\paragraph{Machine Games with Mediators}}
Up to now we have assumed that the only input a machine receives 
is the initial type.  This is appropriate in a
normal-form game, but does not allow us to 
consider
game where
players can communicate with each other and (possibly) with a trusted
mediator. 
\nshortv{In Appendix \ref{sec:mediator} we extend a Bayesian machine game 
to allow for communication with a trusted mediator; roughly speaking a
Bayesian machine game with a mediator is a pair $(G,\F)$, where $G$ is 
a Bayesian machine game (but where the machines involved now are interactive TM) and $\F$ is a mediator. A particular mediator of interest is the 
\emph{communication mediator}, denoted $\C$, which corresponds to
what cryptographers call \emph{authenticated channels} and 
economists call \emph{cheap talk}.
}%
\nfullv{
We now extend Bayesian machine games 
to allow for communication.
For ease of exposition, we assume that all communication passes between
the players and a trusted mediator.  Communication between the players
is modeled by having a trusted mediator who passes along messages
received from the players.
Thus, we think of the players as having reliable 
communication channels to and from a mediator; no other
communication channels are assumed to exist.
The formal definition of a Bayesian machine game with a mediator is
similar in spirit to that of a Bayesian machine game, but now we assume
that the machines are \emph{interactive} Turing machines, that can
also send and receive messages.  We omit the formal definition of an
interactive Turing machine (see, for example, \cite{goldreich01}); roughly
speaking, the machines use a special tape where the message to be sent
is placed and another tape where a message to be received is written.
The mediator is modeled by an interactive Turing  machine that we denote
$\F$.
A \emph{Bayesian machine game with a mediator} (or a mediated Bayesian
machine game) is thus a pair $(G, \F)$, where $G= ([m],\M,\Pr,
\complexity_1, \ldots, \complexity_n, 
u_1, \ldots, u_n)$ is a Bayesian machine game (except that $\M$ here denotes 
a set of \emph{interactive} machines) and $\F$ is an interactive Turing
machine. 

Like machines in Bayesian machine games, interactive machines in a game
with a mediator take as argument a view and produce an outcome.  Since 
what an interactive machine does can depend on the
history of messages sent by the mediator, the message history (or, more
precisely, that part of the message history actually read by the
machine) is also part of the view.  
Thus, we now define a view to be a string $t;h;r$ in $\bit^*;\bit^*;\bit^*$,
where, as before,
$t$ is that part of the type actually read and $r$ is a finite bitstring
representing the string of random bits actually used, and $h$ is a finite
sequence of messages received and read.  
Again, if $v = t;\barw;r$, we take $M(v)$ to be the output of $M$
given the view.

We assume that the system proceeds in synchronous stages; a message sent
by one machine to another in stage $k$ is received by the start of stage
$k+1$.  
More formally, following \cite{ADGH06},
we assume that a \emph{stage}
consists of three phases.  
In the first phase of a stage, each player $i$ sends a message to the
mediator, or, more precisely, player $i$'s machine $M_i$ computes a message
to send to the mediator; machine $M_i$ can also send an
empty message, denoted $\lambda$.  In the second phase, 
the mediator receives the message and 
mediator's machine sends each
player $i$ a message in response (again, the mediator can send an empty
message). 
In the third phase, each player $i$ 
performs an action other than that of sending a message (again, it may
do nothing).
The messages sent and the actions taken can depend on the machine's
message history (as well as its initial type).
We can now define the expected utility of a profile of interactive
machines in a Bayesian machine game with a mediator.   The definition is
similar in spirit to the definition in Bayesian machine games, except
that we must take into account the dependence of a player's actions on
the message sent by the mediator.  
Let $\view_i(\vec{M}, \F, \vec{t}, \vec{r})$ denote the string
$(t_i;\barw_i;r_i)$ 
where $\barw_i$ denotes the messages received by player $i$ if the machine profile is $\vec{M}$, the
mediator uses machine $\F$, the type profile is $\vec{t}$, and
$\vec{r}$ is the profile of random strings used by the players and the
mediator.  
Given a mediated Bayesian machine game $G'=(G,\F)$,
we can define the random variable $u_i^{G',\vec{M}}(\vec{t},\vec{r})$ as
before, except that now $\vec{r}$ must include a random string for the
mediator, and to compute the outcome and the complexity function, $M_j$
gets as an argument $\view_j(\vec{M}, \F, \vec{t}, \vec{r})$, 
since this is the view that machine $M_j$ gets in this setting.  
Finally, we define 
$U_i^{G'}(\vec{M}) = \mathbf{E}_{\Pr^+}[u_i^{G',\vec{M}}]$ as before,
except that now $\Pr^+$ is a distribution on $\T \times
({\{0,1\}^\infty})^{n+1}$ rather than $\T \times
({\{0,1\}^\infty})^{n}$, since we must include a random string for the
mediator as well as the players' machines.
We can define Nash equilibrium and
computationally robust Nash equilibrium in games with mediators as in
Bayesian machine games;
we leave the details to the reader.

Up to now, we have considered only players communicating with a
mediator.  We certainly want to allow for the possibility of players
communicating with each other.  We model this using a
particular mediator that we call the
\emph{communication mediator}, denoted $\C$, which corresponds to
what cryptographers call \emph{secure channels} and 
economists call \emph{cheap talk}.
With this mediator, 
if $i$ wants to send a message to $j$,
it simply sends the message and its intended recipient to the
mediator $\C$. 
The mediator's strategy is simply to forward the messages, and the
identities of the senders, to the intended recipients.  
(Technically, we assume that a message $m$ from $i$ to the mediator with
intended recipient $j$ has the form $m;j$.  Messages not of this form
are ignored by the mediator.)
} %

\paragraph{Repeated and extensive games}
We can extend the above treatment to consider
repeated games (where
information from earlier plays is relevant to later plays), and, more
generally, arbitrary extensive-form games (i.e., games defined by game
trees). We capture an extensive-form game by simply viewing nature as a
mediator;  we allow  utility functions to 
take into account the messages sent by the player to the mediator
(i.e., talk is not necessarily ``cheap''), and also the random coins
used by the mediator.  
In other words, we assume, without loss of generality, that all
communication and signals sent to a player are sent through the
mediator, and that all actions taken by a player are sent as messages to
the mediator. We leave the details of the formal definition to the
reader. 

As shown by the example below, Nash equilibrium in machine games gives a
plausible explanation of observed behavior in the finitely-repeated
prisoner's dilemma.   

\begin{example}\label{xam:pd} 
{\rm Recall that 
in the prisoner's dilemma, there are two prisoners, who
can choose to either cooperate or defect.
As described in the table below, if they both cooperate, they both get
3; if they both defect, then both get -3; if one defects and the other
cooperates, the defector gets 5 and the cooperator gets $-5$.
(Intuitively, the cooperator stays silent, while the defector ``rats
out'' his partner.  If they both rat each other out, they both go to jail.)}
\begin{table}[htb]
\begin{center}
\begin{tabular}{c |  c c}
& $C$ & $D$\\
\hline
$C$ &$(3,3)$ &$(-5,5)$ \\
$D$  &$(5,-5)$  &$(-3,-3)$  \\
\end{tabular}
\end{center}
\end{table}

{\rm It is easy to see that defecting dominates cooperating: no matter what
the other player does, a player is better off defecting than
cooperating.  Thus, ``rational'' players should defect.  And, indeed,
$(D,D)$ is the only Nash equilibrium of this game.  Although $(C,C)$
gives both players a better payoff than $(D,D)$, this is not an
equilibrium.}

{\rm Now consider finitely repeated prisoner's dilemma (FRPD), where
prisoner's dilemma is played for some fixed number $N$ of rounds.  
The only Nash equilibrium is to always defect; this can be seen by a
backwards induction argument.  (The last round is like the one-shot game,
so both players should defect; given that they are both defecting at the
last round, they should both defect at the second-last round; and so on.)
This seems quite unreasonable.  And, indeed, in experiments, people do
not always defect \cite{Axelrod}.  In fact, quite often they cooperate
throughout the 
game.  Are they irrational?  It is hard to call this irrational
behavior, given that the ``irrational'' players do much better than
supposedly rational players who always defect.  }

{\rm There have been many
attempts to explain cooperation in FRPD in the literature; see, for
example, \cite{KMRW,Ney85,PY94}.  
In particular, \cite{Ney85,PY94} demonstrate that if players are
restricted to 
using a finite automaton with bounded complexity, then there 
exist
equilibria 
that allow for cooperation.  However, the
strategies used in those equilibria are quite complex, and require
the use of large automata;\footnote{The idea behind
these equilibria is to force players to remember a short history of the
game, during which players perform random actions; this requires 
the use of many states.} 
as a consequence this approach does not seem to provide a satisfactory
explanation as to why people choose to cooperate.} 
{\rm By using our framework, we can provide a 
straightforward explanation.  
Consider the \emph{tit-for-tat} strategy, which proceeds as follows: a
player cooperates at the first round, and then at round $m+1$, does
whatever his opponent did at round $m$.  Thus, if the opponent cooperated
at the previous round, then you reward him by continuing to cooperate; if he
defected at the previous round, you punish him by defecting.  If both players
play tit-for-tat, then they cooperate throughout the game.  Interestingly,
tit-for-tat does exceedingly well in FRPD tournaments, where computer
programs play each other \cite{Axelrod}.  }

{\rm Now consider a machine-game version of FRPD, where at each round
the player receive as signal the move of the opponent in the previous
rounds before they choose their action. In such a game, tit-for-tat is a
simple program, which needs  
no
memory (i.e., the machine is stateless).
Suppose that 
we charge even a modest amount $\alpha$ for memory usage 
(i.e., stateful machines get a penalty of 
at least $\alpha$, whereas stateless machines get
no penalty), 
that there is  
a discount factor $\delta$, with $0.5 < \delta < 1$, so that if the
player 
gets a reward of $r_m$ in round $m$, his total reward 
over the whole
$N$-round game 
(excluding the complexity penalty)
is taken to be $\sum_{m=1}^N  \delta^m r_m$,
that $\alpha \ge 2\delta^N$, and that $N > 2$.  
In this
case,
it will be a Nash equilibrium for both players to play
tit-for-tat.  
Intuitively, no matter what the cost of memory is (as long as it is
positive), for a sufficiently long game, tit-for-tat is a Nash equilibrium.

To see this, note that the best response to tit-for-tat is to play
tit-for-tat up to 
but not including
the last round, and then to defect.
But following
this strategy requires the player to keep track of the round number,
which requires the use of extra memory.  The extra gain of 2 achieved
by defecting at the last round
will not be worth the cost of keeping track of
the round number as long as $\alpha \geq 2\delta^N$; thus no 
stateful strategy can do better. 
It remains to argue that no stateless strategy (even a randomized
stateless strategy) can do better against tit-for-tat.  
Note that any strategy that defects for the first time at round $k<N$
does at least $6\delta^{k+1}-2\delta^{k}$ worse than tit-for-tat.  It
gains 2 at round 
$k$ (for a discounted utility of $2\delta^k$), but loses at least 6
relative to tit-for-tat in the next round, for a discounted utility of
$6\delta^{k+1}$.
From that point on the best response is to either continue defecting (which at each round leads to a loss of 6), or cooperating until the last round and then defecting (which leads to an additional loss of 2 in round $k+1$, but a gain of 2 in round $N$). 
Thus, any strategy that defects at round $k<N$
does at least $6\delta^{k+1}-2\delta^{k}$ worse than tit-for-tat.

A strategy that defects at
the last round gains $2\delta^N$ relative to tit-for-tat.
Since $N>2$, the probability that a stateless strategy
defects at round $N-1$ or earlier is at least as high as the probability
that it defects for the first time at round $N$. 
(We require that $N>2$ to make sure that there exist some round $k<N$
where the strategy is run on input $C$.) 
It follows that
any stateless strategy that defects 
for the first time in the last round with probability $p$ 
in expectation gains at most $p
(2\delta^N-(6\delta^N - 2\delta^{N-1})) = p\delta^{N-1}(2 - 4\delta)$,
which is negative when $\delta > 0.5$. 
Thus, when $\alpha \geq 2\delta^N$, $N>2$, and $\delta>0.5$, tit-for-tat
is a Nash equilibrium in FRPD. 
(However, also note that depending on the cost of memory,
tit-for-tat may \emph{not} be a Nash equilibrium for sufficiently short
games.)} 
{\rm The argument above can be extended to show that tit-for-tat is a
Nash equilibrium even if there is also a charge for randomness or
computation, as long as there is no computational charge for machines as
``simple'' as tit-for-tat; this follows since adding such extra charges
can only make things worse for strategies other than tit-for-tat.} 
{\rm Also note that even if only one player is 
charged for memory, and memory is free for the other player, then there
is a Nash equilibrium where the bounded player plays tit-for-tat, while
the other player plays the best response of cooperating up to 
but not including
the last
round of 
the game, and then defecting.
\qed}
\end{example}

\paragraph{The revelation principle}
The \emph{revelation principle} is one of the fundamental principles in
traditional implementation theory. A specific instance of it
\cite{Myerson79,F86} stipulates that for every Nash equilibrium in a
mediated games $(\G,\F)$, there exists a different mediator $\F'$ such
that it is a Nash equilibrium for the players to \emph{truthfully}
report their type to the mediator and then perform the action suggested
by the mediator. 
As we demonstrate, this principle no longer holds when we take
computation into account 
(a similar point is made by Conitzer and
Sandholm \citeyear{CS04}, although they do not use our formal model).
The intuition is simple: truthfully reporting 
your type will not be an equilibrium if it is too ``expensive'' to send the
whole type to the mediator.   
For a naive counter example, consider a game where a player get utility
1 whenever its complexity is 0 (i.e., the player uses the strategy
$\bot$) and positive utility otherwise. Clearly, in this game it can
never be an equilibrium to truthfully report your type to any
mediator. This example is degenerate as the players actually never use
the mediator. In the following example, we consider a game with a Nash
equilibrium where the players use the mediator. 

Consider a 2-player game where each player's type is an $n$-bit number.
The type space consists of all pairs of $n$-bit numbers 
that either are the same, or that differ in all but at most $k$
places, where $k \ll n$.  The player receive a utility of 1 if they can
guess correctly whether their types are the same or not, while having
communicated less than $k+2$ bits; otherwise it receives a utility of
0. 
Consider a mediator that upon receiving $k+1$ bits from the players
answers back to both players whether the bits received are identical or
not. 
With such a mediator it is an equilibrium for the players to provide the
first $k+1$ bits of their input and then output whatever the mediator
tells them. 
However, providing the full type is always too expensive (and can thus
never be a Nash equilibrium), no matter what mediator the players have
access to. 

\section{A Computational Notion of Game-Theoretic
Implementation}\label{sec:security} 
In this section we extend the traditional notion of game-theoretic
implementation of mediators to consider computational games. Our aim is
to obtain a notion of implementation that can be used to capture the
cryptographic notion of implementation. For simplicity, we focus on
implementations of mediators that receive a single message from each
player and return a single message to each player (i.e., the mediated
games consist only of a single stage). 

We 
provide a definition 
that captures the
intuition that the machine profile $\vec{M}$ 
implements a
mediator $\F$ if,
whenever a set
of players want to 
to truthfully provide their ``input'' to the mediator $\F$, 
they also want to run $\vec{M}$ using the same inputs.  
To formalize ``whenever'',
we consider what we call \emph{canonical
coalition games},
where each player 
$i$ has a type $t_i$ of the form $x_i;z_i$, where $x_i$ is 
player $i$'s intended ``input'' and $z_i$ consists of some additional
information that player $i$ has
about the state of the world. 
We assume that the input $x_i$ has some fixed length $n$.
Such games are called
\emph{canonical games of input length $n$}.%
\footnote{Note that by simple padding, canonical games represent
a setting where all parties' input lengths are upper-bounded by some
value $n$ that is common knowledge.
Thus, we can represent any game where there are only finite many
possible types as a canonical game for some input length $n$.}
Let $\Mcanon^\F$ denote the machine that, given type $t=x;z$ 
sends $x$ to the  mediator 
$\F$
and outputs as its action
whatever string it receives back from $\F$,
and then halts.
(Technically, we model the fact that $\Mcanon^\F$ is expecting to
communicate with $\F$ by assuming that the mediator $\F$ appends a
signature to its messages, and any messages not signed by $\F$ are
ignored by $\Mcanon^\F$.) 
Thus, the machine $\Mcanon^\F$ ignores the extra information $z$.
Let
$\vec{\Mcanon}^\F$ denote the machine profile where each player uses the
machine $\Mcanon^\F$. 
Roughly speaking, to capture the fact that whenever
the players want to compute 
$\F$,
they also want to run
$\vec{M}$), we require that
if $\vec{\Mcanon}^F$ is an
equilibrium in the game $(G,\F)$ 
(i.e., if it is an 
equilibrium to simply provide the intended input to $\F$ and finally 
output whatever $\F$ replies), running $\vec{M}$ using the intended
input is an equilibrium as well. 
We actually consider a more general notion of implementation:
we are interested in understanding how well equilibrium 
in a set of games with 
mediator $\F$ can be implemented using a machine profile $\vec{M}$
and a possibly different mediator $\F'$.
Roughly speaking, we want that, for every game $G$ 
in some set $\G$ of games,
if $\vec{\Mcanon}^\F$ is an equilibrium in $(G,\F)$, then
$\vec{M}$ is an equilibrium in 
$(G,\F')$.
In particular, we want to understand what degree of robustness $p$ in the
game $(G,\F)$ is required to achieve an $\epsilon$-equilibrium 
in the game 
$(G,\F')$. 
We also require that the equilibrium with mediator $\F'$
be as ``coalition-safe'' as the equilibrium with mediator $\F$. 

\BD [Universal implementation] 
Suppose that $\G$ is a set of $m$-player canonical games, $\Z$ is a set of
subsets of $[m]$, $\F$ and $\F'$ are mediators, $M_1, \ldots, M_m$ are
interactive machines, 
$p: \IN \times \IN \rightarrow \IN$, and $\epsilon: \IN \rightarrow \IR$.
$(\vec{M},\F')$ is a \emph{($\G,\Z,p$)-universal implementation of $\F$
with error $\epsilon$} if, for all $n \in \IN$, 
all games $G \in \G$ with input length $n$, 
all $\Z' \subseteq \Z$
if $\vec{\Mcanon}^\F$ is a $p(n,\cdot)$-robust $\Z'$-safe Nash equilibrium 
in the mediated machine game $(G,\F)$ then
\beginsmall{enumerate}
\item {\em (Preserving Equilibrium)} $\vec{M}$ is a
$\Z'$-safe $\epsilon(n)$-Nash equilibrium in the
mediated machine game $(G,\F')$.  
\item {\em (Preserving Action Distributions)} For each type profile
$\vec{t}$, the  
action profile induced by $\vec{\Mcanon}^\F$ in $(G,\F)$ 
is identically distributed to the action profile induced by
$\vec{M}$ in $(G,\F')$.
\endsmall{enumerate}
\ED

Note that, depending on the class $\G$, our notion of universal
implementation imposes severe restrictions on the complexity of the
machine profile $\vec{M}$. For instance, if $\G$ consists of all games,
it requires that the complexity of $\vec{M}$ is the same as the
complexity of $\vec{\Mcanon}^{\F}$. 
(If the complexity of $\vec{M}$ is higher than that of
$\vec{\Mcanon}^\F$, then we can easily construct a game $G$ by choosing the utilities appropriately such that it is an
equilibrium to run $\vec{\Mcanon}^\F$ in $(G,\F)$, but running $\vec{M}$ is
too costly.)
Also note that if $\G$ consists of games where players strictly prefer 
smaller complexity, then universal implementation requires that
$\vec{M}$ be the optimal algorithm (i.e., the algorithm with the lowest
complexity) that implements the functionality of $\vec{M}$, since
otherwise a player would prefer to switch to the optimal
implementation. 
Since few algorithms algorithms have been shown to be provably optimal
with respect to, for example, the number of computational steps of a
Turing machines, this, at first sight, seems to severely limit the
use of our definition.  
However, if we consider games with ``coarse'' complexity functions, or a
complexity functions where, say, the first $T$ steps are ``free'' (e.g.,
machines that execute less than $T$ steps are assigned complexity 1),
the restrictions above are not so severe. Indeed, it seems quite natural
to assume that a player is indifferent between the ``small'' differences
in computation. 
Our notion of universal implementation is related to a number of other
notions in the literature; we briefly review the most relevant ones here.
\BI 
\item
Our definition of universal implementation
captures intuitions similar 
in spirit to Forges' \citeyear{F90} notion of a \emph{universal
mechanism}.
It differs in one obvious way: our definition considers
\emph{computational} games, where the utility functions
depend on complexity considerations.
As mentioned, Dodis, Halevi and Rabin \citeyear{DHR00} (and more recent
work,  
such as \cite{ADGH06,LMPS04,HT04,ADGH06,GK06,KN08}) consider notions of
implementation where the players are modeled as polynomially-bounded
Turing machines, but do not consider computational games. As such, the
notions considered in these works do not provide any a-priori guarantees
about the incentives of players with regard to computation. 
\item
Recently, Izmalkov, Lepinski and Micali
\citeyear{ILM08}
introduced a strong notion of implementation, called \emph{perfect implementation}.
Although their notion 
does not explicitly consider 
games with computational costs, 
their
notion can be  viewed as a strengthening of ours
under certain assumptions about the cost of computation;
however, their notion can be achieved only 
with the use of strong primitives that cannot be implemented under standard
computational and systems assumptions \cite{LMS05}. 
\item
Our definition is more general than earlier notions of implementation
in that 
we consider 
also universality with respect to (sub-)classes of games $\G$.
This extra generality will be extensively used in the sequel.
\item
Our notion of coalition-safety also differs somewhat from earlier notions.
Note that 
if $\Z$ contains all subsets of players with $k$ or less
players, then universal implementation implies that all
$k$-resilient Nash equilibria and all strong $k$-resilient Nash
equilibria are preserved. 
However, 
unlike the notion of $k$-resilience considered by Abraham et al.
\citeyear{ADGH06,ADH07},
our notion provides a ``best-possible'' guarantee for games that do
not have a $k$-resilient Nash equilibrium.   We guarantee that if a certain
subset $Z$ of players have no incentive to deviate in the mediated game,
then that subset will not have incentive to deviate in the cheap-talk
game; this is similar in spirit to the definitions of \cite{ILM08,LMPS04}.
Note that, in contrast to \cite{ILM08,LMS05},  rather than just allowing
colluding players to communicate only through their moves in the game, 
we allow 
coalitions of players that are controlled by a single
entity; this is equivalent to considering collusions where the colluding
players are allowed to freely communicate with each other.  
In other words, whereas the definitions of \cite{ILM08,LMS05} require 
protocols to be ``signalling-free'', our definition does not 
impose such restrictions.
We believe that
this model is better suited to capturing the security of cryptographic
protocols
in most traditional settings (where signalling is not an issue).
\item
We require only that a Nash equilibrium is preserved when
moving from the game with mediator $\F$ 
to the communication game. 
Stronger notions of implementation require that the equilibrium in the
communication game be a \emph{sequential equilibrium} \cite{KW82}; see,
for example, \cite{Gerardi04,Bp03}.  
Since every Nash equilibrium in
the game with the mediator $\F$ is also a sequential equilibrium, these
stronger notions of implementation actually show that 
sequential equilibrium is preserved when passing from the game with
the mediator to 
the communication game.

While these notions of implementation guarantee that an
equilibrium with the mediator is preserved in the communication game,
they do not guarantee that new equilibria are not introduced in the
latter. 
An even stronger guarantee is provided by Izmalkov, Lepinski and
Micali's  \citeyear{ILM08} notion of perfect
implementation; this notion 
requires a one-to-one
correspondence $f$ between \emph{strategies} in the corresponding games 
such that each player's utility with strategy profile $\vec{\sigma}$ in
the game with the mediator is the same as his utility with strategy
profile $(f(\sigma_1),\ldots,f(\sigma_n))$ in 
the communication game
without the mediator.  Such a correspondence,
called \emph{strategic equivalence} by Izmalkov, Lepinski, and Micali
\citeyear{ILM08}, 
guarantees 
(among other things)
that 
\emph{all} types of equilibria are preserved when passing from one game to 
the other, and that
no new equilibria are introduced in the communication game.
We focus on the simpler notion of implementation, which
requires only that 
Nash equilibria are preserved, and leave open an exploration of more refined notions.
%
\iffalse
%
%
%
since it seems to us the minimal requirement that still  of 
captures the intuition that whenever players 
want to compute $f$,
they also want to execute $\Pi$. 
%
%
%
%
%
%
%
%
%
%
%
%
%
We hope to explore more refined notions in future work.
\fi
\EI

%
%

%
%
%
%
%
%
%
%
%
%
%
%
%
%
%
%
%
%
%
%
\iffalse
Forges must also allow communication at two stages: the \emph{ex ante}
stage, before the players learn their types, and the \emph{interim}
stage, after players learn their types.  We have communication only at
the interim stage.  This makes our notion more widely applicable; in
some settings, it may not be possible for the players to communicate
before learning their types. 
\fi
%
%
%
%
%
%
%
%
%
%
%
%
%
%
%
%
%
%
%
%
%
%
%
%
%
%
%
%
%

%
\paragraph{Strong Universal Implementation}
Intuitively, $(\vec{M},\F')$ universally implements $\F$ if, 
whenever a set of parties want to compute $\F$, then they also
want to run $\vec{M}$ (using $\F'$),
where we take ``wanting to compute $\F$ (resp., run $\vec{M}$)''
to be ``it is an equilibrium to use  $\vec{\Mcanon}^\F$ when playing
with the mediator  $\F$ (resp., to use $\vec{M}$ when playing 
with the mediator $\F'$)''.
We now strengthen this notion to also require that
whenever a subset of the players do \emph{not} want to compute $\F$
(i.e., if they prefer to do ``nothing''), then they also do not 
want to run $\vec{M}$, even if all other players do so.  
\nfullv{Recall that $\bot$ denotes the (canonical) machine that does nothing.}
Recall that
if adversary $Z$ uses $\bot$, then it sends no messages and
writes nothing on all the output tapes of players in $Z$.
\BD [Strong Universal Implementation] 
Let $(\vec{M},\F')$ be a \emph{($\G,\Z,p$)-universal implementation of
$\F$ with error $\epsilon$}.  $(\vec{M},\F')$ is a
\emph{strong ($\G,\Z,p$)-implementation of $\F$} if, for all $n \in \IN$,
all games $G \in \G$ with input length $n$, all $Z \in \Z$, if
$\bot$ is a $p(n,\cdot)$-robust best response to $\Mcanon^\F_{-Z}$ in $(G,\F)$
then
$\bot$ is an $\epsilon$-best response to $\vec{M}_{-Z}$ in $(G,\F')$.
%
\iffalse
%
%
%
$\{Z\}$-safe Nash equilibrium in the mediated machine game $(G,\F)$, then  
$(\vec{\bot}_Z, \vec{M}_{-Z})$ is a
%
%
%
$\{Z\}$-safe $\epsilon(n)$-Nash equilibrium in the
mediated machine game $(G,\F')$. 
\fi
\ED
\iffalse
\Rnote{We should also add a paragraph about why we care about NE even
though cheap talk games seemingly are extensive form games. Old
motivation below.} 
Why not go for SPE or Seq:
\BI
\item As argued in Rubinstein, some agent might prefer a ``cheap'' strategy
on the equilibrium path than ``doing good'' outside of the equilibrium path.
In other words, it might requiring a longer description to do well outside
of the equilibrium path and players might therefore choose to act ``irrationally'' outside the equilibrium path (if they believe that they will never end up there).
\item in a sense, threats become \emph{credible} because it is rational
to use a strategy that in general is cheaper.
\EI
\fi
%

%
%
%
%
%
\section{Relating Cryptographic and Game-Theoretic Implementation}
\label{sec:mainthe}

We briefly recall the notion of precise secure computation
\cite{MP06,MP07,GMW}; more details are given in Appendix~\ref{sec:precsec}.
An \emph{$m$-ary functionality} is
specified by a random process that maps vectors of inputs to 
vectors of outputs (one input and one output for each player). 
That is, formally, 
$f = \union_{n=1}^\infty f^n$, where 
$f^n:((\bitset^n)^m  \times \bitset^\infty) \rightarrow (\bitset^n)^m$
and $f^n (f^n_1, ...,f^n_m)$.  
We often abuse notation and suppress the random bitstring $r$, writing
$f^n(\vec{x})$ or $f(\vec{x})$.  (We can think of $f^n(\vec{x})$ and
$f_i^n(\vec{x})$ as random variables.)
A machine profile $\vec{M}$ \emph{computes $f$} if
for all $n \in N$, all inputs $\vec{x} \in (\bitset^n)^m$ the output
vector of the players after an execution of $\vec{M}$ on input $\vec{x}$ 
(where $M_i$ gets input $x_i$)
is identically distributed to $f^n(\vec{x})$.

As usual, the security of 
a
protocol $\vec{M}$ for computing a function
$f$ is defined by comparing the \emph{real execution} of $\vec{M}$ with 
an \emph{ideal execution} where all players directly talk to a 
trusted third party
(i.e., a mediator) computing $f$. 
Roughly speaking, a machine profile $\vec{M}$ \emph{securely computes}
an $m$-ary functionality $f$ if the real execution of $\vec{M}$
emulates the ideal execution. This is
formalized by requiring that for every real-execution adversary $A$, 
there exists an ideal-execution adversary $\tilde{A}$, called the
\emph{simulator}, such that the real execution of $\vec{M}$ with $A$
is emulated by the ideal execution with $\tilde{A}$.
The job of $\tilde{A}$ is to provide 
appropriate inputs to the trusted party, and
to reconstruct the view of $A$ in the real execution.
This should be done is such a way that no \emph{distinguisher} $D$
can distinguish the outputs of the parties, and the view of the adversary, in the real
and the ideal execution. 
(Note that in the real execution, the view of the adversary is
simply the actual view of $A$ in the execution, whereas in the ideal execution we refer to 
the reconstructed view output by $\tilde{A}$).

The traditional notion of secure
computation \cite{gmw87} requires only that 
the \emph{worst-case} complexity (size and running-time) of
$\tilde{A}$ is 
polynomially related to that of $A$.
Precise secure computation \cite{MP06,MP07} additionally
requires that the running time of the simulator $\tilde{A}$ ``respects''
the running time of the adversary $A$ in an ``execution-by-execution''
fashion: a secure computation is said to have precision $p(n,t)$ if the
running-time of the simulator $\tilde{A}$ (on input security parameter
$n$) is bounded by $p(n,t)$ whenever $\tilde{A}$ outputs a view in which the
running-time of $A$ is $t$. 

In this work we introduce a weakening of the notion of precise secure
computation. The formal definition is given in Appendix
\ref{sec:weakprec}. We here 
highlight some key differences: 
\beginsmall{itemize}
\item The standard definition requires the existence of a simulator for
every $A$, such the real and the ideal execution cannot be distinguished
given any set of inputs and any distinguisher. In analogy with the work
of Dwork, Naor, Reingold, and Stockmeyer~\citeyear{magic}, we change the
order of the 
quantifiers.  We simply require that given any adversary, any input
distribution and any distinguisher, there exists a simulator that tricks
that particular distinguisher, except with probability $\epsilon(n)$; 
$\epsilon$ is called the error of the secure computation.
\item The notion of precise simulation requires that the simulator
\emph{never} exceeds its precision bounds. We here relax this assumption and
let the simulator exceed its bound with 
probability $\epsilon(n)$.
\endsmall{itemize}

\medskip

Additionally, we generalize this notion to consider arbitrary complexity
measures $\complex$ (instead of just running-time),
and to consider \emph{general adversary structures} \cite{HirMau00}
(where the specification of a secure computation includes 
a set $\Z$ of subsets of players such that the adversary is
allowed to corrupt only
the players in one of the subsets in
$\Z$; in contrast, in \cite{gmw87,MP06} only
\emph{threshold adversaries} are considered,
where $\Z$ consists of all subsets up to a pre-specified size $k$.)

We will be most interested in secure computation protocols with 
precision $p$ where $p(n,0)=0$; such functions are called \emph{\pnatural}.
In our setting (where only the
machine $\bot$ has complexity 0), 
this property essentially says that the adversary running $\bot$
in the ideal execution (i.e., aborting---not sending any messages to the
trusted third party and
writing nothing on all the output tapes of players in $Z$) must be a valid
simulator of the adversary running $\bot$ in the real execution.%
\footnote{Although this property traditionally is not required 
by standard definitions of security, all ``natural'' simulators 
satisfy it.} 
Note that the we can always regain the ``non-precise'' notion of secure
computation by instantiating $\complex(M,v)$ with 
the sum of 
the \emph{worst-case} running-time
of $M$ (on inputs of the same length as the input length in $v$) and size of $M$.
Thus, by the classical results of \cite{bgw,GMR,gmw87} it directly follows
that 
there exists weak
$\complex$-precise secure computation protocols with precision
$p(n,t)=poly(n,t)$ 
(where $p$ is {\pnatural})
when $\complex(M,v)$ is 
the sum 
of the worst-case running-time
of $M$ and size of $M$. 
The
results of \cite{MP06,MP07} extend to show the existence of weak
$\complex$-precise secure computation protocols with precision
$p(n,t)=O(t)$ 
(where $p$ is {\pnatural})
when $\complex(M,v)$ is a 
linear
combination of the running-time
(as opposed to just worst-case running-time)
of $M(v)$ and size of $M$. See Appendix \ref{sec:compeq} for more details. 

\subsection{Equivalences}
As a warm-up, we show that ``error-free'' secure
computation, also known as \emph{perfectly-secure computation}
\cite{bgw}, already 
implies the traditional game-theoretic notion of implementation
\cite{F90} (which does not consider computation). To do this, we first
formalize the traditional game-theoretic notion using our notation: Let
$\vec{M}$ be an $m$-player profile of machines. We say that $(\vec{M},
\F')$ is a \emph{traditional game-theoretic implementation} of $\F$ if
$(\vec{M}, \F')$ is a $(\G^{\sf nocomp}, \{ \{1\}, \ldots \{m\}\},
0)$-universal implementation of $\F$ with 0-error, where $\G^{\sf
nocomp}$ denotes the class of all $m$-player canonical machine games
where the utility functions do not depend on the complexity
profile. (Recall that the traditional notion does not consider
computational games or coalition games.) 

We say that a mediator $\F$ \emph{computes the $m$-ary functionality
$f$} if 
for all $n \in N$ and all inputs $\vec{x} \in (\bitset^n)^m$, the output
vector of the players after an execution of $\vec{\Mcanon}$ on input
$\vec{x}$ and with the  
mediator $\F$ is distributed identically to $f(\vec{x})$.

\BP \label{perfectseceq.thm}
Suppose that $f$ is an $m$-ary functionality, 
$\F$ is a
mediator that computes $f$, and
$\vec{M}$ is a machine profile.
Then $(\vec{M}, \C)$ is a traditional game-theoretic implementation of $\F$
if $\vec{M}$ is a (traditional) perfectly-secure computation of $\F$.
\EP
%
\iffalse
Recall that the traditional notion of secure computation implies weak $\complex$-precise secure computation, 
when $\complex(M,v)$ is 
the sum 
of the worst-case running-time
of $M$ and size of $M$. 
Thus, by Proposition~\ref{perfectseceq.thm}, traditional secure computation with $0$-error (a.k.a. \emph{perfect secure computation}) implies traditional game-theoretic implementation.
\medskip
\fi
\BPRF 
We start by showing that running $\vec{M}$ is a Nash equilibrium if
running $\vec{\Mcanon}^{\F}$ with mediator $\F$ is one. 
Recall that the cryptographic notion of error-free secure
computation requires that for every player $i$ and every ``adversarial''
machine 
$M'_i$ controlling player $i$, there exists a ``simulator'' machine $\tilde{M}_i$, such that
the outputs of all players in 
the execution of $(M'_i, \vec{M}_{-i})$
are \emph{identically distributed} to the output of the players
in the execution of $(\tilde{M}_i, \vec{\Mcanon}^{\F}_{-i})$ with
mediator $\F$.\footnote{The follows from the fact that 
perfectly-secure
computation is error-free.} 
%
%
\iffalse
the following two experiments are
identically distributed: The first experiment considers an 
execution of $(M'_i, \vec{M}_{-i})$.
%
%
%
The second experiment considers an 
execution of $(\tilde{M}_i, \vec{\Mcanon}^{\F}_{-i})$ with mediator $\F$.
%
%
%
%
%
\fi
In game-theoretic terms, this means
that every ``deviating'' strategy $M'_i$ in the communication game
can be mapped into a deviating strategy $\tilde{M}_i$ in the mediated game
with the same output distribution for each type, and, hence, the same
utility, since the utility depends only on the type and the 
output distribution; this follows since we require universality only
with respect to games in $\G^{\sf nocomp}$.  
Since no deviations in the mediated 
game can give higher utility than the Nash equilibrium strategy of using $\Mcanon_i^{\F}$, running $\vec{M}$ must also be a Nash equilibrium. 

It only remains to show that $\vec{M}$ and $\vec{\Mcanon}^{\F}$ induce
the same action distribution; this follows directly from the definition
of secure computation by considering an adversary that does not corrupt
any parties.
\EPRF

We now relate our computational notion of implementation to
precise secure computation. 
Our goal is to show that weak precise secure computation is equivalent to 
strong
$\G$-universal implementation for certain natural classes $\G$ of games.
We consider games that satisfy a number of natural restrictions.
If $G =([m], \M,\Pr, \vec{\complex},\vec{u})$ is a 
canonical game with input length $n$,
then 
\nshortv{
$G$ is 
\emph{$\complex$-\gnatural} if $G$ is
(1) \emph{machine universal}: the machine set $\M$ is the set of
Turing machines; 
(2)  \emph{normalized}: the range of $u_Z$ is $[0,1]$ for all subsets
$Z$ of $[m]$;
(3) \emph{monotone}:
for all subset $Z$ of $[m]$, all type profiles $\vec{t}$, action profiles
$\vec{a}$, and all complexity profiles $(c_{Z},\vec{c}_{-Z})$,
$(c'_{Z},\vec{c}_{-Z})$, if $c'_{Z} > c_{Z}$, then
$u_Z(\vec{t}, \vec{a}, (c'_{Z},\vec{c}_{-Z})) \leq u_i(\vec{t}, \vec{a},
(c_{Z},\vec{c}_{-Z}));$
(4) a $\vec{\complex'}$-game:  $\complex_Z=\complex'_Z$ for
all subsets $Z$ of $[m]$.  
}%
\nfullv{
\begin{enumerate}
\item $G$ is \emph{machine universal} if the machine set $\M$ is the set of Turing machines.
\item $G$ is \emph{normalized} if the range of $u_Z$ is $[0,1]$ for all subsets $Z$ of $[m]$; 
\item $G$ is \emph{monotone} if,
for all subset $Z$ of $[m]$, all type profiles $\vec{t}$, action profiles
$\vec{a}$, and all complexity profiles $(c_{Z},\vec{c}_{-Z})$,
$(c'_{Z},\vec{c}_{-Z})$, if $c'_{Z} > c_{Z}$, then
$u_Z(\vec{t}, \vec{a}, (c'_{Z},\vec{c}_{-Z})) \leq u_i(\vec{t}, \vec{a},
(c_{Z},\vec{c}_{-Z})).$
\item $G$ is a $\vec{\complex'}$-game if $\complex_Z=\complex'_Z$ for
all subsets $Z$ of $[m]$.  
\end{enumerate}
}%
Let $\G^{\vec{\complex}}$ denote the class of 
\nfullv{
machine-universal, normalized, monotone, canonical 
$\vec{\complex}$-games. 
}%
\nshortv{
$\complex$-{\gnatural} games.
}%

Up to now we have placed no constraints on the complexity function.  We
do need some minimal constraints for our theorem. For one direction of
our equivalence results (showing that precise secure computation implies
universal implementation), we require 
that honestly running the protocol
should have constant complexity,
and that it be the same with and without a mediator.
More precisely, we say that a complexity function
$\complex$ is $\vec{M}$-\emph{acceptable} if, for every subset $Z$, the
machines $(\Mcanon^\F)^b_Z$ and $M^b_Z$ have the same complexity $c_0$ for all inputs; that is,
$\complex_Z((\Mcanon^\F)^b_Z,\cdot) = c_0$ and $\complex_Z(M^b_Z,\cdot)
= c_0$.%
\footnote{Our results continue to hold if
$c_0$ is a function on the input length $n$ but otherwise does not
depend on the 
view.}
Note that an assumption of this nature is
necessary in order to show that $(\vec{M},\C)$ is a
$(\G^{\vec{\complex}}, \Z, p)$-universal implementation of $\F$.  
As mentioned earlier,
if the complexity of $\vec{M}$ is higher than that of
$\vec{\Mcanon}^\F$, then we can construct a game $G$ 
such that it is an
equilibrium to run $\vec{\Mcanon}^\F$ in $(G,\F)$, but running $\vec{M}$ is
too costly. 
The assumption that $\vec{M}$ and
$\vec{\Mcanon}^\F$ have the same complexity is easily satisfied when 
considering coarse complexity function (where say the first $T$ steps 
of computation are free).
Another way of satisfying this assumption is to consider a
complexity function 
that simply charges $c_0$ for the use of the mediator, where $c_0$ is the complexity of running the protocol. Given 
this view, universal implementation requires only that players want to run $\vec{M}$ as long as they are willing to pay $c_0$ in complexity for talking to the mediator.
(We remark that we can weaken the assumption that $\vec{M}$ and
$\vec{\Mcanon}^\F$ 
have exactly the same complexity, and allow them to have approximately
equal complexities,
by 
considering universality with respect to a slightly more restricted
class of games. We omit the details.) 
For the other direction of our equivalence (showing that universal
implementation implies precise secure computation),
we 
require that certain 
operations, like moving output from one tape to another, 
do not incur any additional complexity.
Such complexity functions are called \emph{\cnatural}; we
provide a formal definition at the beginning of
Section~\ref{sec:proof}.  
%
\iffalse
\nfullv{
%
%
%
%
%
%
%
More precisely, we say that a complexity 
%
%
%
%
%
%
function $\complex$ is 
%
%
%
%
%
%
\emph{{\cnatural}} if,
%
%
%
%
%
%
%
%
%
%
%
%
%
%
%
for every set $Z$ of players, there exists some canonical player
%
%
$i_Z \in Z$ such that the following two conditions hold.
\BE 
\item
%
%
%
%
For every machine $M_Z$ controlling the players in $Z$,
%
%
there exists some machine $M'_Z$ with the same complexity as $M_Z$ 
such that the output of $M'_Z(v)$ is identical to $M_Z(v)$ except that
player $i_Z$ outputs $y;v$, where $y$ is the output of $i_Z$ in the
execution of $M_Z(v)$ (i.e., $M'_Z$ is identical to $M_Z$ with the only
exception being that player $i_Z$ also outputs the view of $M'_Z$). 

%
%
%
%
%
\item For every machine $M'_Z$ controlling players $Z$,
there exists some machine $M_Z$ with the same complexity as $M_Z$ 
such that the output of $M_Z(v)$ is identical to $M_Z'(v)$ 
%
%
%
%
except if player $i_Z$ outputs $y;v'$ for some $v' \in \bitset^*$ 
in the execution of $M_Z'(v)$.  In that case,
player $i_Z$  outputs only $y$ in the execution by $M_Z(v)$; furthermore
%
$M_Z(v)$ outputs $v'$ on its own output tape. 
%
%
%
%
\EE
%
%
%
We remark that our results would still hold if 
%
%
%
%
%
%
we weakened the two conditions above to allow the complexities to be
close, but not necessarily equal, by allowing some overhead in complexity 
%
%
(at the cost of some slackness in the parameters of the theorem). 
}%
\fi
We can now state the connection between secure computation
and game-theoretic implementation.
%
%
%
\iffalse
We say that a mediator $\F$ \emph{computes the $m$-ary functionality
$f$} if 
%
%
%
for all $n \in N$ and all inputs $\vec{x} \in (\bitset^n)^m$, the output
vector of the players after an execution of $\vec{\Mcanon}$ on input
$\vec{x}$ and with the  
%
%
mediator $\F$ is distributed identically to $f(\vec{x})$.
\fi
%
%
%
%
%
%
%
%
%
%

\BT [Equivalence: Information-theoretic case]\label{thm:thm1}
Suppose that $f$ is an $m$-ary functionality, 
$\F$ is a
mediator that computes $f$, 
$\vec{M}$ is a machine profile that computes $f$,
$\Z$ is a set of subsets of $[m]$, $\vec{\complex}$ is 
an $\vec{M}$-acceptable {\cnatural} complexity function, and
$p$ is a 
{\pnatural} precision function.
Then $\vec{M}$ is a 
weak $\Z$-secure computation of $f$ with $\vec{\complex}$-precision $p$
and $\epsilon$-statistical error ``if and only if'' $(\vec{M},\C)$ is a
strong $(\G^{\vec{\complex}}, \Z, p)$-universal implementation of $\F$
with 
error $\epsilon$.%
\footnote{We put ``if and only if'' in quotes, since the result as
stated is only true if there are finitely many possible adversaries. 
If there are infinitely many possible adversaries, we seem to need to use
slightly different $\epsilon$'s in each direction.  
More precise, in the ``if'' direction, we only show that for every $\epsilon'<\epsilon$, it holds that strong universal implementation with error $\epsilon'$
implies secure computation with error $\epsilon$.
See
Appendix~\ref{sec:proof} for details.} 
\ET

In Appendix \ref{sec:compeq} we provide a ``computational'' analogue of
the equivalence theorem above: 
roughly speaking, we show an equivalence between \emph{computational}
precise secure computation and  
strong universal implementation with respect to ``computationally-efficient''
classes of games (i.e., games where 
the complexity and utility functions can be computed by polynomial-size
circuits, and where we  consider only machines whose computations can be
carried out by
polynomial-size circuits).   
As a corollary of this result, we also obtain a game-theoretic
characterization of the ``standard'' (i.e., ``non-precise'') notion of
secure computation; roughly speaking, we show an equivalence between
secure computation and strong universal implementation with respect to
computationally-efficient 
classes of games 
when considering a complexity function $\complex(M,v)$ that is a
combination of the \emph{worst-case} running-time 
of $M$ and the size of $M$. 
%
\iffalse
where all polynomial-time computation is ``free'' (this
is modeled by simply considering the complexity function 
%
%
$\complex^{\bot}$ that assigns complexity 1 to all machines except
$\bot$).
\fi
%
%
%
%
%
%
%
%

%
%
%
\paragraph{Proof overview}
We now provide a high-level overview of the proof of Theorem
\ref{thm:thm1}.
Needless to say, this oversimplified sketch leaves out many crucial
details which complicate the proof.

\para{Weak precise secure computation implies strong universal implementation.} 
At first
glance, it might seem like  
the traditional notion of secure computation of \cite{gmw87} easily implies  
the notion of universal implementation: 
if there exists some (deviating) strategy $A$ 
in the communication game implementing mediator $\F$ that results in a
different distribution over actions than in equilibrium, then the
simulator $\tilde{A}$ for $A$ could be used to obtain the same
distribution; moreover, the running time of the simulator is within a
polynomial of that of $A$. Thus, it would seem like secure computation implies
that any ``poly''-robust equilibrium can be implemented.
However, 
the utility function in the game considers the complexity of each
execution of the 
computation. So, even if the worst-case running time of $\tilde{A}$
is polynomially related to that of $A$, the utility of 
corresponding executions might be quite different.  This difference may
have a significant effect on the equilibrium.
To make the argument go through we need 
a simulation that preserves complexity in an 
execution-by-execution manner.  This is
exactly what precise zero knowledge \cite{MP06} does. 
Thus, intuitively, 
the degradation in computational robustness by a universal implementation 
corresponds to the precision of a secure computation. 

More precisely, to show that a machine profile $\vec{M}$ is a universal
implementation, we need to show that whenever $\Mcanon$ is a $p$-robust
equilibrium in a game $G$ with mediator $\F$, then $\vec{M}$ is an
$\epsilon$-equilibrium (with the communication mediator $\C$). Our proof
proceeds by contradiction: we show that a deviating strategy $M'_Z$ (for
a coalition controlling $Z$) for the $\epsilon$-equilibrium $\vec{M}$
can be turned into a deviating strategy $\tilde{M}_Z$ for the $p$-robust
equilibrium $\vec{\Mcanon}$. 
We here use the fact that $\vec{M}$ is a weak precise secure computation to find the machine $\tilde{M}_Z$; intuitively $\tilde{M}_Z$ will be the simulator for $M'_Z$. 
The key step in the proof is a method for embedding any coalition
machine game $G$ into a distinguisher $D$
that ``emulates'' 
the role of the utility function in $G$. If done appropriately, this ensures 
that the utility of the (simulator) strategy $\tilde{M}_Z$ is close to
the utility of the strategy $M'_Z$, which contradicts 
the assumption
that
$\vec{\Mcanon}$ is an $\epsilon$-Nash equilibrium. 

The main obstacle in embedding 
the utility function of $G$ into a distinguisher $D$ 
is that the utility of a machine $\tilde{M}_Z$ in $G$ depends not only on the types and actions of the players, but also on the complexity of running $\tilde{M}_Z$. In contrast, the distinguisher $D$ does not get the
complexity of $\tilde{M}$ as input (although it gets its output $v$). On
a high level (and oversimplifying), to get around this problem, we let
$D$ compute the utility \emph{assuming} (incorrectly) that $\tilde{M}_Z$
has complexity $c= \complex(M',v)$ (i.e., the complexity of $M'_Z$ in
the view $v$ output by $\tilde{M}_Z$). 
Suppose, for simplicity, that $\tilde{M}_Z$ is \emph{always} ``precise''
(i.e., it always respects the complexity bounds).\footnote{This is an
unjustified assumption, and in the actual proof we actually need to
consider a more complicated construction.} 
Then it follows that (since the complexity $c$ is always close to the
actual complexity of $\tilde{M}_Z$ in every execution) the utility
computed by $D$ corresponds to the utility of some game $\tilde{G}$ that
is at most a $p$-speed up of $G$.
(To ensure that $\tilde{G}$ is indeed  a speedup and not a
``slow-down'', we need to take special care with simulators that
potentially run faster than 
the adversary they are simulating. 
The monotonicity of 
$G$ helps us to circumvent this problem.) Thus, although we are not
able to embed $G$ into the distinguisher $D$, we can embed a related
game $\tilde{G}$ into $D$. This suffices to show that
$\vec{\Mcanon}$ 
is not a Nash equilibrium in $\tilde{G}$, contradicting the assumption
that $\vec{\Mcanon}$
is a $p$-robust Nash equilibrium. A similar argument can be used to show that $\bot$
is also an $\epsilon$-best response to $\vec{M}_{-Z}$ if $\bot$ is a
$p$-robust best response to $\vec{\Mcanon}_{-Z}$, demonstrating that
$\vec{M}$ in fact is a strong universal implementation. 

\para{Strong universal implementation implies weak precise secure
computation.} 
To show that strong universal implementation implies weak precise secure
computation, we again proceed by contradiction. We show how the existence
of a distinguisher $D$ and an adversary $M'_Z$ that cannot be simulated
by \emph{any} machine $\tilde{M}_Z$ can be used to construct a game $G$
for which $\vec{M}$ is not a strong implementation. 
The idea is to have a utility function that assigns high utility to some
``simple'' strategy $M^*_Z$.  In the mediated game with $\F$, no
strategy can get better utility than $M^*_Z$. On the other hand, in the
cheap-talk game, the strategy $M'_Z$ does get higher utility than
$M^*_Z$. As $D$ 
indeed is a function that ``distinguishes'' a mediated execution from a
cheap-talk game, our approach will be to try to embed the distinguisher
$D$ into the game $G$. 
The choice of $G$ depends on whether $M'_Z = \bot$.
We now briefly describe these games.

If $M'_Z = \bot$, then there is no simulator for the
machine $\bot$ that simply halts. In this case, we construct a game $G$
where
using $\bot$ results in
a utility that is determined by running the distinguisher. (Note that
$\bot$
can be easily identified, since it is the only strategy that has
complexity 0.)  
All other strategies instead get some canonical utility $d$, which is
higher than the utility of $\bot$ in the mediated game.
However,
since $\bot$ cannot be ``simulated'', playing $\bot$ in the
cheap-talk game leads to an even higher utility, contradicting 
the assumption
that
$\vec{M}$ is a universal implementation. 

If $M'_Z \neq \bot$,
we construct a game $G'$
in which each strategy other than $\bot$ gets
a utility that is determined by running the distinguisher.  Intuitively,
efficient strategies (i.e., strategies that have relatively low
complexity compared to $M'_Z$) that 
output views on which the distinguisher outputs 1 with high probability get high utility.
On the other hand, $\bot$
gets a utility $d$ that is at least as good as what the
other strategies 
can get in the mediated game with $\F$. 
This makes $\bot$ a best response in the mediated game;  
in fact, we can define the game $G'$ so that it is actually a
$p$-robust best response. 
However, it is not even an $\epsilon$-best-response in the
cheap-talk game: $M'_Z$ gets higher utility, as it receives a view that
cannot be simulated. (The \cnatural\ condition on the
complexity function $\complex$ is used to argue that $M'_Z$ can output
its view at no cost.) 
\subsection{Universal Implementation for Specific Classes of Games}
\label{sec:utilityassumptions}
Our equivalence result might seem like a negative
results. It demonstrates that considering only rational players 
(as opposed to adversarial players)
does not
facilitate protocol design. Note, however, that for the
equivalence to hold, we must consider implementations
universal with respect to essentially \emph{all} games.  
In many settings, it might be reasonable to
consider implementations universal with respect to only certain
subclasses of games; in such scenarios, universal implementations may be
significantly simpler or more efficient, and may also circumvent
traditional lower bounds. 

We list some natural restrictions on classes of games below,
and discuss how such restrictions can be leveraged in protocol design.
These examples illustrate some of the benefits of a fully
game-theoretic notion of security 
that does not rely on the standard cryptographic simulation paradigm.

\begin{description}
\item {\bf Games with punishment:}
Many natural situations can be described as games where players can
choose actions that ``punish'' an individual player $i$. For instance,
this punishment can represent the cost of being excluded from future
interactions. 
Intuitively, games with punishment model situations where players do not 
want to be caught cheating. 
Punishment strategies (such as the grim-trigger strategy in repeated
prisoner's dilemma, where a player defects forever once his opponent
defects once \cite{Axelrod})
are
extensively used in the game-theory literature. 
We give two examples where cryptographic protocol
design is facilitated when requiring only implementations that are
universal with respect to games with punishment. 

\medskip
\noindent {\em Efficiency:}
As observed by Malkhi et al.~\citeyear{fairplay}, and more recently formalized by Aumann
and Lindell \citeyear{aumannlindell}, in situations were players do not want
to be caught cheating, it is easier to construct efficient
protocols. 
Using our framework we can formalize this intuition in a straightforward way. 
For concreteness, consider classes of normalized 2-player games, where 
the expected utility for both players in every Nash equilibrium is $\beta$, but where 
player $1-i$ receives payoff $\alpha< \beta$ if player $i$ outputs the string {\sf punish}.
Assume first 
that $\alpha = 0$ and $\beta = 1/2$, that is, the Nash equilibrium
strategy gives utility 1/2 but 
a ``punished'' player gets 0. 
In such a setting, it is sufficient to come up with protocols which
guarantee that a cheating player is caught with probability $1/2$; to
prevent cheating simply ``punish'' a player that gets caught cheating.  
It follows that the utility of cheating is bounded by $1/2 \times 1 + 1/2 \times 0 = 1/2$.
By the same argument it follows that for general $\alpha, \beta$, it is
sufficient to have a protocol where a cheating player gets caught with
probability $\frac{1-\beta}{1-\alpha}$.  

\medskip
\noindent {\em Fairness:}
It is well-known that, for many functions, secure 2-player computation 
where both players receive output is impossible 
if we require \emph{fairness} (i.e., that either both or neither of the
players receives an output) \cite{goldreich03}.
Such impossibility results can be easily circumvented by considering
universal implementation with respect to games with punishment.
This follows from the fact that although it is impossible to get
secure computation with fairness, the weaker notion of
secure computation with \emph{abort} \cite{GL02} is achievable. 
Intuitively, this notion guarantees
that the only attack possible is one where one of the players prevents
the other player from  
getting its output; this is called an ``abort''.
To get a universal implementations with respect to games with punishment,
it is 
thus sufficient to use 
any (weak) precise secure computation protocol with
abort (see \cite{GL02,MP07}) modified so that players
output {\sf punish} if the other player aborts. 
It immediately follows
that a player can never get a higher utility by aborting
(as this will be detected by the other player, and consequently the
aborting player will be punished). This result
can be viewed as a generalization of the approach of
\cite{DHR00}.\footnote{For this application, it is 
not necessary to use our game-theoretic definition of security. An
alternative way to capture fairness in this setting would be to require
security with respect to the standard (simulation-based) definition with
abort, and additionally fairness (but not
security) with respect to rational agents, according to the definition of
\cite{DHR00,HT04}; this approach is similar to the one used by Kol and Naor
\citeyear{KN08}. Our formalization is arguably
more natural.} 

\item {\bf Games with switching cost:}
Unlike perfectly-secure protocols, computationally-secure protocols
protocols inherently have a non-zero error probability. For
instance, secure 2-player computation can be achieved only with
computational security (with non-zero error probability). 
By our equivalence result, it follows that universal implementations
with respect to the most general classes of 2-player games also
require non-zero error probability. This is no longer the case for
implementations universal with respect to restricted classes of games. 

Consider, for example, a scenario where the complexities of $\vec{M}$
and $\vec{\Mcanon}^\F$ are 1, while all other machines (except
$\bot$) have complexity at least 2. 
As mentioned earlier, every game $G$ where $\vec{M}$ is an
$\epsilon$-Nash equilibrium, can be transformed into 
another game $G'$ where $\vec{M}$ is a true Nash equilibrium: $G'$ is
identical to $G$ except that all strategies with complexity different
than 2 are penalized $\epsilon$. 
This penalty can be though of as the cost of searching for a better
algorithm, or the cost of implementing the new algorithm; such a cost
seems reasonable in situations where the algorithm $\vec{M}$ is freely
available to the players (e.g., it is accessible on a web-page), but any
other strategy requires some implementation cost.  
Of course, in such a setting it might seem unreasonable that the
machine $\bot$ (which does nothing) is penalized. Luckily, to get
universal implementations with respect to two-player games, we do not
require a penalty for using $\bot$: it is easily seen that in all
traditional secure computation protocols, playing $\bot$ can never be a
profitable deviation.\footnote{This follows from the fact that $\bot$ is
a  
perfect simulator for the strategy $\bot$.} 
Thus, it is sufficient to penalize all machine with complexity at least 2.

%
\iffalse
Let $\G^{\vec{\complex}, \epsilon}$ denote the class of machine-universal, normalized, monotone, canonical 
$\vec{\complex}$-games where for every $\vec{t},\vec{a}, \vec{c}$ it holds that $u_Z(\vec{t}, \vec{a}, (1,\vec{c}_{-Z}))> u_Z(\vec{t}, \vec{a}, (c',\vec{c}_{-Z})) + \epsilon(n)$ if the input length in $\vec{t}$ is $n$ and $c'>1$. 
In other words, we consider games where all strategies with complexity at least $2$ get a penalty (the switching cost) of $\epsilon(n)$. This penalty can be though of as the cost of searching for a better algorithm, or the cost of implementing the new algorithm; such a cost seems reasonable in situations when the algorithms $\vec{\Mcanon}^\F$ is ``freely'' available to the players (e.g., it is accessible on a web-page), but any other strategy requires some implementation. Note that the machine $\bot$, that does nothing and has complexity 0, of course, does not get a penalty.

Using the same proof as for theorem \ref{} and \ref{} we can now 
turn universal implementations with error $\epsilon$ into ones with 0-error if considering universality with respect to $\G^{\vec{\complex}, \epsilon}$; to do this we only need to make sure that the ``simulator'' $\bot$ for the machine $\bot$ in the cheap-talk game is error free (as the machine $\bot$ does not have a switching cost), but this holds for all known secure computation protocols.
\fi
\item {\bf Strictly monotone games:} 
In our equivalence results we considered monotone games, where
players never prefer to compute more. It is sometimes reasonable to assume 
that players \emph{strictly} prefer to compute less.
We outline two possible advantages of considering universal implementations with respect to strictly monotone games.

\noindent {\em Error-free implementations:}
Considering universality with respect to only strictly monotone games
gives another approach for achieving error-free implementations. This
seems particularly promising if we consider an idealized model where
cryptographic functionalities (such as one-way functions) are modeled as
black-boxes (see, e.g., the random oracle model of Bellare and Rogaway
\citeyear{BR93}), and the complexity function considers the number of
calls to the cryptographic function. Intuitively, if the computational
cost of trying to break the cryptographic function 
(e.g., we need to evaluate the one-way function at least one more time)
is higher than the
expected gain, it is not worth deviating from the
protocol. 

\noindent {\em Fairness}
One vein of research on secure computation considers protocols for achieving
fair exchanges using \emph{gradual-release} protocols (see
e.g., \cite{BN00}). In a gradual-release protocol, the players are
guaranteed that if at   
any point one player ``aborts'', then the other player(s) can compute
the output within a comparable amount of time (say within twice the
time). By making  
appropriate assumptions about the utility of computation, we can ensure
that  
players never have incentives to deviate.  
Intuitively, if the cost of computing $t$ extra steps is positive, even if 
the other player computes, say $2t$, extra steps, it will never be worth
it for a player to abort.\footnote{One extra assumption that is needed
is that  
players prefer to get the output than not getting it, even if they can
trick other players into computing for a long time. This is required to
ensure that 
a player does not prefer to abort and not compute anything while the
other player attempts to compute the output.} 
\end{description}

\section{Conclusion}\label{sec:conclusion}
We have defined a general approach to taking computation into account in
game theory that subsumes previous approaches, and shown a close
connection between computationally robust Nash equilibria and precise
secure computation.  This opens the door to a number of exciting
research directions, in both  secure computation and game theory.  We
briefly describe a few here:
\beginsmall{itemize}
\item 
%
%
%
\iffalse
It is well-known that secure two-player computation of 
%
many 
%
functions, where both players receive output, is impossible 
if we require 
fairness.
\emph{fairness} (i.e., that either both or neither of the
%
%
players receives an output) (see e.g. \cite{goldreich03}). 
%
It seems that by making weak assumptions about the utility of
computation (such as strict monotonicity---i.e., that a player strictly
prefers to do less computation given the same action and type profile),  
\emph{gradual release of secrets} protocols 
%
%
%
%
(e.g., the techniques of Boneh and Naor \citeyear{BN00}) could be used
to obtain the universal 
implementations with fairness (with respect to appropriate subclasses of
games). 
%
%
%
%
%
%
%
%
%
%
%
%
%
\fi
%
We have illustrated some situations where universal implementations
with respect to restricted classes of games facilitates protocol design.
More generally, we believe that combining
computational assumptions with assumptions about utility will be a
fruitful line of research for secure computation. 
For instance, it is conceivable that difficulties associated with
concurrent executability of protocols could be alleviated by making
assumptions regarding the cost of message scheduling;
the direction of Cohen, Kilian, and Petrank \citeyear{CKP01} (where players
who delay messages are themselves punished with delays) seems relevant
in this regard.

\item Our notion of universal implementation uses Nash equilibrium as
solution concept. 
It is well known that in (traditional) extensive form games (i.e., games
defined by a game tree), a Nash equilibrium might prescribe non-optimal
moves at game 
histories that do no occur 
on the equilibrium path. This can lead to ``empty threats'': 
``punishment'' strategies that 
are non-optimal and thus not credible. Many recent works 
on
implementation 
(see e.g., \cite{Gerardi04,ILM08})
therefore focus on 
stronger solution concepts such as \emph{sequential equilibrium} 
\cite{KW82}. We note that when taking computation into account, the
distinction between 
credible and non-credible threats becomes more subtle: the threat of
using 
a non-optimal strategy in a given history might be credible if, for
instance, 
the overall complexity of the strategy is smaller than any strategy that
is optimal 
at every history. Thus, a simple strategy that is
non-optimal off
the equilibrium path might be preferred to a more complicated (and thus
more
costly) strategy that 
performs better off the equilibrium path (indeed, people often
use non-optimal but simple  
``rules-of-thumbs'' when making decisions).
Finding a good definition of empty threats in games with computation
costs seems challenging. 
\item As we have seen, universal implementation is equivalent to a
variant of precise 
secure computation with reversed quantifier. It would be interesting
to find a notion of implementation that corresponds more closely to the
standard definition 
without a change in the order of quantifier; in particular, whereas the
traditional 
definition of zero-knowledge guarantees \emph{deniability} (i.e., the
property 
that the interaction does not leave any ``trace''), the new one does not.
Finding a game-theoretic definition that also captures deniability seems like an interesting question.

\item As we have seen, Nash equilibria do not always exist
in games with computation.  
This leaves
open the question of what the appropriate solution 
concept
is.  
The issue of finding an appropriate solution concept
becomes of particular interest in extensive-form games.
As mentioned above,
it is more standard in such games to consider
refinements of Nash equilibrium such as sequential equilibrium.
Roughly speaking, in a sequential equilibrium, every player
must make a best response at every information set, where an information
set is a set of nodes in the game tree that the agent cannot
distinguish.  The standard assumption in game theory is that the
information set is given
\emph{exogenously}, as part of the description of the game tree.  As
Halpern \citeyear{Hal15} has argued, an exogenously-given
information set does not always represent the information that an agent
actually has.  The issue becomes even more significant in our framework.
While an agent may have information that allows him to distinguish two
nodes, the computation required to realize that they are different may
be prohibitive, and (with computational costs) an agent can rationally
decide not to do the computation.  This suggests that the information
sets should be determined by the machine.  More generally, in defining
solution concepts, it has proved necessary to reason about players'
beliefs about other players' beliefs; now these beliefs will have to
include beliefs regarding computation.  Finding a good model of such
beliefs seems challenging.
\item A natural next step would be to introduce notions of computation
in the epistemic logic.  
There has already been some work in this direction (see, for example,
\cite{HMT,HMV94,Mos}).  We believe that combining the ideas of this
paper with those of the earlier papers will allow us to get, for
example, a cleaner knowledge-theoretic account of zero knowledge than
that given by Halpern, Moses, and Tuttle \citeyear{HMT}.
\item Finally, it would be interesting to use behavioral experiments to,
for example, determine the ``cost of computation'' in various games (such as
the finitely repeated prisoner's dilemma). 
\endsmall{itemize}
\paragraph{Acknowledgments}
The second author wishes to thank Silvio Micali and abhi shelat for 
many exciting discussions about cryptography and game theory,
and Adam Kalai for enligthening discussions.
We also think Tim Roughgarden for encouraging us to think of conditions
that guarantee the existence of Nash equilibrium.

%
\iffalse
\subsection{new TODO's for abstract}
\BI
\item collusions
\item specific utility function : $U_Z = U_G$ - complexity (or - $complexity^c$)
\item specific ZK game.
\item composition of universal implementation
\EI
\fi
%
%
\fullv{\bibliographystyle{chicagor}}
\shortv{\bibliographystyle{plain}}
\bibliography{z,joe}
\newpage
\appendix

\newpage
\noindent {\Large \bf Appendix}

\section{Precise Secure Computation}\label{sec:precsec}
In this section, we review the notion of precise secure computation \cite{MP06,MP07} which
is a strengthening of the traditional notion of secure computation \cite{gmw87}.
We consider a system where
players are connected through secure (i.e., authenticated and private)
point-to-point channels.
We consider a malicious adversary 
that is allowed to corrupt a subset of the $m$ players before the
interaction begins; these players may then 
deviate arbitrarily from the protocol.
Thus, the adversary is \emph{static}; it cannot corrupt players based on
history.  

An \emph{$m$-ary functionality} is
specified by a random process that maps vectors of inputs to 
vectors of outputs (one input and one output for each player). 
That is, formally, 
$f = \union_{n=1}^\infty f^n$, where 
$f^n:((\bitset^n)^m  \times \bitset^\infty) \rightarrow (\bitset^n)^m$
and $f^n (f^n_1, ...,f^n_m)$.  
We often abuse notation and suppress the random bitstring $r$, writing
$f^n(\vec{x})$ or $f(\vec{x})$.  (We can think of $f^n(\vec{x})$ and
$f_i^n(\vec{x})$ as random variables.)
A machine profile $\vec{M}$ \emph{computes $f$} if
for all $n \in N$, all inputs $\vec{x} \in (\bitset^n)^m$ the output
vector of the players after an execution of $\vec{M}$ on input $\vec{x}$ 
(where $M_i$ gets input $x_i$)
is identically distributed to $f^n(\vec{x})$.\footnote{A common relaxation
requires only that the output vectors are statistically close. 
All our results can be modified to apply also to
protocols that are satisfy only such a ``statistical'' notion of
computation.} 
As usual, the security of protocol $\vec{M}$ for computing a function
$f$ is defined by comparing the \emph{real execution} of $\vec{M}$ with 
an \emph{ideal execution} where all players directly talk to a 
trusted third party
(i.e., a mediator) computing $f$. In particular, we require that the outputs
of the players in both of these executions
cannot be distinguished, and additionally that the view of the adversary
in the real execution can be 
reconstructed by the ideal-execution adversary (called the
\emph{simulator}).   
Additionally, \emph{precision} requires that the running-time of the
simulator in 
each run of the ideal execution is closely related to the running time
of the real-execution adversary in the (real-execution) view output by
the simulator.  

\paragraph{\em The ideal execution}
Let $f$ be an $m$-ary functionality.
Let $\tilde{A}$ be a probabilistic polynomial-time 
machine (representing the ideal-model adversary)
and suppose that $\tilde{A}$ controls the players in $Z \subseteq [m]$.
We characterize the \emph{ideal execution of $f$} given adversary
$\tilde{A}$ using a function
denoted $\ideal_{f,\tilde{A}}$ that maps an input vector $\vec{x}$, 
an auxiliary input $z$, and a 
tuple $(r_{\tilde{A}},r_f) \in
({\bit^\infty})^{2}$ (a random string for the adversary $\tilde{A}$ and
a random string for the trusted third party)
to a triple $(\vec{x},\vec{y},v)$, where $\vec{y}$ is the output vector
of the players $1, \ldots, m$,  
and $v$ is the output of the adversary $\tilde{A}$ on its tape
given input 
$(z,\vec{x},r_{\tilde{A}})$, computed
according to the following three-stage process.

In the first stage, each player $i$ receives its input $x_i$.
Each player $i \notin Z$ 
next sends $x_i$ to the
trusted party.
(Recall that in the ideal execution, there is a trusted third party.)
The adversary $\tilde{A}$ determines the value 
$x_i' \in \bitset^{n}$ a player $i \in Z$ sends to the trusted party.
We assume that the system is synchronous, so the trusted party can tell
if some player does not send a message;  
if player $i$ does not send a message
(or if its message is not in $\bitset^n$)
it is taken to have sent 
$0^n$.
Let $\vec{x}'$ be the vector of values received by the trusted party.
In the second stage, the trusted party
 computes $y_i= f_i(\vec{x}',r_f)$ 
and sends $y_i$ to $P_i$ for every $i\in [m]$.
Finally, 
in the third stage, 
each player $i \notin Z$ outputs the value $y_i$ 
received from the trusted party. 
The adversary $\tilde{A}$ determines the output of the players $i \in
Z$.
$\tilde{A}$ finally also outputs an arbitrary value $v$ (which is
supposed to
be the ``reconstructed'' view of the real-execution adversary $A$).
Let $\view_{f,\tilde{A}}(\vec{x},z,\vec{r})$ denote the  
the view of $\tilde{A}$ in this execution. 
Again, we occasionally abuse notation and suppress the random strings,  
writing $\ideal_{f,\tilde{A}}(\vec{x},z)$ and
$\view_{f,\tilde{A}}(\vec{x},z)$;
we can think of  
$\ideal_{f,\tilde{A}}(\vec{x},z)$ and $\view_{f,\tilde{A}}(\vec{x},z)$ 
as random variables.

\paragraph{\em The real execution}
Let $f$ be an $m$-ary functionality, let $\Pi$ be a
protocol for computing $f$, and let $A$ be a machine 
that controls the same set $Z$ of players as $\tilde{A}$.
We characterize the \emph{real execution of $\Pi$} 
given adversary $A$ 
using a function denoted
$\real_{\Pi,A}$ that maps an input vector $\vec{x}$, an auxiliary
input $z$, and a  
tuple $\vec{r} \in
({\bit^\infty})^{m+1-|Z|}$ ($m-|Z|$ random strings for the players not in
$Z$ and a random string for the adversary $A$),
to a triple $(\vec{x},\vec{y},v)$, where $\vec{y}$ is the output of
players $1, \ldots, m$, and $v$ is the view of $A$ that results
from executing protocol $\Pi$ on inputs $\vec{x}$, when players $i \in
Z$ are controlled  by the adversary $A$, who is given auxiliary input $z$.
As before, we often suppress the vector of random bitstrings $\vec{r}$
and write $\real_{\Pi,A}(\vec{x}, z)$.

\medskip
We now formalize the notion of precise secure computation.
For convenience, we slightly generalize the definition of
\cite{MP06} to consider \emph{general adversary structures}
\cite{HirMau00}.  More precisely, we assume that the
specification of a secure computation protocol includes 
a set $\Z$ of subsets of players, where the adversary is
allowed to corrupt 
only
the players in one of the subsets in
$\Z$; the definition of \cite{MP06,gmw87} considers only
\emph{threshold adversaries}
where $\Z$ consists of all subsets up to a pre-specified size $k$.
We first provide a definition of precise
computation in terms of running time, as in \cite{MP06}, although other
complexity 
functions could be used;  we later consider general
complexity functions. 

We use a 
complexity function $\timec$, which, on
input a machine $M$ and a view $v$, 
roughly speaking, 
gives the number of ``computational steps'' taken by $M$
in the view $v$.  In counting computational steps, we assume a
representation of machines such that a machine $M$, given 
as input 
an encoding of another machine $A$ and an input $x$,
can emulate the computation of $A$
on input $x$ with only linear overhead.
(Note that this is clearly the case for ``natural'' memory-based
models of computation. An equivalent representation is
a universal Turing machine that receives the code it is supposed to run
on one input tape.)
In the following definition, we say that a function is \emph{negligible} if
it is  asymptotically smaller than the
inverse of any fixed polynomial. More precisely, a function
$\nu: \IN \rightarrow \IR$ is negligible if, for all $c > 0$, there
exists some $n_c$ such that
$\nu(n)<n^{-c}$ for all $n > n_c$.

Roughly speaking, a computation is secure if the ideal execution cannot
be distinguished from the real execution.  To make this precise, a 
\emph{distinguisher} is used.  
Formally, a distinguisher gets as input a bitstring $z$, a triple
($\vec{x},\vec{y},v)$ (intuitively, the output of either
$\ideal_{f,\tilde{A}}$ or $\real_{\Pi,A}$ on $(\vec{x},z)$ and some
appropriate-length tuple of random strings) and a random string $r$, and
outputs either 0 or 1.
As usual, we typically suppress the random bitstring and write, for example,
$D(z,\ideal_{f,\tilde{A}}(\vec{x}, z))$ 
or $D(z,\real_{\Pi,A}(\vec{x},z))$.

\BD [Precise Secure Computation]
\label{def:precise.securecomp}
Let $f$ be an $m$-ary function, $\Pi$ a protocol computing $f$, $\Z$ 
a set of subsets of $[m]$,  
$p: \IN\times \IN \rightarrow \IN$, and 
$\epsilon: \IN \rightarrow \IR$.
Protocol $\Pi$ is a \emph{$\Z$-secure computation of $f$ with precision
$p$ 
and $\epsilon$-statistical error} 
if, for all $Z \in \Z$ and every
real-model adversary $A$ that controls the players in $Z$, 
there exists an ideal-model adversary $\tilde{A}$, called the 
\emph{simulator}
that controls the players in $Z$
such that, 
for all $n \in N$, all $\vec{x}=(x_1, \ldots, x_m)\in (\bitset^n)^m$,
and all 
$z \in \bitset^*$, 
the following conditions hold:
\begin{enumerate}
\item For every distinguisher $D$,
$$\Big| 
{\Pr}_U [D(z,\real_{\Pi,A}(\vec{x},z)) = 1]  -
{\Pr}_U [D(z,\ideal_{f,\tilde{A}}(\vec{x}, z))] = 1 \Big|
\leq \epsilon(n);$$
\item ${\Pr}_U [\timec(\tilde{A},\view_{f,\tilde{A}}(\vec{x},z)) 
\leq p (n, \timec(A,\tilde{A}(\view_{f,\tilde{A}}(n,\vec{x},z)) ]) =
1$.%
\footnote{Note that the three occurrences of $\Pr_U$ in the first two
clauses represent slightly different probability measures, although this
is hidden by the fact that we have omitted the superscripts.  
The first occurrence of $\Pr_U$ should be $\Pr_U^{m-|Z|+3}$,
should we are taking the probability over the $m+2-|Z|$ random inputs to
$\real_{f,A}$ and the additional random input to $D$; similarly, the
second and third occurrences of $\Pr_U$ should be $\Pr_U^3$.}
\end{enumerate}
$\Pi$ is a \emph{$\Z$-secure computation of $f$ with precision $p$ and
$(T,\epsilon)$-computational error} if it satisfies the two conditions
above with the adversary $A$ and the distinguisher $D$ restricted to
being computable by a TM with running time bounded by $T(\cdot)$.

Protocol $\Pi$ is 
a \emph{$\Z$-secure computation of $f$ with
statistical precision $p$} if there exists some negligible function
$\epsilon$ such that  
$\Pi$ is a $\Z$-secure computation of $f$ with precision
$p$ and $\epsilon$-statistical error.  Finally, protocol $\Pi$ is 
a \emph{$\Z$-secure computation of $f$ with computational precision $p$}
if for every 
polynomial $T$, there exists some negligible function $\epsilon$ such
that  
$\Pi$ is a $\Z$-secure computation of $f$ with precision $p$ and $(T,\epsilon)$-computational error.
\ED
The traditional notion of secure computation is obtained
by replacing condition 2 with the requirement that the worst-case
running-time of $\tilde{A}$ is polynomially related to the worst-case
running time of $A$.

We will be most interested in secure computation protocols with 
precision $p$ where 
$p(n,0)=0$; such functions are called \emph{\pnatural}.
In our setting (where only the
machine $\bot$ has complexity 0), 
this property essentially says that the adversary running $\bot$
in the ideal execution (i.e., aborting---not sending any messages to the
trusted third party and
writing nothing on all the output tapes of players in $Z$) must be a valid
simulator of the adversary running $\bot$ in the real execution.
The following theorems were provided by Micali and Pass \citeyear{MP07,MP06},
using the results of  
\cite{bgw,gmw87}. Let $\Z_t^m$ denote all the subsets of $[m]$
containing $t$ or less elements. 
%
%
\iffalse
\BT \label{thm:MP06} For every $m$-ary function $f$, there exists a
protocol $\Pi$ that $\Z_{\lceil m/3 \rceil-1}^m$-securely computes $f$
with precision $p(n,t)=O(t)$  
and $\epsilon(n)=2^{-n}$-statistical error.
\ET
\fi
%
\BT \label{thm:MP06}
If $f$ is an $m$-player functionality, then
there exists a {{\pnatural}} precision function
$p$ and a protocol $\Pi$ such that $p(n,t)=O(t)$ and $\Pi$
$\Z_{\lceil m/3 \rceil-1}^m$-securely computes
$f$ with precision $p$ and $0$-statistical error. 
\ET
This result can also be extended to more general adversary structures by
relying 
on the results of \cite{HirMau00}.
\iffalse
{By applying the techniques of \cite{MP06} to the protocols of
\cite{hirtmaurer}, we can generalize Theorem~\ref{thm:MP06} as follows: 
\BT Hirt-Maurer, general adversary structure perfect secure comp.
\ET
\fi
We can also consider secure computation of specific 2-party
functionalities. 

\BT \label{thm2:MP06} 
Suppose that there exists an enhanced trapdoor permutation. 
For every $2$-ary
function $f$ where only one party gets an output (i.e., $f_1(\cdot) =
0$), there exists a 
a {\pnatural} precision function $p$ and 
protocol $\Pi$ such that $p(n,t)=O(t)$ and $\Pi$ $\Z_{1}^2$-securely
computes $f$ 
with computational-precision $p$.
\ET
\cite{MP06} also obtains unconditional results (using statistical
security) for the special case of zero-knowledge proofs. We refer the
reader to \cite{MP06,P06} for more details. 

\subsection{Weak Precise Secure Computation}\label{sec:weakprec}
Universal implementation is not equivalent to precise secure
computation, but to a (quite natural) weakening of it.
\emph{Weak precise secure computation},
which we are about to define, differs from precise secure computation in
the following respects:
\beginsmall{itemize}
\item Just as in traditional definition of zero knowledge
\cite{GMR}, precise zero knowledge
requires that for every adversary, there exists a simulator  
that, on all inputs, produces an
interaction that no distinguisher can distinguish from the real
interaction.
This simulator must work for all inputs and all distinguishers.
In analogy with the notion of ``weak 
zero knowledge'' \cite{magic}, we here switch the order of the
quantifiers and require instead that for every input distribution
$\Pr$ over $\vec{x} \in (\bitset^n)^m$ and $z \in
\bitset^*$, 
and every distinguisher $D$,  
there exists a (precise) simulator that ``tricks'' 
$D$; in essence, we allow there to be a different simulator for each
distinguisher. 
As argued by Dwork et al.~\citeyear{magic}, 
this order of quantification is arguably reasonable when
dealing with concrete security.   To show that a computation is secure
in every concrete setting, it suffices to show that, in every concrete
setting (where a ``concrete setting'' is characterized by an input
distribution and the distinguisher used by the adversary), there is a
simulator.  
\item We further weaken this condition 
by requiring only that 
the probability of the distinguisher outputting 1 on a real view
be (essentially) no higher than the probability of
outputting 1 on a simulated view.
In contrast, the traditional definition requires these probabilities to be
(essentially) equal. 
If we think of the distinguisher outputting 1 as meaning that the
adversary has learned some important feature, then we are saying
that the likelihood of the adversary learning an important feature
in the real execution is essentially no higher than that of the 
adversary learning an important feature in the ``ideal'' computation.  
This condition on the distinguisher is in keeping with the standard
intuition of the role of the distinguisher.

\item We allow the adversary and the simulator to depend not only
on the probability distribution, but also on the particular security
parameter $n$ (in contrast, the definition of \cite{magic} is
\emph{uniform}).  
That is why, when considering weak precise secure computation with
$(T,\epsilon)$-computational error, we require that the adversary $A$
and the simulator $D$ be computable by \emph{circuits} of size at
most $T(n)$ (with a possibly different circuit for each $n$), rather
than a Turing machine with running time $T(n)$.  
Again, this is arguably reasonable in a concrete setting, where the
security parameter is known. 
\item 
We also allow the computation not to meet the precision bounds with a
small probability.  The obvious way to do this is to change the
requirement in the definition of precise secure computation by replacing
1 by $1-\epsilon$, to get 
$${\Pr}_U [\timec(\tilde{A},\view_{f,\tilde{A}}(\vec{x},z)) 
\leq p (n, \timec(A,\tilde{A}(\view_{f,\tilde{A}}(\vec{x},z)) ] 
\ge 1-\epsilon(n),$$ where $n$ is the input length.  
We change this requirement in two ways.
First, rather than just
requiring that this precision inequality hold for all $\vec{x}$ and $z$,
we require that the probability of the inequality holding be at least
$1-\epsilon$ for all distributions
$\Pr$ over $\vec{x} \in (\bitset^n)^m$ and $z \in
\bitset^*$.
The second difference is to 
add an
extra argument to the distinguisher, 
which tells the distinguisher whether the precision requirement is met.  
In the real computation, we assume that the precision requirement is
always met, thus, whenever it is not met, the distinguisher can
distinguish the real and ideal computations.  We still want the
probability that the distinguisher can distinguish the real and ideal
computations to be at most $\epsilon(n)$.  
For example, our definition disallows a scenario where 
the complexity bound is not met with probability
$\epsilon(n)/2$ and the distinguisher can distinguish the computations
with (without taking the complexity bound into account) with probability
$\epsilon(n)/2$.  
\item In keeping with the more abstract
approach used in the definition of robust implementation, the definition
of weak precise secure computation 
is parametrized by the abstract
complexity measure $\complex$, rather than using $\timec$.
This just gives us a more general definition; we can always 
instantiate $\complexity$ to measure running time.
\endsmall{itemize}
%
\iffalse
Since we allow the distinguisher to be any arbitrary
probabilistic function, we 
obtain a notion of \emph{statistical} secure computation. Although it 
is well-known that many two-player functions cannot be computed with 
%
%
statistical security, the \emph{zero-knowledge proof} function 
is an exception, 
%
%
where the \emph{zero-knowledge proof function for
$L$} is the function $f^{L,zk}=(f_1,f_2)$
such that $f_1(x_1,x_2) = \bot$,
%
and
$f_2(x_1, x_2) = 1$
if $x_1=x_2$ and  
$x_1 \in L$ and $0$ otherwise. 
%
%
%
%
%
%
%
Statistically precise zero-knowledge proofs are known for
some nontrivial languages.  For example,
%
%
Micali and Pass 
%
\citeyear{MP06} show
how to obtain 
%
%
%
%
precise zero-knowledge protocols
with error $f_\epsilon(m)=2^{-m}$ and
``time-complexity''-precision $p(m,t) =2t$  
%
%
%
%
%
for
{\sf Graph Non-Isomorphism} and {\sf Quadratic
Non-Residuosity}.
%
\fullv{
\footnote{Interestingly, we do not know whether the
language {\sf Graph Isomorphism} has a precise zero-knowledge proof.} 
}
\fi

%
%
%
%
%
%

%

\BD [Weak Precise Secure Computation]
\label{def:weak.precise.securecomp}
Let $f$, $\Pi$, $\Z$, $p$, and $\epsilon$ be as in the definition of
precise secure computation.
Protocol $\Pi$ is a \emph{weak $\Z$-secure computation of $f$ with
and $\epsilon$-statistical error}
if,
for all $n \in N$, all $Z \in \Z$, all real-execution adversaries $A$
that control the players in $Z$,
all distinguishers $D$, 
and all probability distributions $\Pr$ over $(\bit^n)^{m}\times \bit^*$,
there exists an ideal-execution adversary $\tilde{A}$ 
that controls the players in $Z$
such that
$$\begin{array}{ll}
{\Pr}^+ (\{(\vec{x},z): D(z, \real_{\Pi,A}(\vec{x},z),1) = 1\})\\
\ \ - \, {\Pr}^+ (\{(\vec{x},z): D(z, \ideal_{f,\tilde{A}}(\vec{x},
z), {\sf precise}_{Z,A,\tilde{A}}(n,\view_{f,\tilde{A}}(\vec{x},z)))
= 1\})  
 \leq \epsilon(n),
\end{array}$$
where ${\sf precise}_{Z,A,\tilde{A}}(n,v) = 1$ if and only if
$\complex_Z(\tilde{A},v) \leq p (n,
\complex_Z(A,\tilde{A}(v)))$.%
\footnote{Recall that $\Pr^+$ denotes the product of $\Pr$ and 
$\Pr_U$ (here, the first $\Pr^+$ is actually $\Pr^{+(m+3-|Z|)}$, while
the second is $\Pr^{+3}$).}
$\Pi$ is a \emph{weak $\Z$-secure computation of $f$ with precision $p$ and
$(T,\epsilon)$-computational error} if it satisfies the two conditions
above with the adversary $A$ and the distinguisher $D$ restricted to
being computable by a randomized circuit of size $T(n)$. 
Protocol $\Pi$ is a \emph{$\Z$-weak secure computation of $f$ with
statistical $\vec{\complex}$-precision $p$} if there exists some
negligible function $\epsilon$ such that $\Pi$ is a $\Z$-weak secure
computation of $f$ with precision $p$ and statistical $\epsilon$-error. 
Finally, Protocol $\Pi$ is a \emph{$\Z$-weak secure computation of $f$ with
computational $\vec{\complex}$-precision $p$} if for every polynomial
$T(\cdot)$, there exists some negligible function $\epsilon$ such that
$\Pi$ is a $\Z$-weak secure computation of $f$ with precision $p$ and
$(T,\epsilon)$-computational error. 
\ED

Our terminology suggests that weak precise secure computation is weaker
than precise secure computation.    This is almost immediate from the
definitions if $\complex_Z(M,v) = \timec(M,v)$ for all $Z \in \Z$.  
A more interesting setting considers a complexity measure that can
depend on $\timec(M,v)$ and the size of the description of $M$.
It directly follows by inspection 
that Theorems \ref{thm:MP06} and \ref{thm2:MP06} 
also hold if, for example,
$\complex_Z(M,v) = \timec(M,v) + O(|M|)$ for all $Z \in \Z$, since the
simulators in those results only incur a constant additive overhead.
(This is not a coincidence. As argued in \cite{MP06,P06}, the definition of precise simulation
guarantees the existence of a ``universal'' simulator $S$, with ``essentially'' the same
precision, that works for \emph{every} adversary $A$, 
provided that $S$ also gets the code of $A$; namely given
a real-execution adversary $A$, the ideal-execution adversary
$\tilde{A}=S(A)$.\footnote{This follows by 
considering the simulator $S$ for the universal TM (which receives the code to be executed as auxiliary input).}  
Since $|S|=O(1)$, it follows that $|\tilde{A}| = |S|+|A| = O(|A|)$.)
That is, we have the following variants of Theorems~\ref{thm:MP06} and
\ref{thm2:MP06}:  

\BT 
If $f$ is an $m$-player functionality, 
$\complex(M,v) = \complex_Z(M,v) = \timec(M,v) + O(|M|)$,
then
there exists a {{\pnatural}} precision function
$p$ and a protocol $\Pi$ such that $p(n,t)=O(t)$ and $\Pi$
weak $\Z_{\lceil m/3 \rceil-1}^m$-securely computes
$f$ with $\complex$-precision $p$ and $0$-statistical error. 
\ET

\BT 
Suppose that there exists an enhanced trapdoor permutation, 
and $\complex(M,v) = \complex_Z(M,v) = \timec(M,v) + O(|M|)$.
For every $2$-ary function $f$ where only one party gets an output (i.e., $f_1(\cdot) = 0$), there exists a
a {\pnatural} precision function $p$ and 
protocol $\Pi$ such that $p(n,t)=O(t)$ and $\Pi$ weak $\Z_{1}^2$-securely
computes $f$ with computational $\complex$-precision $p$.
\ET

\section{Proof of Theorem~\protect{\ref{thm:thm1}}}\label{sec:proof}

In this section, we prove Theorem~\ref{thm:thm1}.  Recall that 
for one direction of our main theorem we require that certain 
operations, like moving output from one tape to another, do not
incur any additional complexity.  We now make this precise.

\BD
A complexity
function $\complex$ is 
\emph{{\cnatural}} if,
for every set $Z$ of players, there exists some canonical player
$i_Z \in Z$ such that the following two conditions hold.
\BE 
\item
For every machine $M_Z$ controlling the players in $Z$,
there exists some machine $M'_Z$ with the same complexity as $M_Z$ 
such that the output of $M'_Z(v)$ is identical to $M_Z(v)$ except that
player $i_Z$ outputs $y;v$, where $y$ is the output of $i_Z$ in the
execution of $M_Z(v)$ (i.e., $M'_Z$ is identical to $M_Z$ with the only
exception being that player $i_Z$ also outputs the view of $M'_Z$). 

\item For every machine $M'_Z$ controlling players $Z$,
there exists some machine $M_Z$ with the same complexity as $M_Z$ 
such that the output of $M_Z(v)$ is identical to $M_Z'(v)$ 
except if player $i_Z$ outputs $y;v'$ for some $v' \in \bitset^*$ 
in the execution of $M_Z'(v)$.  In that case,
player $i_Z$  outputs only $y$ in the execution by $M_Z(v)$; furthermore,
$M_Z(v)$ outputs $v'$ on its own output tape. 
\EE
\ED
We remark that our results would still hold if 
we weakened the two conditions above to allow the complexities to be
close, but not necessarily equal, by allowing some overhead in complexity 
(at the cost of some slackness in the parameters of the theorem).

We now prove each direction of Theorem~\ref{thm:thm1} separately, to make
clear what assumptions we need for each part.  
We start with the ``only
if'' direction.  
\iffalse
Before proving it, we provide some 
%
%
intuition  as to \emph{why} the proposition holds.  At first
glance, it might seem like  
the traditional notion of secure computation of \cite{gmw87} easily implies  
the notion of universal implementation: 
%
%
%
%
%
%
%
%
%
%
if there exists some strategy $A$ for the adversary $A$ 
in the communication game implementing mediator $\F$ that results in a
different distribution over actions than in equilibrium, then the
simulator $\tilde{A}$ for $A$ could be used to obtain the same
distribution.  Moreover, the running time of the simulator is within a
polynomial of that of $A$.
%
%
%
%
%
However, this is not good enough for our purposes.  In the utility
function, we consider the complexity of each execution of the
computation, so that even if the worst-case running time of $\tilde{A}$
is polynomially related to that of $A$, the utility of the two
computations might be quite different.
%
%
%
%
%
%
%
%
%
We need 
a simulation that preserves complexity in an 
%
%
execution-by-execution manner.  This is
exactly what the notion of precise zero knowledge of \cite{MP06} does. 
Thus, intuitively, 
the degradation in computational robustness by a universal implementation 
corresponds to the precision of a secure computation. 
(Needless to say, this oversimplified sketch leaves out many crucial
%
%
details that complicate the proof.)
\fi
%
The following result strengthens the ``only if''
direction, by requiring only that $\vec{\complex}$ is  
$\vec{M}$-acceptable
(and not necessarily {\cnatural}).
\BT
\label{prop1}
Let $\vec{M}, f, \F,\Z$ be as in the statement of
Theorem~\ref{thm:thm1}, and let 
$\vec{\complex}$ be an 
$\vec{M}$-acceptable complexity function. 
If $\vec{M}$ is a weak $\Z$-secure computation of $f$ with
$\complex$-precision $p$ and error $\epsilon$, then $(\vec{M},\C)$ is a
strong
$(\G^{\vec{\complex}}, \Z, p)$-universal implementation of $\F$ with
error $\epsilon$. 
\ET%

\BPRF
Suppose that $\vec{M}$ is a weak $\Z$-secure computation of $f$ with
$\vec{\complex}$-precision $p$ and $\epsilon$-statistical error.
Since $\vec{M}$ computes $f$, for every
game $G\in \G^{\vec{\complex}}$, the action profile induced by $\vec{M}$
in  
$(G,\C)$ 
is identically distributed to the action profile induced by
$\vec{\Mcanon}^\F$ in $(G,\F)$.
We now show that 
$(\vec{M}, \C)$ is a $(\G^{\vec{\complex}}, \Z, p)$-universal implementation of $\F$ with error $\epsilon$.
\BCM 
\label{clm1}
$(\vec{M}, \C)$ is a $(\G^{\vec{\complex}}, \Z, p)$-universal implementation of $\F$ with error $\epsilon$.
\ECM
\BPRF
Let $G\in \G^{\vec{\complex}}$ be a 
game with input length $n$ 
such that $\vec{\Mcanon}^\F$ is a $p(n,\cdot)$-robust $\Z$-safe equilibrium in
$(G,\F)$. We show that $\vec{M}$ is a $\Z$-safe $\epsilon(n)$-equilibrium in $(G,\C)$.
Recall that this is equivalent to showing 
that no coalition of players $Z \in \Z$ can increase their utility by more than
$\epsilon(n)$ by deviating from their prescribed strategies.
In other words, for all $Z \in \Z$ and machines $M'_Z$,
we need to show that
$$
U_Z^{(G,\C)}(M'_Z,\vec{M}_{-Z}) \leq
U_Z^{(G,\C)}(M^b_Z,\vec{M}_{-Z}) + \epsilon(n).$$
Suppose, by way of contradiction, that there exists some machine $M'_Z$
such that 
\begin{equation}
\label{eq1}
U_Z^{(G,\C)}(M'_Z,\vec{M}_{-Z}) > 
U_Z^{(G,\C)}(M^b_Z,\vec{M}_{-Z}) + \epsilon(n).
\end{equation}
We now obtain a contradiction by showing that there exists some other
game $\tilde{G}$ that is at most a $p(n,\cdot)$-speedup of $G$ and a machine
$\tilde{M}_Z$ such that 
\begin{equation}
U_Z^{(\tilde{G},\F)}(\tilde{M}_Z,\vec{\Mcanon}^\F_{-Z}) > 
U_Z^{(\tilde{G},\F)}((\Mcanon^\F)^b_{Z}, \vec{\Mcanon}^\F_{-Z}).
\end{equation}
This contradicts 
the assumption
that $\Mcanon^\F$ is a $p$-robust equilibrium.

Note that
the machine $M'_Z$ can be viewed as a real-execution adversary controlling
the players in $Z$. 
The machine $\tilde{M}_Z$ will be defined as the simulator for $M'_Z$
for some appropriately defined input distribution $\Pr$ on $\T \times
\bit^*$ and 
distinguisher $D$. 
Intuitively, $\Pr$ will 
be the type distribution in the game $G$ (where $z$ is nature's type),
and the distinguisher $D$ will capture the
utility function  $u_Z$. 
There is an obvious problem with using the distinguisher to capture the
utility function: the 
distinguisher outputs a single bit, whereas the
utility function outputs a real. 
To get around this problem, we define a probabilistic distinguisher that
outputs 1 with a probability that is determined by the 
expected utility;
this is possible since the game is normalized, so the expected
utility is guaranteed to be in $[0,1]$.
We also cannot quite use the same distribution for the machine $M$ 
as for the game $G$.  The problem is that, if  $G\in \G$ is a canonical game
with input length $n$, the types in $G$ have the form $x;z$, where $x
\in \bit^n$.  The protocol $\Mcanon^\F_i$ in $(G,\C)$ ignores the $z$, and
sends the mediator the $x$.  On the other hand, in a secure computation,
the honest players provide their input (i.e., their type) to the
mediator.  Thus, we must convert a type $x_iz_i$ of a player $i$
in the game $G$ to a type $x$ for $\Mcanon^\F_i$.

More formally, we proceed as follows.  
Suppose that $G$ is a canonical game with input length $n$, and the type
space of $G$ is $\T$.
Given $\vec{t} = (x_1z_1, \ldots, x_nz_n, t_N) \in T$, define
$\vec{t}^D$ by taking  
$t^D_i = x_1z_i$ if $i \in Z$, 
$t^D_i = x_i$ if $i \notin Z$, and 
$t^D_N = t_N;z1;\ldots;zm$.
Say that $(\vec{x},z)$ is
\emph{acceptable} if 
there is some (necessarily unique) 
$\vec{t} \in \T$
$z = t_1^D;\ldots;t_n^D;t_N^D$ and $\vec{x} = (t^D_1, \ldots, t^D_m)$.
If $(\vec{x},z)$ is acceptable, let $\vec{t}_{\vec{x},z}$ be the element
of $T$ determined this way.
If $\Pr_G$ is the probability distribution over types,
$\Pr(\vec{x},z) = \Pr_G(\vec{t}_{\vec{x},z})$ if $(\vec{x},z)$ is
acceptable, and $\Pr(\vec{x},z) = 0$ otherwise.

\begin{sloppypar}
Define the probabilistic distinguisher $D$ 
as follows:
if $\iprecise=0$ or $(\vec{x},z)$ is not acceptable, 
then $D(z,(\vec{x},\vec{y},\iview),\iprecise) = 0$; otherwise
$D(z,(\vec{x},\vec{y},\iview),\iprecise) = 1$ with probability
$u_Z(\vec{t}_{\vec{x},z},\vec{y}, \complex_Z(M'_Z, view), \vec{c_0}_{-Z})$.
\end{sloppypar}

Since we can view $M'_Z$ as a real-execution adversary controlling the
players in $Z$, the definition of weak precise secure computation
guarantees that, for the distinguisher $D$ and the distribution $\Pr$
described above, there exists a simulator $\tilde{M}_Z$
such that
\begin{equation}
\label{eq11}
\begin{array}{ll}
&{\Pr}^+ (\{(\vec{x},z): D(z, \real_{\vec{M},M'_Z}(\vec{x},z),1) = 1\})\\
&\ \ - \, {\Pr}^+ (\{(\vec{x},z): D(z, \ideal_{f,\tilde{M}_Z}(\vec{x},
z), {\sf
precise}_{Z,M'_Z,\tilde{M}_Z}(n,\view_{f,\tilde{M}}(\vec{x},z)) =
1\}) 
\leq \epsilon(n).
\end{array}
\end{equation}
We can assume without loss of generality that if $\tilde{M}_Z$ sends no
messages and outputs nothing, then $\tilde{M}_Z = \bot$. 

We next define a new complexity function $\vec{\tilde{\complex}}$ that,
by construction, will be at most a $p(n,\cdot)$-speedup of
$\vec{\complex}$. Intuitively, 
this complexity function will consider the speedup required to make up
for the ``overhead'' of the simulator $\tilde{M}_{Z}$ 
when simulating $M'_Z$. To ensure that the speedup is not too large, we
incur it only 
on views where 
the simulation by $\tilde{M}_Z$ is ``precise''.
Specifically,
let the complexity function $\vec{\tilde{\complex}}$ be identical to 
$\vec{\complex}$, except that 
if ${\sf precise}_{Z,M'_Z,\tilde{M}_Z}(n,\tilde{v}) = 1$
and $\complex(\tilde{M}_Z,\tilde{v}) \geq \complex(M'_Z,v)$,
where $v$ is the view output by $\tilde{M}_Z(\tilde{v})$,
then 
$\tilde{\complex}_Z(\tilde{M}_Z,\tilde{v}) = \complex_Z(M'_Z,v)$.
(Note that $\tilde{v}$ is a view for the ideal execution.  $M'_Z$ runs in
the real execution, so we need to give it as input the view output by
$\tilde{M}_Z$ given view $\tilde{v}$, namely $v$.  Recall that the
simulator $\tilde{M}_Z$ is trying to reconstruct the view of $M'_Z$.
Also, note that we did not define
$\tilde{\complex}_Z(\tilde{M}_Z,\tilde{v}) = \complex_Z(M'_Z,v)$ if 
$\complex(\tilde{M}_Z,\tilde{v}) < \complex(M'_Z,v)$, for then
$\tilde{\complex}_Z$ would not be a speedup of $\complex_Z$.
Finally, we remark that we rely on the fact that the precision
function $p$ is {\pnatural}, and so $p(n,0) = 0$.  This ensures
that $\tilde{\complex}_Z$ assigns 0 complexity only to $\bot$, as is
required for a complexity function.)
By construction,
$\tilde{\complex}_Z$ is at most a $p(n,\cdot)$-speedup of $\complex_Z$.
Let $\tilde{G}$ be identical to $G$ except that the complexity function
is $\vec{\tilde{\complex}}$. 
It is immediate that 
$\tilde{G}$ is at most a $p(n,\cdot)$-speedup of $G$.  

We claim that it follows from the definition of $D$ that 
\begin{equation}\label{eq:U}
U_Z^{(\tilde{G},\F)}(\tilde{M}_Z,\vec{\Mcanon}^\F_{-Z}) \geq 
{\Pr}^+ (\{(\vec{x},z): D(z, \ideal_{f,\tilde{M}_Z}(\vec{x}, z), {\sf
precise}_{Z,M'_Z,\tilde{M}_Z}(n,\view_{f,\tilde{M}}(\vec{x},z))) = 1\}).
\end{equation}
To see this, let $a_Z(\vec{t},\vec{r})$ (resp.,
$a_i(\vec{t},\vec{r})$) denote the output that $\tilde{M}_Z$ places
on the output tapes of the players in $Z$ (resp., the output of player
$i \notin Z$) when the strategy profile
$(\tilde{M}_Z,\Mcanon^\F_{-Z})$ is used with mediator $\F$, 
the type profile is $\vec{t}$, and 
the random strings are $\vec{r}$.  (Note that these outputs
are completely determined by
$\vec{t}$ and $\vec{r}$.)  Similarly, let
$\iview_{\tilde{M}_Z}(\vec{t},\vec{r})$ and 
$\iview_{\Mcanon^\F_i}(\vec{t},\vec{r})$ and 
denote the views of the adversary and player $i \notin Z$ in
this case. 

Since each type profile $\vec{t} \in T$ is $t_{\vec{x},z}$ for some
$(\vec{x},z)$, and $\Pr_G(t_{\vec{x},z}) = \Pr(\vec{x},z)$, we have
$$\begin{array}{lll}
&U_Z^{(\tilde{G},\F)}(\tilde{M}_Z,\vec{\Mcanon}^\F_{-Z})\\
= &\sum_{\vec{t},\vec{r}} \Pr^+_G(\vec{t},\vec{r})
u_Z(\vec{t}, (a_Z(\vec{t}, \vec{r}),
\vec{a}_{-Z}(\vec{t},\vec{r})),
(\tilde{\complex}_Z(\tilde{M}_Z,\iview_{\tilde{M}_Z}(\vec{t},\vec{r})),
\vec{\tilde{\complex}}_{-Z}(\Mcanon^\F_{-Z},\iview_{-Z}(\vec{t},\vec{r}))))
\\  
=  &\sum_{\vec{x},z,\vec{r}} \Pr^+(\vec{x},z,\vec{r})  
u_Z(\vec{t}_{\vec{x},z}, (a_Z(\vec{t}_{\vec{x},z},\vec{r}),
\vec{a}_{-Z}(\vec{t}_{\vec{x},z},\vec{r})),
(\tilde{\complex}_Z(\tilde{M}_Z,
\iview_{\tilde{M}_Z}(\vec{t}_{\vec{x},z},\vec{r})), 
\vec{c_0}_{-Z})). 
\end{array}
$$
In the third line, we use the fact that
$\tilde{\complex}_i(\Mcanon^\F_i,\iview_i(\vec{t},\vec{r})) =
\complex_i(\Mcanon^\F_i,\iview_i(\vec{t},\vec{r})) =  
c_0$ for all $i \notin Z$, since $\complex$ is $\vec{M}$-acceptable.
Thus, it suffices to show that, for all $\vec{x}$, $z$, and $\vec{r}$, 
\begin{equation}\label{eq:D}
\begin{array}{ll}
  &u_Z(\vec{t}_{\vec{x},z}, (a_Z(\vec{t}_{\vec{x},z},\vec{r}),
\vec{a}_{-Z}(\vec{t}_{\vec{x},z},\vec{r})),
\tilde{\complex}_Z(\tilde{M}_Z, \iview_{\tilde{M}_Z}(\vec{t}_{\vec{x},z},\vec{r})),
\vec{c_0}_{-Z}) \\
\ge &\Pr_U (D(z, \ideal_{f,\tilde{M}_Z}
(\vec{x}, z,\vec{r}), {\sf
precise}_{Z,M_Z',\tilde{M}_Z}(n,\view_{f,\tilde{M}}(\vec{x},
z,\vec{r}))) = 1).
\end{array}
\end{equation}
This inequality clearly holds if ${\sf
precise}_{Z,M_Z',\tilde{M}_Z}(n,\view_{f,\tilde{M}}(\vec{x},z),\vec{r})
= 0$, 
since in that case 
the right-hand side is 0.\footnote{Note that we here rely on the fact
that $G$ is $\complex$-\gnatural\ and hence normalized, so that the range of $u_Z$
is $[0,1]$.} 
Next consider the case when 
${\sf
precise}_{Z,M_Z',\tilde{M}_Z}(n,\view_{f,\tilde{M}_Z}(\vec{x},z,\vec{r})) 
= 1$. In this case, by the definition of $D$, the right-hand side equals
$$ u_Z(\vec{t}_{\vec{x},z}, (a_Z(\vec{t}_{\vec{x},z},\vec{r}),
\vec{a}_{-Z}(\vec{t}_{\vec{x},z},\vec{r})),
(\complex_Z(M'_Z, v_Z(\vec{t}_{\vec{x},z},\vec{r})),\vec{c_0}_{-Z})),$$
where $v_Z(\vec{t}_{\vec{x},z},\vec{r})) = \tilde{M}_Z(\iview_{\tilde{M}_Z}
(\vec{t}_{\vec{x},z},\vec{r}))$ (i.e., the view output by $\tilde{M}_Z$).
By the definition of $\tilde{\complex}$, it follows that when 
$\complex_Z(\tilde{M}_Z, \iview_{\tilde{M}_Z}(\vec{t}_{\vec{x},z},\vec{r}))
\geq \complex_Z(M'_Z, v_Z(\vec{x},z,\vec{r}))$
and 
${\sf
precise}_{Z,M_Z',\tilde{M}_Z}(n,\view_{f,\tilde{M}_Z}(\vec{x},z,\vec{r})) 
= 1$,
then 
$\tilde{\complex}_Z(\tilde{M}_Z,
\iview_{\tilde{M}_Z}(\vec{t}_{\vec{x},z}, \vec{r})) = \complex_Z(M_Z',
v_Z(\vec{t}_{\vec{x},z}))$, 
and (\ref{eq:D}) holds with $\ge$ replaced by =.
On the other hand, when $\complex_Z(\tilde{M}_Z,
\iview_{\tilde{M}_Z}(\vec{t}_{\vec{x},z},\vec{r})) 
< \complex_Z(M'_Z, v_Z(\vec{x},z,\vec{r}))$, then
$\tilde{\complex}_Z(\tilde{M}_Z, \iview_{\tilde{M}_Z}(\vec{t}_{\vec{x},z}, \vec{r})) = \complex_Z(\tilde{M}_Z, \iview_{\tilde{M}_Z}(\vec{t}_{\vec{x},z}, \vec{r}))$, and thus
$\complex_Z(M'_Z, v_Z(\vec{x},z,\vec{r})) >
\tilde{\complex}_Z(\tilde{M}_Z,
\iview_{\tilde{M}_Z}(\vec{t}_{\vec{x},z}, \vec{r}))$; (\ref{eq:D}) then
holds by the monotonicity of $u_Z$.

Similarly, we have that 
\begin{equation}\label{eq:real}
{\Pr}^+(\{(\vec{x},z): D(z, \real_{\vec{M},M'_Z}(\vec{x},z),1) = 1\}) =
U_Z^{(G,\C)}(M'_Z,\vec{M}_{-Z}). 
\end{equation}
In more detail, a similar argument to that for (\ref{eq:U}) shows that
it suffices 
to show that, for all $\vec{x}$, $z$, and $\vec{r}$, 
$$\begin{array}{ll}
&u_Z(\vec{t}_{\vec{x},z}, a_Z(\vec{t}_{\vec{x},z},\vec{r}),
\vec{a}_{-Z}(\vec{t}_{\vec{x},z},\vec{r}),
{\complex}_Z(M'_Z, \iview_{M'_Z} (\vec{t}_{\vec{x},z}, \vec{r})),
{\complex}_{-Z}({M}_{-Z}, \iview_{M_{-Z}}(\vec{t}_{\vec{x},z}, \vec{r})))\\
= &\Pr_U(D(z, \real_{\vec{M},{M}'_Z}(\vec{x},
z,\vec{r}), 1)
= 1),   
\end{array}
$$
where $a_Z(\vec{t},\vec{r})$, $a_i(\vec{t},\vec{r})$,
$\iview_{M'_Z}(\vec{t},\vec{r})$, 
and
$\iview_{M_i}(\vec{t},\vec{r})$ are
appropriately defined outputs and views in an execution of
$(M'_Z,\vec{M'}_Z)$. 
Here we use the fact that $\vec{C}$ is $\vec{M}$-acceptable, so that 
${\complex}_{-Z}({M}_{-Z}, \iview_{M_{-Z}}(\vec{t}_{\vec{x},z},
\vec{r})) = \vec{c_0}_{-Z}$.

It now follows immediately from (\ref{eq11}), (\ref{eq:U}), and
(\ref{eq:real}) that  
$$
U_Z^{(\tilde{G},\F)}(\tilde{M}_Z,\vec{\Mcanon}^\F_{-Z}) \geq
U_Z^{(G,\C)}(M'_Z,\vec{M}_{-Z}) - \epsilon(n).
$$
Combined with (\ref{eq1}), this yields
\begin{equation}
\label{eq4}
U_Z^{(\tilde{G},\F)}(\tilde{M}_Z,\vec{\Mcanon}^\F_{-Z}) >
U_Z^{(G,\C)}(M^b_Z,\vec{M}_{-Z}).
\end{equation}
Since $\vec{M}$ and $\F$ both compute $f$ (and thus must have the same
distribution over outcomes), it follows that 
\begin{equation}
\label{eq5}
U_Z^{(G,\C)}(M^b_Z, \vec{M}_{-Z}) = U_Z^{(G,\F)}((\Mcanon^\F)^b_Z,
\vec{\Mcanon}^\F_{-Z}) = U_Z^{(\tilde{G},\F)}((\Mcanon^\F)^b_Z,
\vec{\Mcanon}^\F_{-Z}). 
\end{equation}
For 
the last equality, recall that $\tilde{G}$ is identical to $G$ except
for the complexity of $\tilde{M}_Z$ (and hence the utility of strategy
profiles involving $\tilde{M}_Z$).  Thus, the last equality 
follows
once we show
that $(\Mcanon^\F)^b_Z \neq \tilde{M}_Z$.  This follows from the various
technical assumptions we have made. If
$(\Mcanon^\F)^b_Z = \tilde{M}_Z$, then $\tilde{M}_Z$ sends no messages
(all the messages sent by $\Mcanon^\F$ to the 
communication mediator are ignored, since they have the wrong form), and
does not output anything (since messages from the communication mediator
are not signed by $\F$).  Thus, $\tilde{M}_Z$ acts like $\bot$.  By
assumption, this means that $\tilde{M}_Z = \bot$, so $\tilde{M}_Z \ne 
(\Mcanon^\F)^b_Z$.
From (\ref{eq4}) and (\ref{eq5}), we conclude that
$$
U_Z^{(\tilde{G},\F)}(\tilde{M}_Z,\vec{\Mcanon}^\F_{-Z}) > 
U_Z^{(\tilde{G},\F)}((\Mcanon^\F)^b_Z, \vec{\Mcanon}^\F_{-Z}),$$
which gives the desired contradiction.
\EPRF
\medskip

It remains to show that $(\vec{M}, \C)$ is also a \emph{strong}
$(\G^{\vec{\complex}}, 
\Z, p)$-universal implementation of $\F$ with error
$\epsilon$.  
That is, if $\bot$ is a $p(n,\cdot)$-best response to
$\vec{\Mcanon}^\F_{-Z})$ in  
$(G,\F)$
then $\bot$ an $\epsilon$-best response to $\vec{M}_{-Z}$ in
$(G,\C)$.
Suppose, by way of contradiction, that there exists some $M'_Z$ such that
\begin{equation}
U_Z^{(G,\C)}(M'_Z,\vec{M}_{-Z}) > 
U_Z^{(G,\C)}(\bot,\vec{M}_{-Z}) + \epsilon(n).
\end{equation}
It follows using the same proof as in Claim \ref{clm1} that there exists
a game $\tilde{G}$ that is at most a $p(n,\cdot)$ speedup 
of $G$ and a machine $\tilde{M}_Z$ such that 
\begin{equation}
U_Z^{(\tilde{G},\F)}(\tilde{M}_Z,\vec{\Mcanon}^\F_{-Z}) > 
U_Z^{(\tilde{G},\F)}(\bot,\vec{\Mcanon}^\F_{-Z}).
\end{equation}
But this contradicts the assumption that $\bot$ is a $p(n,\cdot)$-robust best response to
$\vec{\Mcanon}^\F_{-Z}$ in $(G,\F)$.
\EPRF
We now prove the ``if'' direction of Theorem~\ref{thm:thm1}.
For this direction, we need the assumption that $\vec{\complex}$ is
{\cnatural}.
Moreover, we get a slightly weaker implication: we show only that
for every $\epsilon, \epsilon'$ such that $\epsilon'<\epsilon$ it holds that
strong universal implementation with error $\epsilon'$ implies weak
secure computation with error $\epsilon$. After proving this result,
we introduce some additional restrictions on $\complex$
that suffice to prove the implication for the case when 
$\epsilon' = \epsilon$. 

\BT\label{prop2}
Suppose that $\vec{M}, f, \F,\Z$ are as above, $\epsilon'<\epsilon$, 
and $\vec{\complex}$ is
an
$\vec{M}$-acceptable {\cnatural} complexity function. 
If $(\vec{M},\C)$ is a 
strong $(\G^{\vec{\complex}}, \Z, p)$-universal
implementation of $\F$ with error $\epsilon'$, then $\vec{M}$ is a  
weak $\Z$-secure computation of $f$ with $\complex$-precision $p$ and
error $\epsilon$. 
\ET
\BPRF
Let $(\vec{M}, \C)$ be a $(\G^{\vec{\complex}},\Z, p)$-universal
implementation of $\F$ with error $\epsilon(\cdot)$. 
We show that $\vec{M}$ $\Z$-securely computes $f$ with
$\complex$-precision $p(\cdot,\cdot)$ and error $\epsilon(\cdot)$.
Suppose, by way of contradiction, that there exists some $n \in \IN$, a
distribution $\Pr$ on $(\bit^n)^{m}\times\bit^*$, a subset $Z \in \Z$, a
distinguisher $D$, and a machine $M'_Z\in \M$
that controls the players in $Z$
such that for all
machines $\tilde{M}_Z \in \M$, 
\begin{equation}
\label{eq7}
\begin{array}{ll}
{\Pr}^+(\{(\vec{x},z): D(z, \real_{\vec{M},M'_Z}(\vec{x},z),1) = 1\}) \\
\ \ \ \ -\; 
{\Pr}^+ (\{(\vec{x},z): D(z, \ideal_{f,\tilde{M}_Z}(\vec{x}, z), {\sf
precise}_{Z,M'_Z,\tilde{M}_Z}(n,\view_{f,\tilde{M}_Z}(\vec{x},z) = 1))) 
> \epsilon(n).
\end{array}
\end{equation}
To obtain a contradiction we consider two cases: $M'_Z = \bot$ or
$M'_Z \neq \bot$.

\para{Case 1: $M'_Z = \bot$.}
We define a game $G\in \G^{\vec{\complex}}$ such that
$\vec{\Mcanon}^\F_{-Z}$  
is a $p$-robust $\Z$-safe equilibrium in the game $(G,\F)$, 
and show that 
$$U_Z^{(G,\C)}(M'_Z, M_{-Z}) > U_Z^{(G,\C)}(M^b_Z, \vec{M}_{-Z}) + \epsilon(n),$$ 
which contradicts the assumption that $\vec{M}$ is a 
$(\G,\Z,p)$-universal implementation of $\F$.  

Intuitively, $G$ is such that the strategy $\bot$ (which is the only 
one that has complexity 0) gets a utility that is determined by 
the probability with which the distinguisher $D$ outputs 1 
(on input the type and action profile). On the other hand, all other strategies (i.e., all strategies with positive complexity) get the same utility $d$.
If $d$ is selected so that the probability of $D$ outputting 1 in
$(G,\F)$ is at most $d$, it follows that $\vec{\Mcanon}^\F$ is a
$p$-robust Nash equilibrium. 
However, $\bot$ will be a profitable deviation in $(G,\C)$.

In more detail, we proceed as follows.
Let 
\begin{equation}
\label{eq8.1}
d = {\Pr}^+ (\{(\vec{x},z): D(z, \ideal_{f,\bot}(\vec{x}, z), 1) = 1\}).
\end{equation}

Consider the game $G = ([m], \M, \Pr, \vec{\complex}, \vec{u})$,
where $u_{Z'}(\vec{t}, \vec{a}, \vec{c}) = 0$ for all $Z' \neq Z$
and  
$$
u_Z((t_1, \ldots, t_m, t_N), \vec{a}, (c_Z, \vec{c}_{-Z}))
=\left\{
\begin{array}{ll}
{\Pr}_U(D(t_N, ((t_1, \ldots, t_m), \vec{a}, \lambda), 1) =1) &\mbox{if
$c_Z=0$}\\
d &\mbox{otherwise,}
\end{array}\right.$$
where, as before, $\Pr_U$ is the uniform distribution on
$\bitset^\infty$ ($D$'s random string).
It follows from the definition of $D$ (and the fact that only
$\bot$ has complexity 0) that, for all games $\tilde{G}$ that are 
speedups  of $G$ and 
all machines $\tilde{M}_Z$,
$
U_Z^{(\tilde{G},\F)}(\tilde{M}_Z, \vec{\Mcanon}^\F_{-Z}) \le d$.

Since $M^b_Z \neq \bot$
(since $\complex$ is $\vec{M}$-acceptable and 
$\vec{M}$ has complexity $c_0>0$),
we conclude that $\vec{\Mcanon}^\F$ is a $p$-robust $Z$-safe Nash equilibrium.
In contrast (again since $M^b_Z \neq \bot$),
$
U_Z^{(G,\C)}(M^b_Z, \vec{M}_{-Z}) = d.$
But, since $M'_Z = \bot$ by assumption, we have
$$
\begin{array}{lll}
&U_Z^{(G,\C)}(\bot, \vec{M}_{-Z}) \\ = 
&U_Z^{(G,\C)}(M_Z', \vec{M}_{-Z}) \\ 
= &{\Pr}^+(\{(\vec{x},z): D(z, \real_{\vec{M},M'_Z}(\vec{x},z),1) =
1\}) \\ > &d + \epsilon(n) 
&\mbox{[by (\ref{eq7}) and (\ref{eq8.1})],}
\end{array}
$$
which is a contradiction.

\para{Case 2: $M'_Z \neq \bot$.}
To obtain a contradiction, we first show that, 
without loss of generality, $M'_Z$ lets one of the players in $\Z$ 
output the view of $M'_Z$.
Next, we define a game $G\in \G^{\vec{\complex}}$ such that 
$\bot$ is a $p(n,\cdot)$-best response to $\vec{\Mcanon}^\F_{-Z}$ in
$(G, \F)$. 
We then show that
$$
U_Z^{(G,\C)}(M'_Z, M_{-Z}) \geq U_Z^{(G,\C)}(\bot, \vec{M}_{-Z}) +
\epsilon(n),$$    
which contradicts the assumption that $\vec{M}$ is a strong
$(\G,\Z,p)$-universal implementation of $\F$ with error $\epsilon' < \epsilon$.
To prove the first step, note that 
by the first condition in the definition of {\cnatural},
there exists some canonical player $i_Z \in Z$ and a machine $M''_Z$ 
controlling the players in $Z$ with the same complexity as $M'_Z$ such that 
the output of $M''_Z(v)$ is identical to $M'_Z(v)$, except that
player $i_Z$ outputs $y;v$, where $y$ is the output of $i_Z$ in the
execution of $M'_Z(v)$.
We can obtain a counterexample with $M''_Z$ just as well as with
$M'_Z$
by considering the distinguisher $D'$
which is defined identically to $D$, except that if $y_{i_Z} = y;v$, then 
$D'(z, (\vec{x},\vec{y}, v), precise) = D(z, (\vec{x},\vec{y'}, v),
precise)$, where $\vec{y}'$ is identical to $\vec{y}$ except that
$y'_{i_Z} = y$. 
Consider an adversary $\tilde{M}'_Z$. We claim that
\begin{equation}
\label{eq71}
\begin{array}{ll}
{\Pr}^+(\{(\vec{x},z): D'(z, \real_{\vec{M},M''_Z}(\vec{x},z),1) = 1\}) \\
\ \ \ \ -\; 
{\Pr}^+ (\{(\vec{x},z): D(z, \ideal_{f,\tilde{M}'_Z}(\vec{x}, z), {\sf
precise}_{Z,M'_Z,\tilde{M}'_Z}(n,\view_{f,\tilde{M}'_Z}
=(\vec{x},z)))  =   1\}) > \epsilon(n).
\end{array}
\end{equation}
By definition of $D$, it follows that 
\begin{equation}\label{eq72}
\begin{array}{ll}
{\Pr}^+(\{(\vec{x},z): D(z, \real_{\vec{M},M'_Z}(\vec{x},z),1) = 1\}) 
= {\Pr}^+(\{(\vec{x},z): D'(z, \real_{\vec{M},M''_Z}(\vec{x},z),1) = 1\}).
\end{array}
\end{equation}
By the second condition of the definition of \cnatural,
there exists a machine $\tilde{M}_Z$ 
with the same complexity as $\tilde{M}'_Z$ such that 
the output of $\tilde{M}_Z(v)$ is identical to $M'_Z(v)$ except that
if player $i_Z$ outputs $y;v'$ for some $v \in \bitset^*$ in the execution
by $M'_Z(v)$, then 
it outputs only $y$ in the execution by $\tilde{M}_Z(v)$; furthermore
 $\tilde{M}_Z(v)$ outputs $v'$.
It follows that 
\begin{equation}\label{eq73}
\begin{array}{ll}
&{\Pr}^+ (\{(\vec{x},z): D(z, \ideal_{f,\tilde{M}'_Z}(\vec{x}, z), {\sf 
precise}_{Z,M''_Z,\tilde{M}'_Z}(n,\view_{f,\tilde{M}'}(\vec{x},z))) = 1\}) 
\\= 
&{\Pr}^+ (\{(\vec{x},z): D'(z, \ideal_{f,\tilde{M}_Z}(\vec{x}, z), {\sf
precise}_{Z,M'_Z,\tilde{M}_Z}(n,\view_{f,\tilde{M}}(\vec{x},z))) = 1\}). 
\end{array}
\end{equation}
Equation (\ref{eq71}) is now immediate from (\ref{eq7}), (\ref{eq72}), and
(\ref{eq73}).  
Thus, we can assume without loss of generality that $M'_Z$ is such that,
when $M'_Z$ has view $v$,
the output of
player $i_Z$ is $y; v$. 
We now define a game $G\in \G^{\vec{\complex}}$ such that
$\bot$ is a $p(n,\cdot)$-robust best response to $\vec{\Mcanon}^\F_{-Z}$
in $(G,\F)$, but  
$\bot$ is not an $\epsilon$-best response to $\vec{M}_{-Z}$ in $(G,\C)$.
Intuitively, $G$ is such that simply playing
$\bot$
guarantees a high payoff. However, if a coalition
controlling $Z$ can provide a view that cannot be ``simulated'', then it
receives an even higher payoff. 
By definition, it will be hard to find such a view in the mediated
game. However, by our assumption that $\vec{M}$ is not a secure computation
protocol, it is possible for the machine controlling $Z$ to obtain such
a view in the cheap-talk game. 

In more detail, we proceed as follows.
Let 
\begin{equation}
\label{eq8}
d = \sup_{\tilde{M_Z} \in \M}
{\Pr}^+ (\{(\vec{x},z): D(z, \ideal_{f,\tilde{M}_Z}(\vec{x}, z), {\sf
precise}_{Z,M'_Z\tilde{M}_Z}(n,\view_{f,\tilde{M}_Z}(\vec{x},z))) = 1\}).
\end{equation}
Consider the game $G = ([m], \M, \Pr, \vec{\complex}, \vec{u})$,
where $u_{Z'}(\vec{t}, \vec{a}, \vec{c}) = 0$ for all $Z' \neq Z$
and  
$$
u_Z(\vec{t}, \vec{a}, \vec{c})
=\left\{
\begin{array}{ll}
d &\mbox{if $c_Z = 0$}\\
{\Pr}_U(D(t_N, ((t_1, \ldots, t_m), \vec{a}, v), 1) =1) &\mbox{if
$a_{i_Z}=y;v$, $0 <  p(n,c_Z) \le p(n,\complex_Z(M'_Z,v))$}\\
0 &\mbox{otherwise.}
\end{array}\right.$$
Clearly, 
$G \in \G^{\vec{\complex}}$.

\BCM  
$\bot$ is a 
$p(n,\cdot)$-robust best response to $\vec{\Mcanon}^\F_{-Z}$ in $(G,\F)$.
\ECM
\BPRF 
Suppose, by way of contradiction, that there exists some game
$\tilde{G}$ 
with complexity profile $\vec{\tilde{C}}$ 
that is at most a $p(n,\cdot)$-speedup of $G$ and a machine
$M^*_Z$  
such that
\begin{equation}\label{eq17}
U_Z^{(\tilde{G},\F)}(M^*_Z,\vec{\Mcanon}^\F_{-Z}) > 
U_Z^{(\tilde{G},\F)}(\bot, \vec{\Mcanon}^\F_{-Z}).
\end{equation}
It is immediate from (\ref{eq17}) that $M_Z^* \ne \bot$.  
Thus, it follows from the definition of complexity functions that
$\tilde{\complex}(M_Z^*,v) \ne 0$ for all $v \in \bit^*$.
That means that when calculating
$u^{(\tilde{G},\F)}(M^*_Z,\Mcanon^\F_{-Z})$, 
the second or third conditions in
the definition of $u_Z$ must apply.  
Moreover, the second condition applies on type $(\vec{x},z)$ only if
$a_{i_Z}$ has the form $y;v$ 
and $0 < p(n,\tilde{\complex}_Z(M_Z^*,\iview_{M_Z^*}(\vec{x},z)) \le
p(n,\complex_Z(M'_Z,v))$.  Since $\tilde{\complex}$ is at most a
$p$-speedup of $\complex$, the latter condition implies that 
$0 < \complex_Z(M_Z^*,\iview_{M_Z^*}(\vec{x},z)) \le
p(n,\complex_Z(M'_Z,v))$.  
Hence,
$U_Z^{(\tilde{G},\F)}(M^*_Z,\vec{\Mcanon}^\F_{-Z})$ is at most
%
%
\iffalse
But, since $\vec{\complex}$ is $\vec{M}$-strongly natural,
%
and since a computation that has positive complexity cannot 
be sped up to have complexity 0, 
this can happen only 
%
%
%
%
if $M^*_Z$ is either $(\Mcanon^\F)^b_{Z}$ or $\vec{M}$, or a variant of them 
required to satisfy clause 2(a) or 2(b) in the definition of strongly natural.
%
%
%
In fact, if $M'_Z$ has complexity 0, and player $i_Z$ ever
In fact, if $M^*_Z$ has complexity 0, and player $i_Z$ ever
%
%
%
%
%
%
%
outputs a $n$-bit string, then $i_Z$ must always output an $n$-bit
%
string; but in this case, the utility of $M^*_Z$ will always be
$d$. 
\fi
%
%
%
%
%
%
%
%
%
%
%
%
%
%
%
%
%
%
%
%
%
%
%
%
$${\Pr}^+ (\{(\vec{x},z): D(z, \ideal'_{f,M^*_Z}(\vec{x}, z), {\sf
precise}_{Z,M'_Z,M^*_Z}(n,\view'_{f,M^*_Z}(\vec{x},z))) = 1\}),$$ 
where
 $\ideal'$ is defined identically to $\ideal$, 
except that $y_{i_{Z}}$ 
(the output of player $i_Z$) is parsed as $y;v$, and 
$\view'_{f,M^*_Z}(\vec{x},z)))$ is $v$.

Since $\complex_Z(\bot,v) = 0$ for all $v$, the definition of $u_Z$
guarantees 
that $U_Z^{(\tilde{G},\F)}(\bot, \vec{\Mcanon}^\F_{-Z}) = d$.  
It thus follows from (\ref{eq17}) that
$$U_Z^{(\tilde{G},\F)}(M^*_Z,\vec{\Mcanon}^\F_{-Z}) > d.$$
Thus,
$${\Pr}^+ (\{(\vec{x},z): D(z, \ideal'_{f,M^*_Z}(\vec{x}, z), {\sf
precise}_{Z,M'_Z,M^*_Z}(n,\view_{f,M^*_Z}(\vec{x},z))) = 1\}) > d.$$ 
The second condition of the definition of {\cnatural} complexity implies
that there must 
exist some $M^{**}_Z$ such that
$${\Pr}^+ (\{(\vec{x},z): D(z, \ideal_{f,M^{**}_Z}(\vec{x}, z), {\sf
precise}_{Z,M'_Z,M^{**}_Z}(n,\view_{f,M^{**}_Z}(\vec{x},z)) = 1\})
> d,$$ 
which contradicts (\ref{eq8}).
Thus, $\vec{\Mcanon}^\F$ is a $p$-robust $\Z$-equilibrium
of $(G,\F)$.
\EPRF
Since $(\vec{M},\C)$ is a strong $(\G^{\vec{\complex}}, \Z, p)$-universal
implementation of $\F$ with error $\epsilon$, and
$\bot$ is a $p(n,\cdot)$-robust best response to $\vec{\Mcanon}^\F_{-Z}$ in
$(G,\F)$, it must be
the case that $\bot$ is an $\epsilon$-best response to $\vec{M}_{-Z}$ in $(G,\C)$.
However, by definition of $u_Z$, we have that
$$
\begin{array}{ll}
&U_Z^{(G,\C)}(M'_Z, \vec{M}_{-Z}) \\ 
= &\sum_{\vec{t},\vec{r}} \Pr^+(\vec{t},\vec{r})
u_Z(\vec{t}, M'_Z(\iview_{\tilde{M}_Z}(\vec{t},\vec{r})), 
\vec{M}_{-Z}(\iview_{\vec{M}_{-Z}}(\vec{t},\vec{r})),
\complex_Z(M'_Z,
\iview_{\tilde{M}_Z}(\vec{t},\vec{r}),
\vec{\complex}_{-Z}(\vec{M}_{-Z},
\iview_{\vec{M}_{-Z}}(\vec{t},\vec{r})))\\
= &\Pr^+(\{(\vec{x},z): D(z, \real_{f,M'_Z}(n,\vec{x}, z), 1) =
1\}) \\ 
\end{array}
$$
where $\iview_{\tilde{M}_Z}(\vec{t},\vec{r})$ (resp., $\iview_{\vec{M}_{-Z}}(\vec{t},\vec{r})$)
denotes the view of $\tilde{M}_Z$ when the strategy profile
$(\tilde{M}_Z,\vec{M}_{-Z})$ is used with mediator $\C$.
The second equality follows from the fact that player $i_Z$ outputs the
view of $M'_Z$. %
Recall that (\ref{eq7}) holds (with strict inequality) for \emph{every}
machine $\tilde{M}_Z$. It follows that 
\begin{equation}
\label{eq10}
\begin{array}{ll}
&U_Z^{(G,\C)}(M'_Z, \vec{M}_{-Z}) \\ 
= &{\Pr}^+(\{(\vec{x},z): D(z, \real_{\vec{M},M'_Z}(\vec{x},z),1) = 1\}) \\
\geq &\sup_{\tilde{M}\in M}{\Pr}^+ (\{(\vec{x},z): D(z, \ideal_{f,\tilde{M}_Z}(\vec{x}, z), {\sf
precise}_{Z,M'_Z,\tilde{M}_Z}(n,\view_{f,\tilde{M}_Z}(\vec{x},z) = 1))) 
+ \epsilon(n) \\
= &d + \epsilon(n) 
\end{array}
\end{equation}
where the last equality follows from (\ref{eq8}).

Since $U_Z^{(G,\C)}(\bot, \vec{M}_{-Z}) = d$,
this is a contradiction. 
This completes the proof of the 
theorem.
\EPRF
\medskip
\noindent
Note that if the set 
$$S=\{ {\Pr}^+ (\{(\vec{x},z): D(z, \ideal_{f,\tilde{M}_Z}(\vec{x}, z), {\sf
precise}_{Z,M'_Z,\tilde{M}_Z}(n,\view_{f,\tilde{M}_Z}(\vec{x},z))) = 1) : \tilde{M} \in \M \}$$ 
has a maximal element $d$, then by (\ref{eq7}), equation (\ref{eq10})
would hold with strict inequality, and thus theorem \ref{prop2} would hold  
even if $\epsilon' = \epsilon$.
We can ensure this by introducing some additional technical (but
natural) restrictions on $\complex$. For instance, suppose that
$\complex$ is such that  
for every complexity bound $c$, the number of machines that have
complexity at most $c$ is finite, i.e., for every $c \in \IN$, there
exists some constant $N$ such that $|\{M \in \M : \exists v \in
\bitset^* \; \complex(M,v)\leq c\}| = N$. Under this assumption $S$ is
finite and thus has a maximal element. 

\section{A Computational Equivalence Theorem}\label{sec:compeq}
To prove a ``computational'' analogue of our equivalence theorem (relating computational precise secure computation and universal implementation), we need to introduce some further restrictions on the complexity functions, and the classes of games considered.
\BI
\item A (vector of) complexity functions $\vec{\complex}$ is 
\emph{efficient} if each function is computable by a (randomized)
polynomial-sized circuit. 
\item A secure computation game $G =([m], \M,\Pr,
\vec{\complex},\vec{u})$ with input length $n$ is said to be
\emph{$T(\cdot)$-machine universal} if  
\BI
\item the machine set $\M$ is the set of Turing machines implementable by $T(n)$-sized randomized circuits, and 
\item $\vec{u}$ is computable by a $T(n)$-sized circuits. 
\EI
Let $\G^{\vec{\complex}, T}$ denote the class of $T(\cdot)$-machine
universal, normalized, monotone, canonical $\vec{\complex}$-games.  
\EI

\BT [Equivalence: Computational Case]\label{thm:eqcomp}
Suppose that $f$ is an $m$-ary functionality, $\F$ is a
mediator that computes $f$, 
$\vec{M}$ is a machine profile that computes $f$,
$\Z$ is a set of subsets of $[m]$, $\vec{\complex}$ is 
an $\vec{M}$-acceptable {\cnatural} complexity function, and
$p$ is a 
{\pnatural} efficient precision function.
Then $\vec{M}$ is a 
weak $\Z$-secure computation of $f$ with computational $\vec{\complex}$-precision $p$
if and only if, for every polynomial $T$, there exists some negligible
function $\epsilon$ 
such that $(\vec{M},\C)$ is a
strong $(\G^{\vec{\complex},T}, \Z, p)$-universal implementation of $\F$
with error $\epsilon$. 
\ET

\BPRF
We again separate out the two directions of the proof, to show which
assumptions are needed for each one.

\BT
Let $\vec{M}, f, \F,\Z$ be as above, and let $\vec{\complex}$ be an 
$\vec{M}$-acceptable natural {\cnatural} efficient complexity function, and
$p$ a {\pnatural} precision function.
If $(\vec{M},\C)$ is a weak $\Z$-secure computation of $f$ with
computational $\complex$-precision $p$, then for every polynomial $T$,
there exists some negligible function $\epsilon$ such that $\vec{M}$ is
a strong $(\G^{\vec{\complex},T}, \Z, p)$-universal implementation of
$\F$ with error $\epsilon$. 
\ET
\BPF
The proof follows the same lines as that of
Theorem \ref{prop1}.  
Assume that $\vec{M}$ computes $f$ 
with computational $\vec{\complex}$-precision $p$.
Since $\vec{M}$ computes $f$, it follows that for every polynomial $T$
and game $G \in \G^{\vec{\complex}, T}$, the action profile
induced by $\vec{M}$ in $(G,\C)$ is identically distributed to the
action profile induced by 
$\vec{\Mcanon}^\F$ in $(G,\F)$.
We now show that, for every polynomial $T$, there exists some negligible
function $\epsilon$ such that $(\vec{M}, \C)$ is a $(\G^{\vec{\complex},
T}, \Z, p)$-universal implementation of $\F$ with error
$\epsilon$. Assume, by way of contradiction, that 
there exists polynomials $T$ and $g$ and infinitely many $n\in N$ such
that 
the following conditions hold: 
\BI
\item there exists some game $G\in \G^{\vec{\complex},T}$ with input length $n$
such that $\vec{\Mcanon}^\F$ is a $p(n,\cdot)$-robust $\Z$-safe equilibrium in
$(G,\F)$;
\item there exists some machine $M'_Z \in \M$ such that 
\begin{equation}
U_Z^{(G,\C)}(M'_Z,\vec{M}_{-Z}) > 
U_Z^{(G,\C)}(M^b_Z,\vec{M}_{-Z}) + \frac{1}{g(n)}.
\end{equation}
\EI
It follows using the same proof as in Proposition \ref{prop1} that this
contradicts the weak secure computation property of $\vec{M}$. In fact,
to apply this proof, we only need to make sure that the distinguisher $D$
constructed can be implemented by a polynomial-sized circuit. However,
since by our assumption $\vec{\complex}$ is efficient and 
$\vec{u}$ is $T(\cdot)$-sized computable, it follows that $D$
can be constructed efficiently. Strong universal implementation follows
in the same way. 
\EPF
\BT
Let $\vec{M}, f, \F,\Z$ be as above, let $\vec{\complex}$ be a
$\vec{M}$-acceptable {\cnatural} efficient complexity function, and let
$p$ be an efficient homogeneous precision function. 
If, for every polynomial $T$, there exists some negligible function
$\epsilon$ such that $(\vec{M},\C)$ is a $(\G^{\vec{\complex},T}, \Z,
p)$-universal implementation of $\F$ with error $\epsilon$, then
$\vec{M}$ is a  
weak $\Z$-secure computation of $f$ with computational
$\complex$-precision $p$. 
\ET
\BPF
Assume by way of contradiction that there exist polynomials
$T$ and $g$, infinitely many $n\in N$, a distribution $\Pr$
on $(\bit^n)^{m}\times\bit^*$, a subset $Z \in \Z$, a 
$T(n)$-sized distinguisher $D$, and a $T(n)$-sized machine $M'_Z\in \M$
that controls the players in $Z$
such that for all 
machines $\tilde{M}_Z$,
\begin{equation}
\begin{array}{ll}
{\Pr}^+(\{(\vec{x},z): D(z, \real_{\vec{M},M'_Z}(\vec{x},z),1) = 1\}) \\
\ \ \ \ -\; 
{\Pr}^+ (\{(\vec{x},z): D(z, \ideal_{f,\tilde{M}_Z}(\vec{x}, z), {\sf
precise}_{Z,M'_Z,\tilde{M}_Z}(n,\view_{f,\tilde{M}_Z}(\vec{x},z) = 1))) 
> \frac{1}{g(n)}.
\end{array}
\end{equation}

As in the proof of Theorem \ref{prop2} we construct a game $G$ that
contradicts the assumption 
that $\M$ is a strong universal implementation.
The only thing that needs to be changed is that we need to prove that the game $G$ constructed is in $\G^{\vec{\complex},T'}$ for some polynomial $T'$.
That is, we need to prove that $\vec{u}$ can be computed by poly-sized
circuits (given than $D$ is poly-size computable). We do not know how to
show that the actual utility function $u_Z$ constructed in the proof of
Proposition \ref{prop2} can be made efficient. However, for each
polynomial $g'$, we can approximate it to within an additive factor of
$\frac{1}{g'(n)}$ using polynomial-sized circuits (by repeated
sampling); this is 
sufficient to show that there exists some $T'$ and some polynomial $g''$
such that that $\vec{M}$ is not a $\frac{1}{g''(n)}$-equilibrium in
$(G,\C)$. 
\EPF
\EPRF
\paragraph{Relating Universal Implementation and ``Standard'' Secure Computation}
We note that Theorem \ref{thm:eqcomp} also provides a game-theoretic
characterization of the ``standard'' (i.e., ``non-precise'') notion of
secure computation.
We simply consider the complexity function
$\worstcase(v)$ that is the sum of the
worst-case running-time
of $M$ on inputs of the same length as in the view $v$, and the size of
$M$. 
With this complexity function, the definition of weak precise secure computation reduces to the traditional notion 
of weak secure computation without precision
(or, more precisely, with ``worst-case'' precision just as in the traditional
definition). Given this complexity function, the precision of a secure
computation protocol becomes the traditional ``overhead'' of the
simulator (this is also called \emph{knowledge tightness} \cite{gmw1}). 

%
%
\iffalse
We simply consider the complexity function
$\complex^{\bot}$ that assigns complexity 1 to all machines (except to
$\bot$). With this complexity function, the ``precision''
requirement in the definition of weak precise secure computation 
%
%
%
%
%
is satisfied almost vacuously, and the definition reduces to the
traditional notion 
of weak secure computation without precision.\footnote{The
definition is not completely vacuous, since we still require  
that that the adversary running $\bot$
in the ideal execution (i.e., aborting---not sending any messages to the
trusted third party and
writing nothing on all the output tapes of players in $Z$) must be a valid
%
simulator of the adversary running $\bot$ in the real execution. As
%
%
%
%
mentioned before, although this property is not part of the
standard definition of security, all natural protocols in fact satisfy
it.} 
\fi

%
Roughly speaking, 
``weak secure computation'' with overhead $p$ (where $p$ is a homogeneous
function) is thus equivalent to 
strong $(G^{\vec{\worstcase},poly},p)$-universal implementation with
negligible error. 
\end{document}